\documentclass[twocolumn,preprintnumbers,amsmath,amssymb,pra]{revtex4}

\usepackage{graphicx}
\usepackage{amssymb}
\usepackage{esvect}
\usepackage{enumerate}
\usepackage{mathptmx}      
\usepackage{latexsym}

\begin{document}

\newtheorem{defi}{Definition}
\newtheorem{assu}{Assumption}
\newtheorem{state}{Statement}
\newtheorem{hypo}{Hypothesis}

\title{Evolution equations from an epistemic treatment of time}

\author{Per \"{O}stborn}
\affiliation{Division of Mathematical Physics, Lund University, S--221 00 Lund, Sweden}

\begin{abstract}
Relativistically, time $t$ may be seen as an observable, just like position $r$. In quantum theory, $t$ is a parameter, in contrast to the observable $r$. This discrepancy suggests that there exists a more elaborate formalization of time, which encapsulates both perspectives. Such a formalization is proposed in this paper. The evolution is described in terms of sequential time $n\in \mathbf{\mathbb{N}}$, which is updated each time an event occurs. Sequential time $n$ is separated from relational time $t$, which describes distances between events in space-time. There is a space-time associated with each $n$, in which $t$ represents the knowledge at time $n$ about temporal relations. The evolution of the wave function is described in terms of the parameter $\sigma$ that interpolates between sequential times $n$. For a free object we obtain a Stueckelberg equation $\frac{d}{d\sigma}\Psi(r_{4},\sigma)=\frac{ic^{2}\hbar}{2\langle \epsilon\rangle}\Box\Psi(r_{4},\sigma)$, where $r_{4}=(r,ict)$. Here $\sigma$ describes the time $m$ passed from the start of the experiment at time $n$ and the observation  at time $n+m$. The parametrization is assumed to be natural in the sense that $\frac{d}{d\sigma}\langle t\rangle=1$, where $\langle t\rangle$ is the expected temporal distance between the events that define $n$ and $n+m$. The squared rest energy $\epsilon_{0}^{2}$ is proportional to the eigenvalue $\tilde{\sigma}$ that describes a ‘stationary state’ $\Psi(r_{4},\sigma)=\psi(r_{4},\tilde{\sigma})e^{i\tilde{\sigma}\sigma}$. The Dirac equation follows as a ‘square root’ of the stationary state equation from the condition $\tilde{\sigma}>0$, which is a consequence of the directed nature of $n$. The formalism thus implies that all observable objects have non-zero rest mass, including elementary fermions. The introduction of $n$ releases $t$, so that it can be treated as an observable with uncertainty $\Delta t$.
\end{abstract}

\maketitle

\section{Introduction}
\label{intro}

Special relativity makes it necessary to give up the idea that there is a universally valid measure of temporal intervals. General relativity reinforces this conclusion. It makes the temporal interval between a given pair of events as measured by two observers depend not only on their relative state of motion, but also on their positions in a gravitational field. Mathematically speaking, the theory is generally covariant.

The facts that there is no universal measure of time, and that spatial and temporal coordinates may be mixed in transformations that leave the form of physical law invariant, have led a number of theorists to promote the idea that time should be abandoned altogether as a fundamental concept in physics \cite{barbour,rovelli}.

This idea is gaining traction since no one has yet been able to make the notion of time in general relativity conform with that in quantum theory. In the context of quantum gravity research this is called \emph{the problem of time} \cite{isham,kuchar}. Of course, the simplest way out of this deadlock is to say that there is no time and therefore no problem.

Actually, this is hinted at by a straightforward attempt to express a quantum mechanical evolution equation for the wave function of the entire universe. The result is the Wheeler-DeWitt equation, which has the form of a steady state equation with zero energy, corresponding to a static universe \cite{dewitt}. However, some physicists suspect that this conclusion is invalid, arguing that quantum mechanics can be applied only to proper subsets of the world, never to the universe as a whole \cite{smolin}. The same conclusion is reached in a recent reconstruction of quantum mechanics from epistemic principles \cite{ostborn1}.

However, it is hard to make the notions of time in relativity and quantum theory go together even if we restrict our interest to experimental contexts with limited size. Time $t$ may be called an \emph{observable} in relativity theory, since it depends on the state of motion and position of the observer who measures its value. More precisely, the difference $\Delta t=t’-t$ between two points $(r,t)$ and $(r',t')$ in a given coordinate system $\{r,t\}$ corresponds to the temporal distance between two events that an observer moving along a given world-line is able to measure. In contrast, time $t$ is a universally defined \emph{parameter} that evolves the state in quantum theory. It cannot be called an observable since it is not associated with any self-adjoint operator, and its value is not subject to any Heisenberg uncertainty.

This discrepancy suggests that time should either be removed from the fundamental description of Nature, or there should exist a more elaborate formalization of the concept of time, which conforms to relativity as well as to quantum theory. In this paper we explore the second possibility.

This possibility was explored already in 1941 by Ernst Stueckelberg \cite{stueckelberg1}. He parametrized trajectories in space-time according to $(r(\lambda),t(\lambda))$ in order to allow world lines that bend back and forth in time $t$ as $\lambda$ increases, thus modelling annihilation and pair production of particles and anti-particles. Such a parametrization may resolve the tension between relativity and quantum theory, since $\lambda$ can play the role of a quantum mechanical evolution parameter, and $t$ the role as a relativistic observable. In 1942 Stueckelberg published a quantum mechanical version of his formalism \cite{stueckelberg2}, introducing a relativistic wave function $\Psi(r_{4},\lambda)$ obeying a covariant evolution equation $d\Psi/d\lambda=A\Psi$, where the temporal component of the four-position $r_{4}$ is allowed to display Heisenberg uncertainty, just like the spatial components. In Stueckelberg's own words \cite{stueckelberg1}:

\begin{quote}
\emph{Le proc\'{e}d\'{e} de quantification de [Schr\"{o}dinger] peut alors \^{e}tre mis sous une forme o\`{u} l'espace et le temps interviennent d'une fa\c{c}on enti\`{e}rement sym\'{e}trique.}
\end{quote}

Several researchers have developed Stueckelberg's formalism further \cite{fanchi1,fanchi2,horwitz1,horwitz2,land}. However, there is still no generally accepted physical motivation for the introduction of the additional evolution parameter $\lambda$. Stueckelberg's original rationale is questionable on the grounds that a continuous particle trajectory $(r(\lambda),t(\lambda))$ that bends back and forth in time by necessity contain sections where it leaves its local light cone. If we forbid such bending, on the other hand, we can write $\lambda=\lambda(t)$, so that $\lambda$ loses its independent role and should be eliminated from the formalism.

Some physicists try instead to reserve an independent role for $\lambda$ as a moving `now' inside a fixed space-time \cite{pavsic}. The parameter $\lambda$ is commonly related to the proper time along the world line of a given particle \cite{fanchi2,fanchi3}. Some researchers try to generalize this notion so that the same invariant $\lambda$ can describe the evolution of many particles, making it similar to the external, absolute time of Newtonian mechanics \cite{horwitz1,horwitz2,horwitz3,land}. Some authors argue that it is possible to construct a clock that measures the value of $\lambda$ \cite{fanchi3}, whereas others argue that $\lambda$ is not observable \cite{horwitz3}. In any case, the parameter $\lambda$ must either be relativistically invariant, or play such a part in the formalism that relativistic transformations of spatio-temporal coordinates do not apply to it.

The starting point in the present paper for a formalization of time that conforms to both relativity and quantum theory is the distinction between time as a directed ordering of events and time as an observed measure of the distance between events.

Theorists who try to remove time from the fundamental formalism often do so since they argue that the second aspect of time loses its absolute meaning in general relativity, as discussed above. However, the more primitive notion of time as an ordering of events does retain its meaning. One may even say that it is built into relativity theory, since the metric has a fixed signature in which one of the four axes of space-time is assigned the opposite sign as compared to the other three. The trajectories of all massive objects are constrained to move within the local light cones in a given direction along this particular axis, whereas they may wiggle back and forth along the other three axes. This feature corresponds to the flow of time, making it possible to order events along a world-line in a linear sequence. As emphasized already by Eddington, relativity makes an absolute distinction between time and space in the sense that the relation between a pair of events is either time-like or space-like. "It is not a distinction between time and space as they appear in a space-time frame, but a distinction between temporal and spatial relations." \cite{eddington}

Some theorists argue that time cannot play any fundamental role in physics since an external clock is needed to measure time \cite{rovelli}. Therefore temporal intervals cannot be defined in the universe as a whole, but only for small parts of the world that are monitored from the outside. However, we will make the case that the more primitive notion of time survives this problem as well, that it is nevertheless possible in principle to order all events in the universe.

Tim Maudlin \cite{maudlin0,maudlin1} argues along similar lines that the directed nature of time and the possibility to order all events along a temporal axis should be taken as a fundamental postulate in the scientific description of the world. He even tries to reconstruct geometry from mathematical postulates based on linear ordering \cite{maudlin2}. Lee Smolin also subscribes to the idea that rather than removing time, we should give it greater emphasis in our attempts to understand the physical world \cite{smolin}. Instead of relying on a sequential ordering of all events to achieve a universal definition of time, he argues that evolving laws of nature create the proper notion of time that is independent of external clocks.

Here we build on the epistemic perspective on time introduced in a recent reconstruction of quantum mechanics \cite{ostborn1}. Sequential time $n$ is updated each time the potential knowledge of some observer changes. From this ansatz it is possible to construct a universal ordering of events. At each time $n$ there is knowledge about a set of present and past events, as well as the spatial and temporal relations between them, quantified by $x$ and $t$. This knowledge may be fuzzy, meaning that relational time $t$ becomes an observable associated with an uncertainty $\Delta t$, in the same way as $x$ is associated with a Heisenberg uncertainty $\Delta x$.

In well-defined experimental contexts $C$ of limited size it is often possible to make adjustments to the experimental setup so that the sequential time $m$ passed between the initiation of the experiment at time $n$ and the collection of results at time $n+m$ changes. These adjustments can be parametrized by a continuous parameter $\sigma$. We get a family of contexts $C(\sigma)$. We express evolution equations as derivatives with respect to $\sigma$ of quantities that describe the state of the experimental context. In so doing we are able to get a new perspective on the Dirac equation and the operators that are associated with observables in quantum mechanics. The parameter $\sigma$ plays a similar role in the present formalism as the parameter $\lambda$ in Stueckelberg's theory \cite{stueckelberg1,stueckelberg2,fanchi1,fanchi2,horwitz1,horwitz2,land,pavsic,fanchi3,horwitz3}.

From the physical perspective, we identify relational time $t$ as the aspect of time used in relativity theory, whereas the evolution parameter $\sigma$ associated with sequential time $n$ is the aspect of time used to express quantum mechanical evolution equations.

From the philosophical perspective, relational time $t$ encodes the temporal relations known at a certain time $n$ between present events and memories of past events, or between different memories of past events. To vary sequential time $n$ means to transcend the knowledge about the world at a certain time. Therefore $n$ is adequate to express the physical laws responsible for the evolution of the physical state from one moment of time to the next. This conceptually coherent since physical law by definition transcends the individual physical states it applies to.

What we do, in essence, is to expand the fundamental physical description of the `now' so that it contains information not only about the state of the world at that given moment $n$, but also partial information about the past. Such an expansion is justifiable only if we adopt an epistemic approach to physics, in which the present physical state of the world corresponds to the present knowledge about the world. That knowledge contains information about the past in the form of memories. In contrast, in a realistic model of the world, the information about the past in the form of memories is just a function of the present physical state of the brain \cite{hartle}. The memories become secondary, and the past does not become a necessary part of the description of the present.

From the subjective point of view the expanded notion of the `now' is the more natural one. Suppose that you listen to music. The appreciation of harmonies, and the emotional response they give rise to in the present, depends crucially on memories of sounds in the immediate past, to the extent that the music would cease to exist without these memories. That is, each present state of the listener contains both the present and the past in a crucial way; each fleeting `now' encoded by $n$ can be unfolded to an entire temporal axis $t$. At the formal level of physical description, the very perception of a sound at a given moment relies on sensory recording during an extended period of time, since such a temporal interval of time is needed to determine the frequencies that define the sound that we hear at a given moment. 

The aim of this paper is demonstrate that this perspective makes it possible to formalize the concept of time in such a way that the physical formalism becomes easier to motivate and more coherent. For example, the Dirac equation emerges from first principles, as well as the familiar momentum and energy operators. In addition to the conjugate pairs of quantities $(x,p)$ and $(t,E)$ it adds the pair $(\sigma,\epsilon_{0}^{2})$, where $\epsilon_{0}$ is the rest energy. The evolution parameter $\sigma$ is closely associated to the flow of sequential time $n$. The claimed fact that $n$ corresponds to a universal ordering of events is mirrored by the fact $\epsilon_{0}$ is a universal, observer independent number that can be associated to any object. The fact that the flow of time $n$ never stops is mirrored by the fact that in the present picture there is no object with zero rest energy.

At a more general level, this paper is part of a series that explore the consequences of the use of a certain set of epistemic assumptions as the basis of our physical model of the world. In the first paper of this series \cite{ostborn1}, the basic formalism of quantum mechanics was derived from such assumptions, and it was claimed that the conceptually well-defined starting point enables a better understanding than usual of the components of the formalism, and its domain of validity. In the present paper, it is claimed that the same epistemic approach makes it possible to understand better the roles played by time in our present physical models, and to resolve some open problems related to time. More papers in this series will follow. The aim of the overall project is thus to show that the philosophical approach I have chosen is physically fruitful, that the strict epistemic assumptions might be used to resolve some open physical problems and give clearer motivations for some established physical facts.

However, the aim is not to prove that this philosophical approach is right. I think this is impossible in principle. Rather, the efforts are motivated by the simple idea that we should choose the philosophical approach that is able to explain the most physics with the least number of assumptions. A credible approach should explain every aspect of quantum theory in its full generality, and possibly extend it, so that new predictions can be extracted. What I have done therefore is to mount the epistemic horse to ride it as far as I can along that road. Other people may mount the de Broglie-Bohm horse, the Many-worlds horse - or some other horse in the stable - to see if they can go even further. This set of papers may be seen as a challenge to those who prefer other philosophical approaches to physics.

The present paper is organized as follows. In Section \ref{concepts} we review the construction of quantum mechanics that lies at the root of the present work. The appproach to time that we choose is discussed in some detail in Section \ref{times}. We introduce the evolution parameter $\sigma$ in Section \ref{evolp}, and in Section \ref{parmeterlaw} we argue why $\sigma$ is up to the task to express physical law in well-defined experimental contexts. Section \ref{wavef} describes the wave function from the present perspective, and establishes the relationship between obervable properties and self-adjoint opertators that act on this wave function. In section \ref{evoleq} this machinery is used to derive a new evolution equation, and to motivate the Dirac equation. The properties rest energy, energy and momentum are defined in the process, and the corresponding operators are identified. Finally, in Sections \ref{basiccon} and \ref{discuss} we discuss some basic consequences of the approach, and put it into perspective.

\section{Concepts and formalism}
\label{concepts}

The present study builds on a recent paper \cite{ostborn1}. We refer to that paper for an elaborate presentation of the ideas, the formalization of these ideas, and for some results that are used in the present study. Even so, we feel the need to provide here a brief overview of the approach, in order to make the present study reasonably self-contained.

The physical state $S(n)$ of the world at sequential time $n$ is described as the set of exact states $Z$ that is not excluded by the collective potential knowledge at time $n$. An exact state $Z$ corresponds to a state of complete knowledge of the world, in which the number of objects is precisely known, as well as the values of their internal and relational properties. It is argued that knowledge is always incomplete, so that $S(n)=\{Z\}_{n}$ always contains several elements. We may write $S(n)\subset\mathcal{S}$, where $\mathcal{S}$ is the state space.

Time $n$ is updated according to $n\rightarrow n+1$ whenever the collective potential knowledge changes. Such a change means that the physical states $S(n)$ and $S(n+1)$ can be subjectively distinguished, so that

\begin{equation}
S(n)\cap S(n+1)=\varnothing.
\end{equation}

We may write

\begin{equation}
S(n+1)\subseteq u_{1}S(n),
\label{sevolution}
\end{equation}
where the evolution operator $u_{1}$ is defined by the condition that $u_{1}S(n)\subseteq\mathcal{S}$ is the smallest set for which (\ref{sevolution}) is always fulfilled, and $S(n)\cap u_{1}S(n)=\varnothing$. Physical law can thus be expressed in part as a mapping $u_{1}:\mathcal{P}(\mathcal{S})\rightarrow \mathcal{P}(\mathcal{S})$ from the power set $\mathcal{P}(\mathcal{S})$ of state space $\mathcal{S}$ to itself.

We argue that we cannot reduce the evolution to an element-wise mapping $u_{1}:\mathcal{S}\rightarrow \mathcal{S}$. That is, the exact states $Z$ are not in the domain of $u_{1}$. The reason is that these states are unphysical if we regard them individually, since knowledge is always incomplete. The idea that physical law cannot be properly described by referring to entities or distinctions that are unknowable in principle is promoted to the principle of \emph{explicit epistemic minimalism}. A well-known example of this principle is that a particle that is ejected towards a double slit in such a way that it is forever unknowable which slit it passes cannot be properly described \emph{as if} it passes one slit or the other. Another example is that the exchange of two identical particles cannot be properly described \emph{as if} it were a physical operation leading to a new state.

We distinguish between the state $S$ of the entire world and the state $S_{O}\subset\mathcal{S}$ of an object $O$ that is a proper subset of the world. We may define $S_{O}$ as the set of exact states $Z$ of the world that are consistent with the perception of $O$. If $O_{1}$ and $O_{2}$ are two objects present in the same world, then

\begin{equation}
S_{O1}\cap S_{O2}\neq\varnothing,
\label{overlap}
\end{equation}
since the perceptions of these two objects must be consistent with each other, and

\begin{equation}
S\subseteq  (S_{O1}\cap S_{O2}).
\end{equation}

We may represent the object states in another way, as subsets $S_{OO}\subset\mathcal{S}_{O}$ of the object state space $\mathcal{S}_{O}$. The elements of this state space are exact states $Z_{O}$ of \emph{an object} rather than exact states $Z$ of \emph{the entire world}. We argue that this makes a qualitative difference since the world cannot be properly described as an object. This difference is reflected in qualitatively different relations between the members of a set $\{S_{O}\}$ of object states as compared to a corresponding set $\{S_{OO}\}$.

In particular, we may have

\begin{equation}
S_{OO1}\cap S_{OO2}=\varnothing,
\label{nooverlap}
\end{equation}
in contrast to (\ref{overlap}). This relation simply means that the two objects $O_{1}$ and $O_{2}$ are subjectively distinguishable. If $S_{OO1}(n)\cap S_{OO1}(n+1)=\varnothing$, then a perceived change of object $O_{1}$ defines the temporal update $n\rightarrow n+1$. There must be such an object $O_{1}$ at all times $n$ in order to get $S(n)\cap S(n+1)=\varnothing$. There must also be another object $O_{2}$ for which $S_{OO2}(n)\cap S_{OO2}(n+1)\neq\varnothing$. This relation means that $O_{2}$ does not knowably change as $n\rightarrow n+1$. 

Such objects $O_{2}$ uphold the identity of the world during the temporal update. When something changes something else must stay the same in order to make it epistemically meaningful to relate what comes after with what was before. If a leave falls from a tree, the immobile stem upholds the identity of the world. If the tree is subsequently cut down, the fallen leave resting on the ground upholds the identity of the world.

Let us define \emph{identity} and \emph{identifiability} a bit more precisely. We say that object $O$ is identifiable at time $n$ when

\begin{equation}
S_{OO}(n)\cap S_{OO}(n+1)\neq\varnothing.
\label{identifiable}
\end{equation}

\begin{figure}[tp]
\begin{center}
\includegraphics[width=80mm,clip=true]{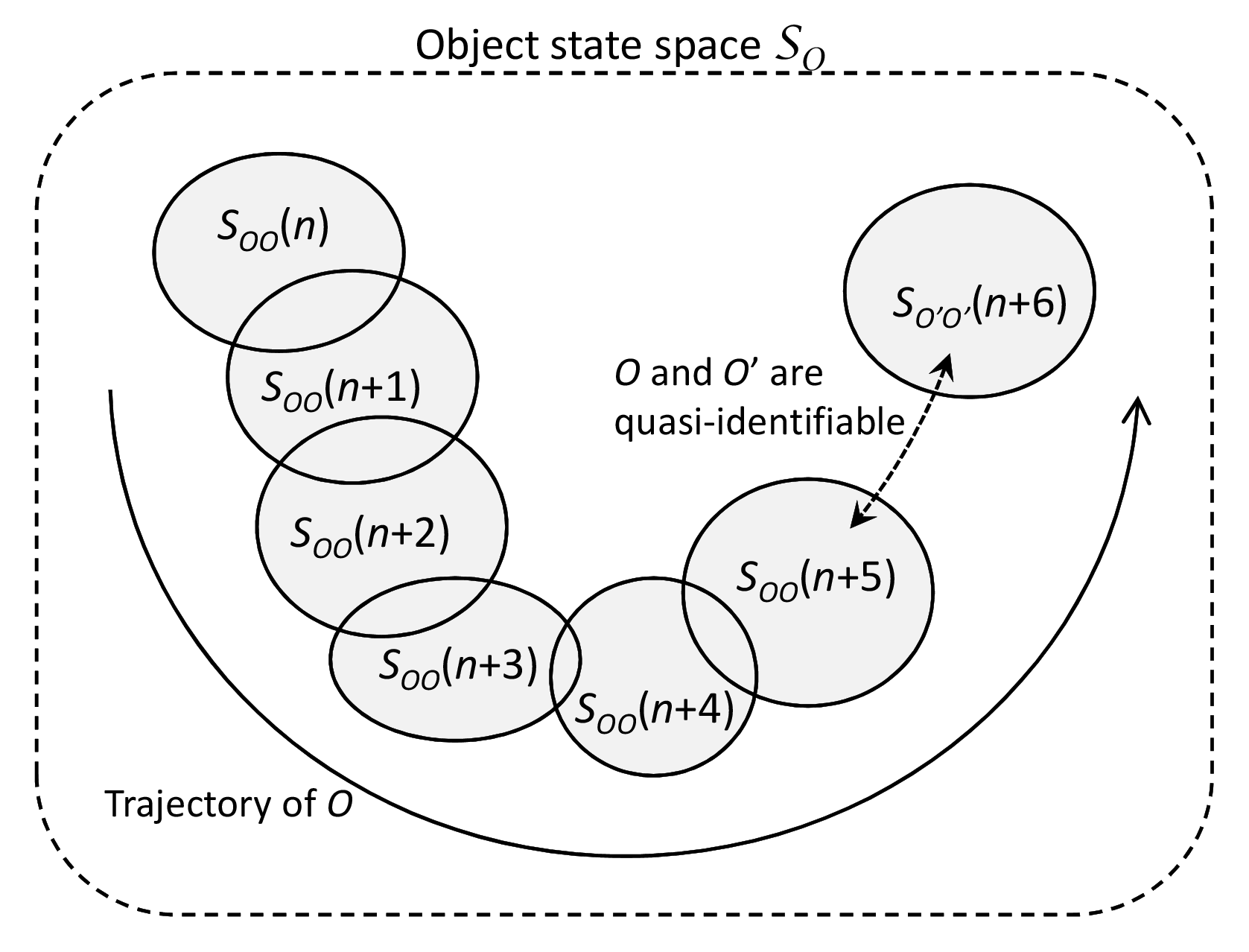}
\end{center}
\caption{The evolution of the state $S_{OO}$ of an object $O$ that is identifiable in the time interval $[n,n+5]$. The object $O'$ at time $n+6$ cannot be identified with object $O$ at time $n+5$, and is therefore given a separate name. If $O$ can nevertheless be modeled as being composed of identifiable quasiobjects which may evolve so that $S_{OO}(n+6)=S_{O'O'}(n+6)$, then we may make the identification $O=O'$ and say that $O$ is quasi-identifiable in the time interval $[n,n+6]$.}
\label{Figure1}
\end{figure}

If object $O$ is identifiable at all times $n, n+1,\ldots,n+m$, then it is said to be identifiable in the time interval $[n,n+m]$. This condition gives meaning to the statement that we track \emph{the same} object from time $n$ to time $n+m$ (Fig. \ref{Figure1}).

The object $O_{1}$ that defines the temporal update $n\rightarrow n+1$ is not identifiable at time $n$. We can nevertheless say that it preserves its identity after the change at time $n+1$. This is so since it is assumed that it is possible to model an object as being composed of identifiable \emph{quasiobjects}. A quasiobject is an object that is not directly perceived, but is used to express in an efficient and general way the physical law that governs the evolution of the objects that are actually perceived. Such an abstract object is identifiable if it can be modelled as being identifiable in the sense of (\ref{identifiable}).

Elementary particles are identifiable quasiobjects, but a large object like the sun can be a temporary identifiable quasiobject if there is nobody at the other side of the world who sees it after sunset. We can still say that it is the same sun that rises the next morning since it can be modeled as a collection of identifiable elementary particles.

We defined the evolution operator $u_{1}$ in relation to (\ref{sevolution}). Similarly, we may define the object evolution operator $u_{O1}:\mathcal{P}(\mathcal{S}_{O})\rightarrow \mathcal{P}(\mathcal{S}_{O})$ so that

\begin{equation}
S_{OO}(n+1)\subseteq u_{O1}[S(n)]S_{OO}(n),
\label{objectevo}
\end{equation}
and $u_{O1}[S(n)]S_{OO}(n)\subseteq\mathcal{S}_{O}$ is the smallest set for which (\ref{objectevo}) is always fulfilled. We have to let $u_{O1}$ depend on the state $S(n)$ of the entire world, since any object $O$ is always related to the world to which it belongs. The overwhelming success of the reductionistic approach in science makes it possible to assume that $u_{O1}$ can always be expressed in terms of its action on a set of elementary particles. Then it attains a general form which is independent of time $n$ and the particular object state $S_{OO}$ on which it acts.

According to the discussion about identifiability (Fig. \ref{Figure1}), an object $O$ does not need to undergo a perceivable change at all temporal updates $n\rightarrow n+1$, but may change only at an arbitrary sequence of moments $n, n+m, n+m', \ldots$, where $1\leq m<m'<\ldots$. Therefore it is meaningful to introduce the $m$-step evolution operator

\begin{equation}
u_{Om}\equiv u_{O1}[u_{m-1}S(n)]u_{O1}[u_{m-2}S(n)]\ldots u_{O1}[u_{1}S(n)]u_{O1}[S(n)],
\label{uomdef}
\end{equation}
where $u_{m}\equiv(u_{1})^{m}$. It tells us all that can be predicted about the object state $S_{OO}(n+m)$ at time $n$.

\begin{figure}[tp]
\begin{center}
\includegraphics[width=80mm,clip=true]{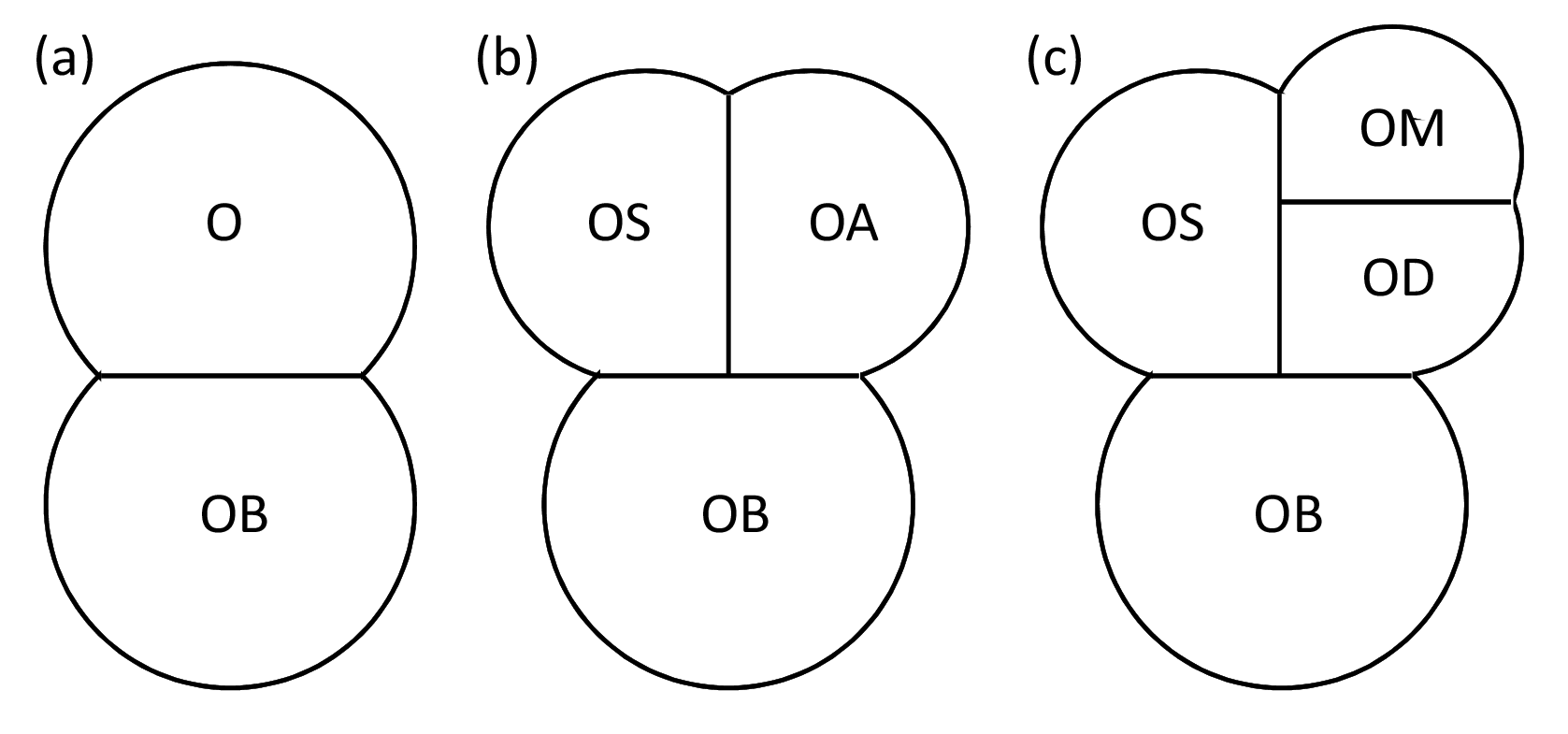}
\end{center}
\caption{Objects that have to be or may be parts of an experimental context $C$. (a) The observed object $O$ and the body of an observer $OB$ are necessary parts. (b) The observed object may be divided into a specimen $OS$ and an apparatus $OA$ with which we decide a property of the specimen. (c) When the specimen is a deduced quasiobject, the apparatus can be divided into a machine $OM$ and a detector $OD$, where the subjective change in the state of the detector defines the outcome of the experiment.} 
\label{Figure2}
\end{figure}

In this paper, we will try to motivate evolution equations for objects rather than for the entire world. Such equations are the ones that can be compared to empirical data in controlled experiments. In such a situation the focus of study is an object that is a proper subset of the world, since an external experimental equipment is needed, and at least one external observer (Fig. \ref{Figure2}). More precisely, we will focus on the evolution of a \emph{specimen} within an experimental context $C$ of a precise but rather general kind. We will now describe what we mean by such a context $C$.

An experiment is performed in order to determine a property of an object that is unknown to begin with. In our vocabulary, the potential knowledge about the object is incomplete at the start of the experiment at time $n$, meaning that there is such a property $P$ whose value $p$ is not precisely known at time $n$.

\begin{figure}[tp]
\begin{center}
\includegraphics[width=80mm,clip=true]{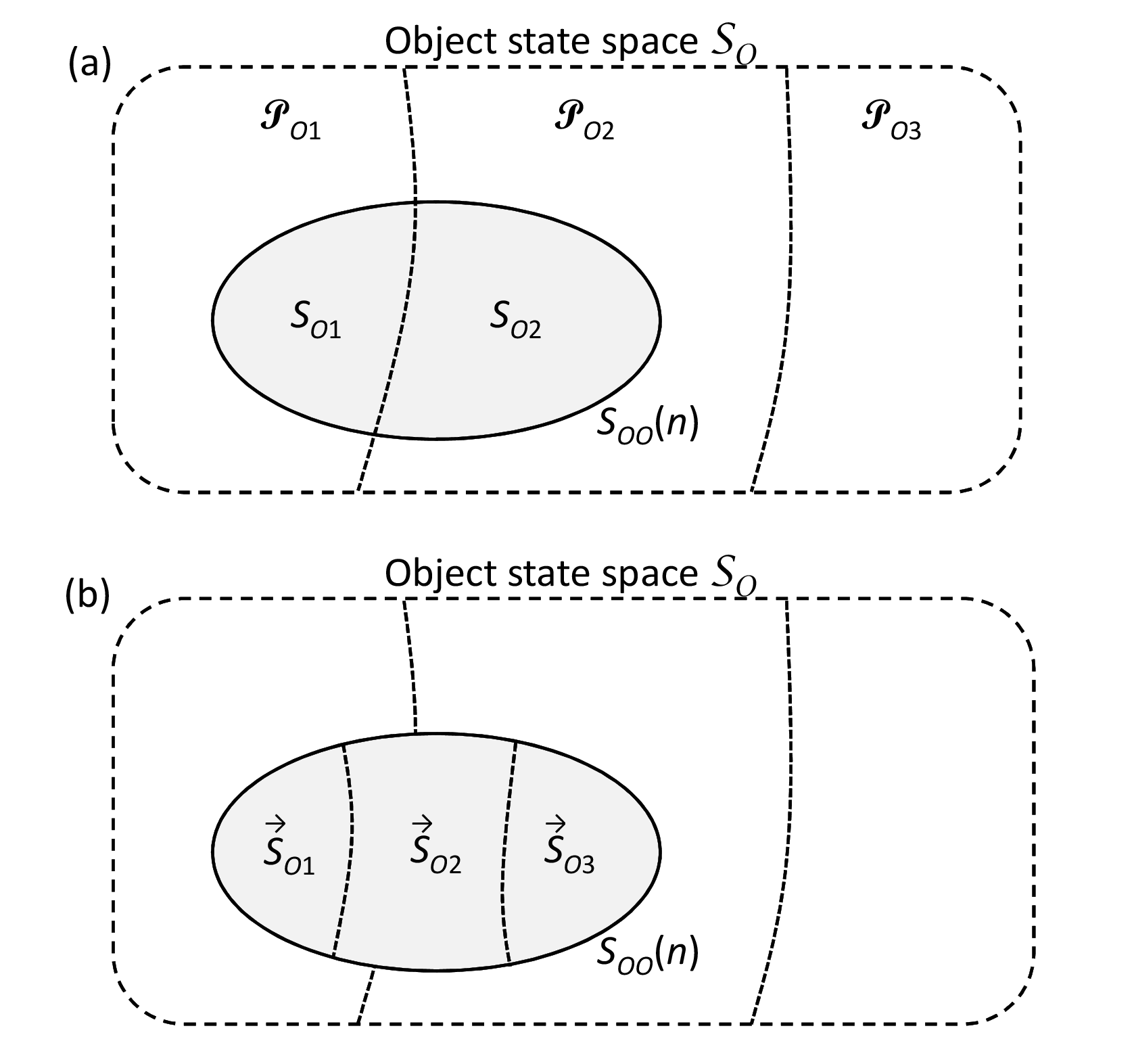}
\end{center}
\caption{(a) An object state $S_{OO}(n)$ for which it can be excluded that the value of $P$ is $p_{3}$, but it is uncertain whether it is $p_{1}$ or $p_{2}$. The corresponding present alternatives $S_{O1}$ and $S_{O2}$ fulfill $S_{O1}\cup S_{O2}=S_{OO}(n)$. (b) In an experimental context $C$ designed to determine an uncertain value we define instead the set $\{\vec{S}_{Oj}\}$ of future alternatives such that the value $p_{j}$ will reveal itself at some later time $n+m$ if $S_{OO}\subseteq\protect\vv{S}_{Oj}$.} 
\label{Figure3}
\end{figure}

Suppose that $P$ has three possible values $p_{1}$, $p_{2}$ and $p_{3}$. Precise knowledge about $P$ means that two of the these three values can be excluded. In contrast, in the state $S_{OO}(n)$ shown in Fig. \ref{Figure3}(a) only $p_{3}$ can be excluded. This leaves two alternatives $S_{O1}$ and $S_{O2}$ corresponding to the cases that the value of $P$ is $p_{1}$ or $p_{2}$, respectively. We may define $S_{Oj}$ as $S_{Oj}\equiv S_{OO}(n)\cap\mathcal{P}_{j}$, where the \emph{property value space} $\mathcal{P}_{Oj}\subseteq\mathcal{S}_{O}$ is the set of exact object states $Z_{O}$ for which the value of $P$ is $p_{j}$.  

In practice, we cannot get to know the value of $P$ at the very same time $n$ as we ask the question and initiate an experiment to find it out, but only at some later time $n+m$. We define a \emph{future alternative} $\vv{S}_{Oj}\subset S_{OO}(n)$ such that if the object state were a subset of $\vv{S}_{Oj}$, then the present physical state $S(n)$ and physical law $u_{O1}[S(n)]$ guarantees that the value of $P$ would turn out to be $p_{j}$ sooner or later [Fig. \ref{Figure3}(b)]. That is, there would be an integer $m\geq 1$ such that we get to know the value $p_{j}$ of $P$ at time $n+m$.

Two alternatives $S_{Oj}$ and $S_{Oj'}$ are disjoint by definition, but two future alternatives $\vv{S}_{Oj}$ and $\vv{S}_{Oj'}$ may or may not overlap. If the momentum of a particle turns out to be $p_{j}$ at time $n+m$, it may turn out to be $p_{j'}$ at a later time $n+m'$. We will, however, focus on experimental contexts $C$ in which the observed properties are defined contextually so that any two future alternatives $\vv{S}_{Oj}$ and $\vv{S}_{Oj'}$ are indeed disjoint, as shown in Fig. \ref{Figure3}(b). For example, the observed momentum may be defined as the reading of a certain part of the apparatus $OA$ (Fig. \ref{Figure2}) placed at a certain location, so that it can only measure the momentum of a given particle once.

A set $\{\vv{S}_{Oj}\}$ of such disjoint future alternatives is called \emph{complete} if

\begin{equation}
S_{OO}(n)=\bigcup_{j}\vv{S}_{Oj},
\end{equation}
and each alternative has the potential to be realized. We require that there is such a complete set of alternatives for each property $P$ of the specimen $OS$ that is observed within the experimental context $C$. Further, there must be at least one such set of alternatives for which we know at the start of the experiment at time $n$ that one alternative will, by physical necessity, be realized before some time $n+\check{m}$, revealing the value $p_{j}$ of property $P$. This means that each experiment that corresponds to a context $C$ has a definite outcome within a known finite time limit $\check{m}$ known in advance.

There may be other sets of disjoint future alternatives defined for another property $P'$ within $C$ for which we know at time $n$ that no alternative will ever be realized. This situation occurs, for instance, in a double-slit experiment resulting in an interference pattern where $p_{j}'$ corresponds to the fact that the particle passes slit $j$.

\begin{figure}[tp]
\begin{center}
\includegraphics[width=80mm,clip=true]{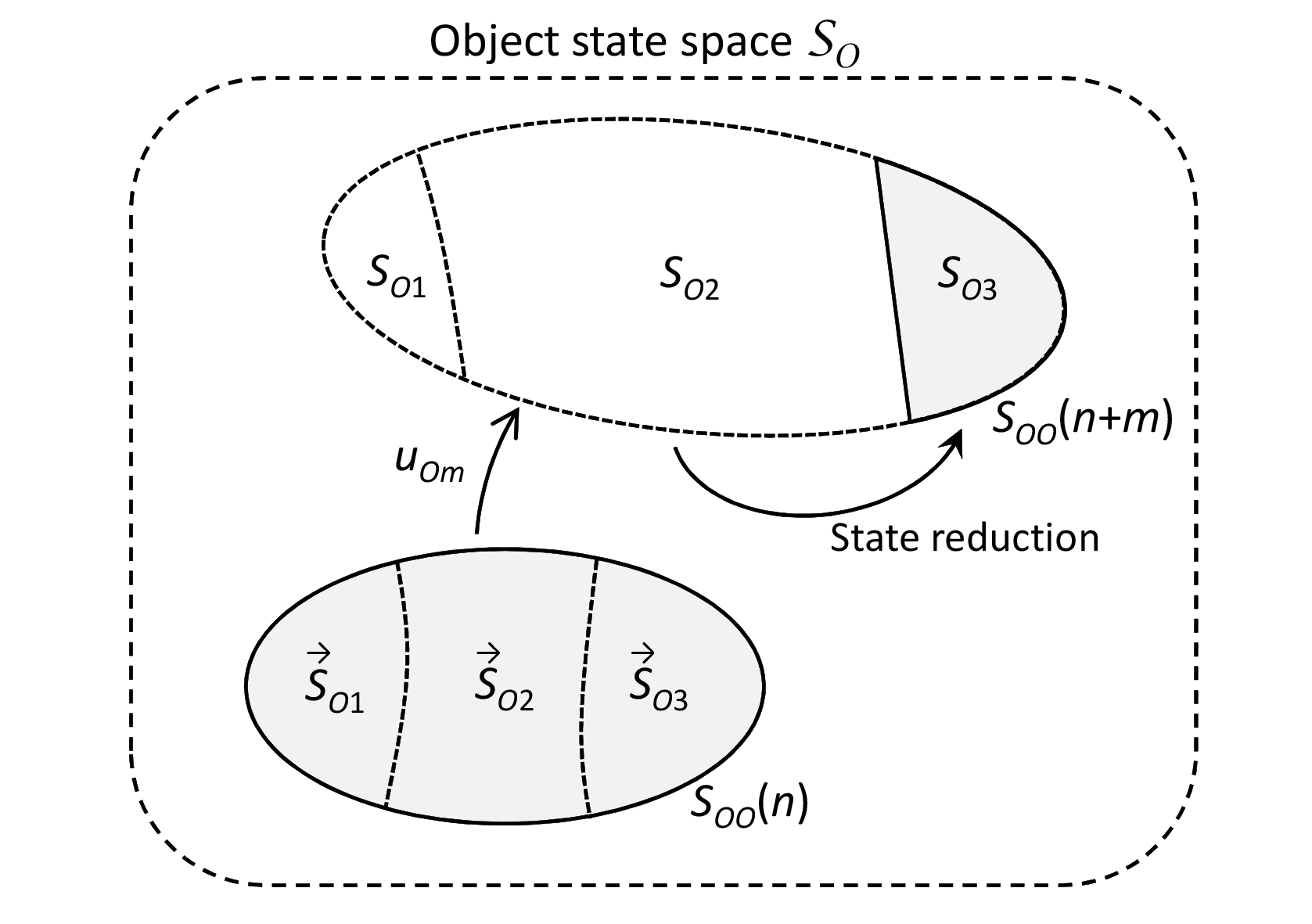}
\end{center}
\caption{At time $n$ it is uncertain which value $p_{1}$, $p_{2}$ or $p_{3}$ of property $P$ will be observed. At time $n+m$ the value $p_{3}$ reveals itself. A state reduction $u_{Om}S_{OO}(n)\rightarrow S_{OO}(n+m)=S_{O3}$ occurs. Compare Fig. \ref{Figure3}.} 
\label{Figure4}
\end{figure}

The observation of value $p_{j}$ at time $n+m$ corresponds to a \emph{state reduction}, as shown in Fig. \ref{Figure4}. A state reduction occurs at time $n+m$ when

\begin{equation}
S_{OO}(n+m)\subset u_{O1}S_{OO}(n+m-1),
\end{equation}
where we have dropped the dependence of $u_{O1}$ on $S(n)$ for brevity. In our experimental context we may write

\begin{equation}
u_{Om}S_{OO}(n)\rightarrow S_{OO}(n+m)=S_{j}\subset u_{Om}S_{OO}(n)
\end{equation}
at the moment $n+m$ when the value $p_{j}$ of $P$ is observed.

The occurrence of state reductions introduce a two-fold indeterminism into the present description of physical law. First, there is no law that dictates exactly \emph{when} the state of a given object will reduce. Second, there is no law that tells exactly \emph{how} its state will reduce. If there were laws of both these kinds, then $u_{1}$ and $u_{O1}$ should be redefined according to (\ref{sevolution}) and (\ref{objectevo}) to include them, and there would be no state reductions at all. That is to say, physical law is indeterministic if and only if state reductions may occur.

The concept of a state reduction can be used to put the related concept of a \emph{wave function collapse} in a partially new light. In particular, the object state does not need to be reduced to the extent that a perfectly sharp value of a property is revealed, corresponding to the fact that the wave function does not need to collapse all the way down to a delta spike. Rather, the sudden knowledge increase associated with the state reduction amounts in general just to the exclusion of more values of $P$ than was previously possible.
This is illustrated in Figs. \ref{Figure3} and \ref{Figure4} by the fact that the alternatives $S_{Oj}$ or $\vv{S}_{Oj}$ do not correspond to exact object states $Z_{O}$ or infinitely thin slices of the object state $S_{OO}$.

In the case of a continuous property such as momentum, the set of alternative values $\{p_{j}\}$ that may be observed within the context $C$ correspond instead to a bin of continuous values within some interval $\Delta p_{j}=[p_{j},p+\delta p_{j})$ which cannot be excluded by the observation with limited resolution. This is a general observation. Regardless the underlying structure of the values of a property $P$, the set of values $\{p_{j}\}$ that may be observed within an experimental context $C$ is always finite, corresponding to a finite set of future alternatives $\{\vv{S}_{Oj}\}$. From the epistemic perspective, two values $p_{j}$ and $p_{j'}$ must be distinguishable, meaning that they form a countable set. Further, the set must be finite because a physical apparatus $OA$ that records the values has finite extension and should have the potential to reveal any of the values $\{p_{j}\}$ within a finite time $m$.

It can be argued that The Hilbert space formalism of quantum mechanics, with Born's rule to calculate the probability to observe any given value $p_{j}$ in the predefined set $\{p_{j}\}$, is the only algebraic formalism that can represent almost all kinds of experimental contexts $C$ of the type introduced above \cite{ostborn1}. The argument depends crucially on the epistemic assumption that a physical theory that relies on entities or distinctions that are unknowable in principle gives rise to wrong predictions. This assumption excludes, for example, a conventional probabilistic description of the double slit experiment, in which the probability $q_{ij}$ that the particle passes slit $i$ and then hits the detector screen at position $x_{j}$ is $q_{ij}=q(i)q(x_{j}|i)$. This is so since the epistemic assumption forbids us to say that the particle passed one slit or the other if it is forever outside potential knowledge which slit it actually passed, making the probability $q(i)$ that the particle passes slit $i$ undefined.

In the present approach there is no such thing as a universal Hilbert space, applying to the whole world at all times. Instead, Hilbert spaces $\mathcal{H}_{C}$ with evolving state vectors $\overline{S}_{C}$ are defined as representations of given contexts $C$ and their evolving states during the course of the experiment. As such, their role is to provide an efficient algebraic description of the evolution of a limited specimen $OS$ during a limited period of time (Fig. \ref{Figure2}). Since the number of observable values within the context $C$ is finite, we can always choose a finite dimension $D_{\mathcal{H}}$ of the associated Hilbert space $\mathcal{H}_{C}$. More precisely, we can always choose $D_{\mathcal{H}}\leq M\times M'\times\ldots$ if the set of properties $\{P,P',\ldots\}$ is defined within the context $C$ with corresponding sets of future alternatives $\{\{\vv{S}_{Oj}\},\{\vv{S}_{Oj'}\},\ldots\}$, where $\{\vv{S}_{Oj}\}$ has $M$ elements, $\{\vv{S}_{Oj'}\}$ has $M'$ elements, and so on.

\begin{figure}[tp]
\begin{center}
\includegraphics[width=80mm,clip=true]{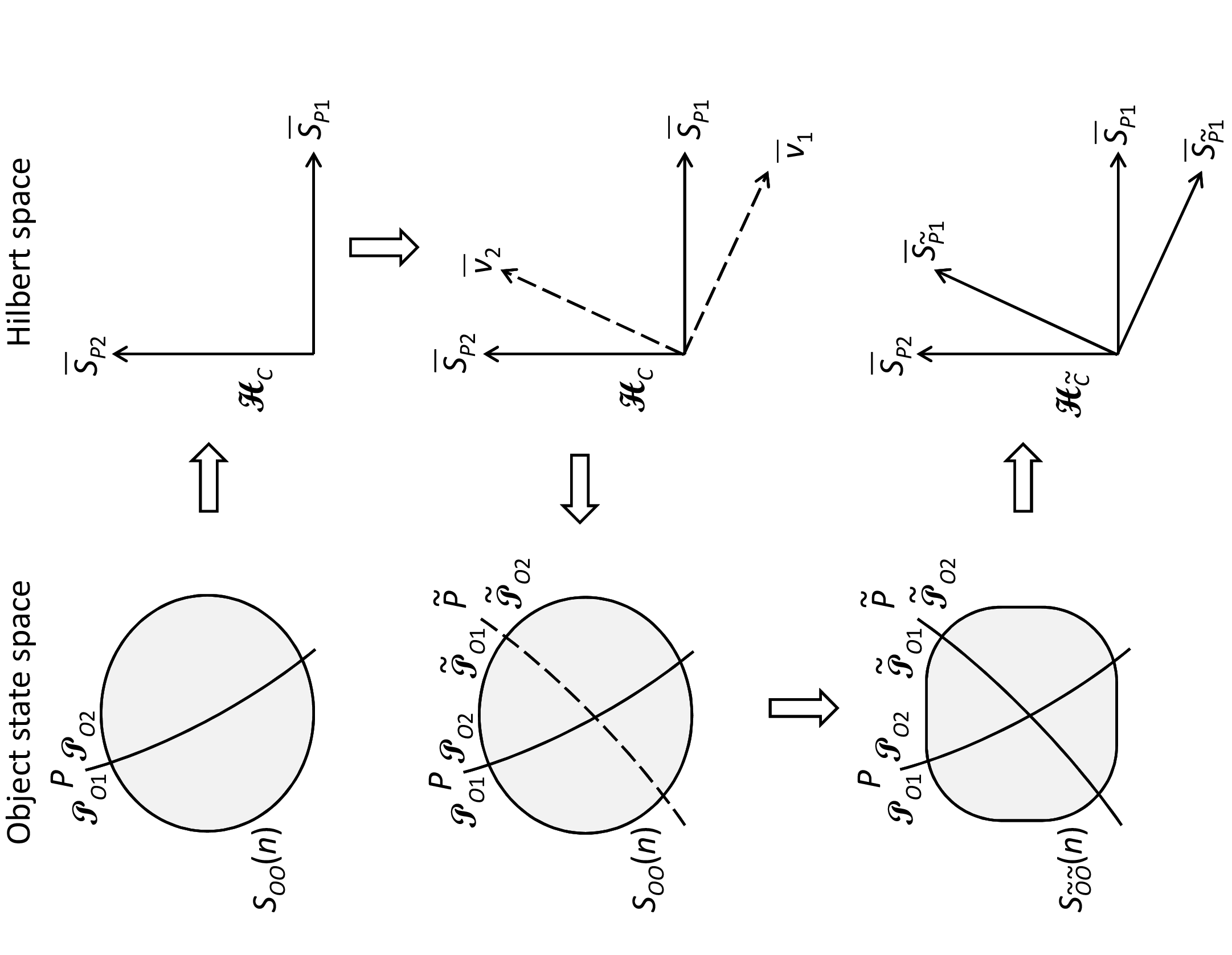}
\end{center}
\caption{In a context $C$, described by its initial object state $S_{OO}(n)$, in which property $P$ is observed with possible values $\{p_{1},p_{2}\}$, we may define a Hilbert space $\mathcal{H}_{C}$ with basis $\{\overline{S}_{P1},\overline{S}_{P2}\}$. We may define another orthonormal basis $\{\overline{v}_{1},\overline{v}_{2}\}$ in $\mathcal{H}_{C}$. Such a basis can always be associated with the complete set of eigenvectors of a self-adjoint operator $\overline{\tilde{P}}$ that corresponds to another property $\tilde{P}$ with possible values $\{\tilde{p}_{1},\tilde{p}_{2}\}$. In another context $\tilde{C}$ in which $\tilde{P}$ is observed after $P$ we can identify $\overline{v}_{j}=\overline{S}_{\tilde{P}j}$ in the corresponding Hilbert space $\mathcal{H}_{\tilde{C}}$. This context $\tilde{C}$ is described by its initial object state $S_{\tilde{O}\tilde{O}}(n)$.}
\label{Figure5}
\end{figure}

We define a certain \emph{property value state} $S_{Pj}$ such that the state of the specimen $OS$ (Fig. \ref{Figure2}) is $S_{Pj}$ whenever we know that the value of its property $P$ is $p_{j}$. This value is defined contextually according to the resolving power of the experiment, and may correspond to a bin of several possible values of the property, as discussed above. The Hilbert space $\mathcal{H}_{C}$ is constructed in such a way that there is a one-to-one correspondence

\begin{equation}
S_{Pj}\subset \mathcal{S}_{O}\leftrightarrow\overline{S}_{Pj}\subset \mathcal{H}_{C}.
\label{pvspace}
\end{equation}
The subspaces $\overline{S}_{Pj}$ and $\overline{S}_{Pk}$ associated with two different values $p_{j}$ and $p_{k}$ of $P$ are orthogonal (Fig. \ref{Figure5}).

It is possible to associate in a unique way one self-adjoint operator $\overline{P}$ with domain $\mathcal{H}_{C}$ to each property $P$ observed within the experimental context $C$ such that

\begin{equation}
\overline{P}\overline{S}_{Pj}=p_{j}\overline{S}_{Pj}
\label{propop}
\end{equation}
for each $j$.

Suppose that two properties $P$ and $P'$ are defined within the context $C$. We call them \emph{simultaneously knowable} if and only if it is possible to have $S_{OO}\subseteq \mathcal{P}_{Oj}$ and $S_{OO}\subseteq \mathcal{P}_{Oj'}'$ for each pair of indices $(j,j')$, where the property value spaces $\mathcal{P}_{Oj}$ and $\mathcal{P}_{Oj'}'$ are defined in relation to Fig. \ref{Figure3}(a). The existence of pairs of properties that are \emph{not} simultaneously knowable is related to the assumption that potential knowledge is always incomplete, meaning that we cannot know the values of all properties at the same time. It can be shown that the two property operators $\overline{P}$ and $\overline{P}'$ commute if and only if $P$ and $P'$ are simultaneously knowable. That is, these operators behave exactly as the operators that we associate with observables in quantum mechanics.

We have argued that we can associate a self-adjoint property operator $\overline{P}$ to each property $P$ with an associated complete set of future alternatives $\{\vv{S}_{Oj}\}$ defined within the experimental context $C$. Conversely, we can associate such a contextual property $\tilde{P}$ to each self-adjoint operator $\overline{\tilde{P}}: \mathcal{H}_{C}\rightarrow\mathcal{H}_{C}$ with a complete set of eigenvectors. However, this statement requires a qualification. In the context $C$ there is a predefined set $\{P,P',\ldots\}$ of observed properties with associated self-adjoint operators $\{\overline{P},\overline{P}',\ldots\}$ acting in $\mathcal{H}_{C}$. If we define another self-adjoint operator $\overline{\tilde{P}}\not\in\{\overline{P},\overline{P}',\ldots\}$ that acts in $\mathcal{H}_{C}$, this operator does not correspond to a property that is actually observed within the context $C$. It can only be associated with a property $\tilde{P}$ observed in \emph{another} experimental context $\tilde{C}$, in which $\tilde{P}$ is observed after the properties $\{P,P',\ldots\}$ (Fig. \ref{Figure5}). The property $\tilde{P}$ defined by the operator $\overline{\tilde{P}}$ will not be simultaneously knowable with any of the other contextual properties $\{P,P',\ldots\}$ defined within $\tilde{C}$.

\section{Sequential and relational time}
\label{times}

A core idea in this study is that the concept of time should be separated into two aspects, which we call sequential and relational time. We will argue that if we incorporate both of them into the physical formalism it becomes more coherent, and it becomes easier to motivate some parts of physical law. In this section we define the roles of these two aspects of time, describe their qualities, and motivate why they should be separated.

\begin{figure}[tp]
\begin{center}
\includegraphics[width=80mm,clip=true]{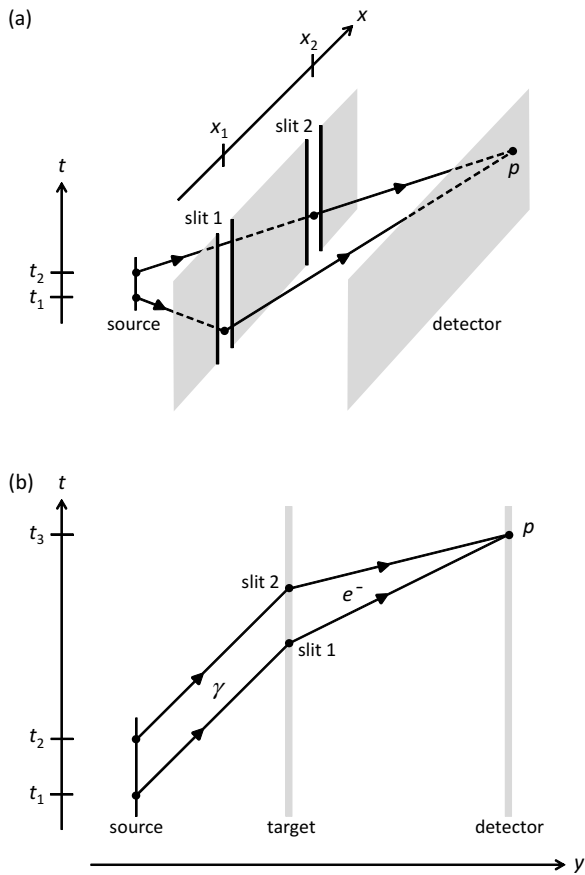}
\end{center}
\caption{(a) The double-slit experiment demonstrates not only interference of spatially separated paths, passing the slits at $x_{1}$ or $x_{2}$, but also interference of temporally separated paths, starting from the source at two different times $t_{1}$ and $t_{2}$. (b) Schematic illustration of a double-slit experiment with interference of paths with temporal separation only. Each of two laser pulses emitted at times $t_{1}$ and $t_{2}$ may ionize a single atom in a stationary target during two `temporal slits'. The two possibilities interfere when the emitted electron is detected.}
\label{Figure6}
\end{figure}

The current physical formalism is unsatisfactory not only in the sense that time is treated differently in quantum mechanics and general relativity, but also in the sense that the treatment of time is unsatisfactory in quantum mechanics itself. This becomes apparent if we consider the double slit experiment and add a vertical temporal axis $t$ to the standard picture [Fig. \ref{Figure6}(a)].

Assume 1) that a single object hits the detector screen at a point $p$ off the symmetry axis of the experimental setup, 2) that the speed of the object on its path from the source to the screen is known, and 3) that information about which slit the object passes is outside potential knowledge. Then there is interference between the two alternative paths. But the two paths correspond to two different departure times from the source. Thus there is not only spatial interference between paths departing from the two slits located at positions $x_{1}$ and $x_{2}$, but also temporal interference between paths departing from the source at times $t_{1}$ and $t_{2}$.

In other words, two possible but unobservable timings $t_{1}$ or $t_{2}$ of the event that the object is emitted from the source contributes to the probability that the object is detected at the point $p$ on the detector screen at a later time $t_{3}$. For the event of object emission we must clearly allow a temporal Heisenberg uncertainty $\Delta t\geq t_{2}-t_{1}$, just as we allow a spatial uncertainty $\Delta x\geq x_{2}-x_{1}$ for the event that the object passes a slit. Generally speaking we must allow both temporal and spatial Heisenberg uncertainties $\Delta t$ and $\Delta x$ in order to describe interference in the double-slit experiment in a satisfactory way.

This is not the case in Shr\"odinger equations where the same variable $t$ that measures timings of events is also used as a precisely defined evolution parameter, for which we must set $\Delta t=0$. This fact suppresses the inherent symmetry between the spatial and temporal aspects of interference. To restore it we need to introduce another kind of continuous evolution parameter $\sigma$ to express differential evolution equations, thus releasing $t$ from this task and allowing it to display the desired uncertainty $\Delta t$.

Interference of two possible but unobservable timings of a single event has been demonstrated more explicitly in another kind of experiment \cite{wollenhaupt,lindner}, emphasizing this need. The basic idea behind these experiments is illustrated in Figure \ref{Figure6}(b). In the experiment by Wollenhaupt \emph{et al.} \cite{wollenhaupt} two ultrashort laser pulses emitted at times $t_{1}$ and $t_{2}$ create a pair of temporal slits with separation $\Delta t=t_{2}-t_{1}$ at which an atom may be ionized, emitting an electron with a continuous energy spectrum. Interference between the two possible emission times of the single electron implies that the probability that an electron with given energy will be detected oscillates as a function of $\Delta t$. For a given time delay $\Delta t$ it also means that the probability to detect an electron oscillates as a function of its energy.

Having argued for the need for a continuous evolution parameter $\sigma$, we will argue next that it is reasonable to relate $\sigma$ to the discrete sequential time $n$ discussed in the previous section, since we expressed the general evolution operators $u_{1}$ and $u_{O1}$ as operators taking us from the physical state $S(n)$ or $S_{OO}(n)$ to the state $S(n+1)$ or $S_{OO}(n+1)$, respectively (see (\ref{sevolution}) and (\ref{objectevo})).

Before we do that, let us characterize sequential time $n$ in more detail, and then define its relationship with relational time $t$. To be able to use $n$ to express the evolution of the physical state it must be defined in such a way that everybody can agree on the ordering of events, since each event corresponds to an update $n\rightarrow n+1$.

Denote by $PK(n)$ the potential knowledge that corresponds to the physical state $S(n)$ \cite{ostborn1}. This potential knowledge can be decomposed into the potential knowledge $PK^{k}(n)$ of different subjects $k$ according to $PK(n)=\bigcup PK^{k}(n)$. Each event $e$ corresponds to a subjective change potentially perceived by a subject $k$, and may thus be expressed as $e^{k}$.

A subjective change perceived by another subject $k'$ corresponds to another event $e^{k'}$. These two events can sometimes be associated with a change of a single object $O$ that is observed by both subjects. This possibility follows from the basic assumption that two subjects may perceive the same object, reflecting the hypothesis that we all live in the same world.

We have assumed that all perceived objects $O$ can be modeled as a collection of identifiable quasiobjects such as elementary particles (Section \ref{concepts}). If so, each event $e^{k}$ corresponds to a change of a given object $O$ that preserves its identity in the process. Symbolically, we may identify the event with the object states just before and after the change, together with the physical state $S(n-1)$ that defines its context:

\begin{equation}
e^{k}\leftrightarrow\{S_{OO}(n-1),S_{OO}(n),S(n-1)\},
\label{eventdef}
\end{equation}
where the assumed distinct change of $O$ means that $S_{OO}(n-1)\cap S_{OO}(n)=\varnothing$.

We assume the inherent potential of each subject $k$ to order the events that occur to her temporally. She may judge that two events occur to her at the same time, meaning that she perceives two objects $O_{a}$ and $O_{b}$ that change simultaneously. We may express her ordering as

\begin{equation}
e_{1}^{k}\succ e_{2}^{k}\succ \{e_{3a}^{k},e_{3b}^{k}\}\succ e_{4}^{k}\succ\ldots,
\label{personalorder}
\end{equation}
where events within brackets occur at the same time.

If subject $k$ can send a message to $k'$ that $e^{k}$ has occurred such that $k'$ receives it before or at the same time as $e^{k'}$ occurs, then we say that $e^{k}$ occurs before $e^{k'}$. In that case $e^{k}$ corresponds to a temporal update $n\rightarrow n+1$ and $e^{k'}$ to a subsequent temporal update $n'\rightarrow n'+1$ with $n'>n$. If such a message cannot be sent $e^{k}$ and $e^{k'}$ may or may not occur at the same sequential time. In the case they do, the two events in the set $\{e^{k},e^{k'}\}$ together correspond to the same temporal update $n\rightarrow n+1$. We assume that the question whether $e^{k}$ and $e^{k'}$ occur at the same time or not has a definite answer, even though it is impossible to check it empirically by means of messaging. We can then express a universal temporal ordering of events of the form

\begin{equation}
e_{1}^{k}\succ e_{1}^{k'}\succ e_{2}^{k}\succ\{e_{3a}^{k},e_{3b}^{k}\}\succ e_{2}^{k'}\succ\{e_{4}^{k},e_{3}^{k'}\}\succ\ldots,
\label{eventorder}
\end{equation}
where two successive symbols $\succ$ correspond to two successive sequential times $n$ and $n+1$.

Two events $e^{k}$ and $e^{k'}$ for which $e^{k}\succ e^{k'}$ cannot be associated with a change of one single object perceived by two different subjects, but a pair of simultaneous events $\{e^{k},e^{k'}\}$ can sometimes be associated with such a situation.

\begin{figure}[tp]
\begin{center}
\includegraphics[width=80mm,clip=true]{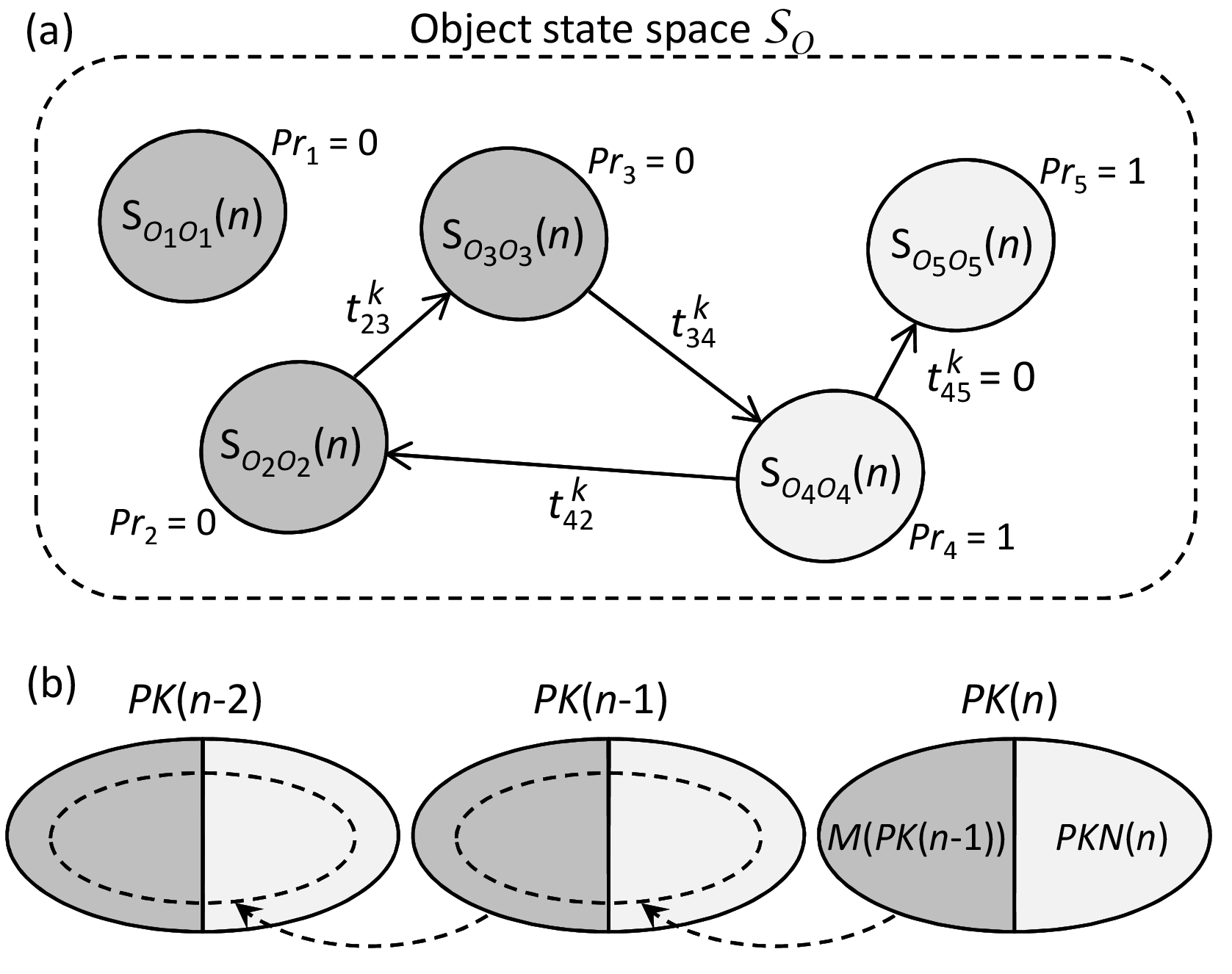}
\end{center}
\caption{The roles of sequential time $n$ and relational time $t$. (a) At any given time $n$ the state of potential knowledge $PK(n)$ that corresponds to the physical state $S(n)$ contains several objects $O_{j}$ with physical object states $S_{O_{j}O_{j}}(n)$. Some of these are perceptions of the present (light grey), and some are memories of the past (dark grey). The temporal distances $t_{ij}^{k}$ measured by subject $k$ relate any two such objects $O_{i}$ and $O_{j}$ and the knowledge of their values is part of $PK^{k}(n)$ and thus of $PK(n)$. (b) The sequential time labels different states of potential knowledge $PK(n)$. The set of memories $M(PK(n-1))\subseteq PK(n)$ at time $n$ consists of all dark grey objects $O_{j}$ in panel (a). This set refers back to the preceding state $PK(n-1)$. The memories may be imperfect, as illustrated by the dashed ovals being subsets of $PK(n-1)$ and $PK(n-2)$, respectively. The backward references define a unique directed ordering of instants $n$, reflecting the perceived flow of time.}
\label{Figure7}
\end{figure}

An event $e^{k}$ and the corresponding temporal update $n-1\rightarrow n$ is always associated with the change of the state of an object $O$ according to (\ref{eventdef}). Its state $S_{OO}(n)$ just after the event corresponds to the present state of the object at time $n$. We may also have memories at time $n$ of its preceding state $S_{OO}(n-1)$. That state is \emph{not} a part of the physical state $S(n)$. Rather, the memory corresponds to another object $O'$ that is part of $S(n)$. Its state at time $n$ may be written $S_{O'O'}(n)=M(S_{OO}(n-1))$, where $M(\ldots)$ denotes the potential memory of the state within brackets. (We note however that we may often identify $O=O'$ in the sense discussed in relation to Fig. \ref{Figure1}). We see that some objects perceived at time $n$ are present objects and some are memories of past objects. We may thus define a binary \emph{presentness property} $Pr$ that applies to all objects such that $Pr=1$ for present objects and $Pr=0$ for memories of past objects, as illustrated in Fig. \ref{Figure7}(a).

The value of $Pr$ defines the object state $S_{OO}$ together with the values of an array of other properties, such as charge and position \cite{ostborn1,ostborn2}. Properties may be \emph{internal} or \emph{relational}. Presentness and charge are internal properties that refer to the object $O$ itself. Position, on the other hand, is a relational property which can only be defined epistemically as a relation between $O$ and a set of other obejcts. It takes the form of a set of spatial distances between $O$ and a set of reference objects that form a coordinate system.

As discussed above, the knowledge about the value of any property $P$ may be incomplete, meaning that there are more than one value which is not excluded by the potential knowledge about the object. This goes for the presentness property as well. Imagine, for example, that you look at a tree and contemplate the leaves rustling in the wind. You know that the leaves move along given trajectories, but you cannot tell which of their positions belong to the present and which belong to the immediate past. The experience is temporally holistic, in a sense.

We see that the potential knowledge $PK(n)$ that corresponds to the physical state $S(n)$ may be divided into two parts

\begin{equation}
PK(n)=PKN(n)\cup M(PK(n-1)),
\end{equation}
where $PKN(n)$ corresponds to the knowledge at time $n$ about objects with $Pr=1$, and $M(PK(n-1))$ corresponds to the knowledge at time $n$ about objects with $Pr=0$ [Fig. \ref{Figure7}(b)]. The fact that the value of $Pr$ may be uncertain can be represented as the possibility that $PKN(n)\cap M(PK(n-1))\neq\varnothing$.

The potential memories $M(PK(n-1))$ of the preceding sequential time $n-1$ may be perfect or imperfect, so that we may write

\begin{equation}
M(PK(n-1))\subseteq PK(n-1)
\label{sloppymemory}
\end{equation}
if we disregard changes in the presentness attribute $Pr$ of some objects when going from $PK(n-1)$ to $M(PK(n-1))$. The knowledge labeled by $M(PK(n-1))$ corresponds by definition to \emph{proper} memories, meaning that there is, by definition, an arrow pointing from  $M(PK(n-1))$ to $PK(n-1)$, as shown in Fig. \ref{Figure7}(b). In the same way there is an arrow pointing from $M(PK(n-2))$ to $PK(n-2)$. In this way a set of arrows is defined that creates a unique directed ordering of the elements in the set $\{PK(n)\}_{n}$ according to

\begin{equation}
\ldots\succ PK(n-2)\succ PK(n-1)\succ PK(n)\succ\ldots
\label{statechain}
\end{equation}
that mirrors the inherent ordering of the time instants $n$, and the ordering of events expressed in (\ref{eventorder}).

It must be stressed that the possibility to order the states of knowledge $PK(n)$ has to be assumed. The ordering cannot be observed empirically by any subject, since the perceptions of any subject at any time $n$ is limited by definition to those objects that make up $PK(n)$. In other words, it is not possible, perception-wise, to transcend $PK(n)$ and observe the chain of states (\ref{statechain}) from outside in the manner shown in Fig. \ref{Figure7}(b). The assumption that there is such a chain can only be justified if it turns out to be helpful in the construction of a coherent model of the physical world and physical law. We hope to convince the reader in the following that this is indeed the case. The transcendence involved is closely related to the transcendence involved in the leap of induction with which we arrive at generally valid physical laws from a finite set of observations.

Even if the ordering of events and states of potential knowledge $PK(n)$ is unambiguous in this transcendental sense, the perceived ordering of past events may become blurred as time passes, just as we argued that it may be ambiguous from the subjective point of view whether a perceived object belongs to the present or to the past. If the perceived ordering of states of knowledge were indeed unambiguous, we would be able to write

\begin{equation}\begin{array}{rcl}
M(PK(n-1)) & = & M(PKN(n-1)\cup M(PK(n-2))\\
& = & M(PKN(n-1))\cup M(M(PK(n-2)))
\end{array}
\label{preserveorder}
\end{equation}
for each $n$. If so, we would have

\begin{equation}\begin{array}{rcl}
PK(n) & = & PKN(n)\cup \\
& & M(PKN(n-1))\cup \\
& & M^{2}(PKN(n-2))\cup\\
& & \ldots\\
& & M^{m}(PKN(n-m))\cup\\
& & \ldots
\end{array}
\label{perfectorder}
\end{equation}
and the ordering would be possible to read in state of knowledge $PK(n)$ from its decomposition into one set for each previous time, each marked with the exponent $m$ of the `memory hierarchy' $M^{m}$ that tells us how far into history the associated memory of the present state $PKN(n-m)$ should be placed.

However, the second equality in the relation (\ref{preserveorder}) is not necessarily fulfilled, so that the sequential time $m$ passed sine a given event is not necessarily imprinted in the collective memory of all subjects at time $n$. We will argue next that it is in fact impossible for any subject to keep track of the number $m$. Let the events $e^{k}_{i}$ and $e^{k}_{j}$ perceived by subject $k$ define the starting point and the ending point such an attempt. After the event $e^{k}_{i}$ subject $k$ must keep track of all changes $e^{k'}$ of all objects perceived by all other subjects $k'$ in the universe. She cannot know the number of such objects. It may very well be infinite. The number $m$ of events $e^{k'}$ that happens in the universe between $e^{k}_{i}$ and $e^{k}_{j}$ is therefore unknowable to $k$ and possibly infinite (see (\ref{personalorder})). The impossibility to know the value of $m$ for any given subject $k$ disqualifies it as a universal physical measure of temporal distance. To give it that role would violate the principle of \emph{explicit epistemic minimalism}, discussed in Section \ref{intro}.

Instead, we must introduce the \emph{relational time} $t_{ij}$ that estimates the temporal distance between any two objects $O_{i}$ and $O_{j}$ that are part of $PK(n)$. The value of $t_{ij}$ may be uncertain just like for any other property, to the extent that the temporal distance between two events may be completely unknown. The distance $t_{ij}$ is determined by counting how many reference objects are placed temporally in between $O_{i}$ and $O_{j}$. Of course, these reference objects correspond to the successive tickings of a clock. (This notion of defining distances between the values of a given property of two objects by putting reference objects in between them is elaborated upon in Ref. \cite{ostborn1}.)

There is no universal clock that all subjects can perceive. Therefore we must allow that different subjects $k$ and $k'$ use different clocks, and place different numbers of tickings between the same pair of events. In other words, we should attach a label $k$ to the distance $t_{ij}^{k}$ and allow that $t_{ij}^{k}\neq t_{ij}^{k'}$. The knowledge about the value of $t_{ij}^{k}$ becomes an object that is part of the potential knowledge $PK^{k}(n)$ of subject $k$ at time $n$, referring to the states $S_{O_{i}O_{i}}(n)$ and $S_{O_{j}O_{j}}(n)$ at the very \emph{same} time $n$. In this way the relational time distance $t_{ij}^{k}$ becomes a \emph{perceived} property at a given time $n$ in contrast to the \emph{transcendent} sequential time distance $m$ that relates \emph{different} sequential times $n-m$ and $n$ in (\ref{perfectorder}). We have argued that it is impossible know $m$. It is not even meaningful to speak about any quantifiable uncertainty of its value. For this reason sequential time $n$ cannot be called a property.

The clock $c^{k}$ used by subject $k$ to measure $t_{ij}^{k}$ can be described as an object $O_{c}^{k}$ that changes at a sequence of times $n,n+m,n+m',\ldots$, forming a corresponding sequence of events $e^{k}_{n},e^{k}_{n+m},e^{k}_{n+m'},\ldots$. Since it is impossible to determine the numbers $m,m',\ldots$, there is no inherent way to say whether the clock $c^{k}$ ticks at regular intervals or not. All subject $k$ can do is to count the number of ticks she perceives herself between the events $e^{k}_{i}$ and $e^{k}_{j}$, events that can be associated with a pair of object states $\{S_{O_{i}O_{i}}, S_{O_{j}O_{j}}\}$. The uncertainty of $t_{ij}^{k}$ may stem either from the fact that the memories of the past ticks may be imperfect, as expressed by (\ref{sloppymemory}), or from the use of a crude clock which places only a few tickings between a typical pairs of events. A comparison with a hypothetical more refined clock then defines an uncertainty.

Even though each value $t_{ij}^{k}$ may be more or less uncertain, its role as a temporal measure makes it possible to state a set of relations that selected sets of values $\{t_{ij}^{k}\}$ must fulfill. These relations constitute \emph{conditional knowledge} in the sense introduced in Ref. \cite{ostborn1}. For each subject $k$ and each set of objects $\{O_{1},O_{2},\ldots,O_{N}\}$ with arbitrary labeling we have

\begin{equation}
0=t_{12}^{k}+t_{23}^{k}+\ldots+t_{N-1,N}^{k}+t_{N1}^{k}.
\label{circlezero}
\end{equation}
In the case $N=2$ we get $0=t_{ij}^{k}+t_{ji}^{k}$ for any two objects $O_{i}$ and $O_{j}$, making it clear that (\ref{circlezero}) reflects the directed nature of time. The case $N=3$ is illustrated by the cyclic set of temporal distances $\{t_{23}^{k},t_{34}^{k},t_{42}^{k}\}$ shown in in Fig. \ref{Figure7}(a). In addition, we must require

\begin{equation}
t_{ij}^{k}=0
\end{equation}
for each subject $k$ whenever we have $Pr=1$ for both objects $O_{i}$ and $O_{j}$, as illustrated in Fig. \ref{Figure7}(a).

\begin{figure}[tp]
\begin{center}
\includegraphics[width=80mm,clip=true]{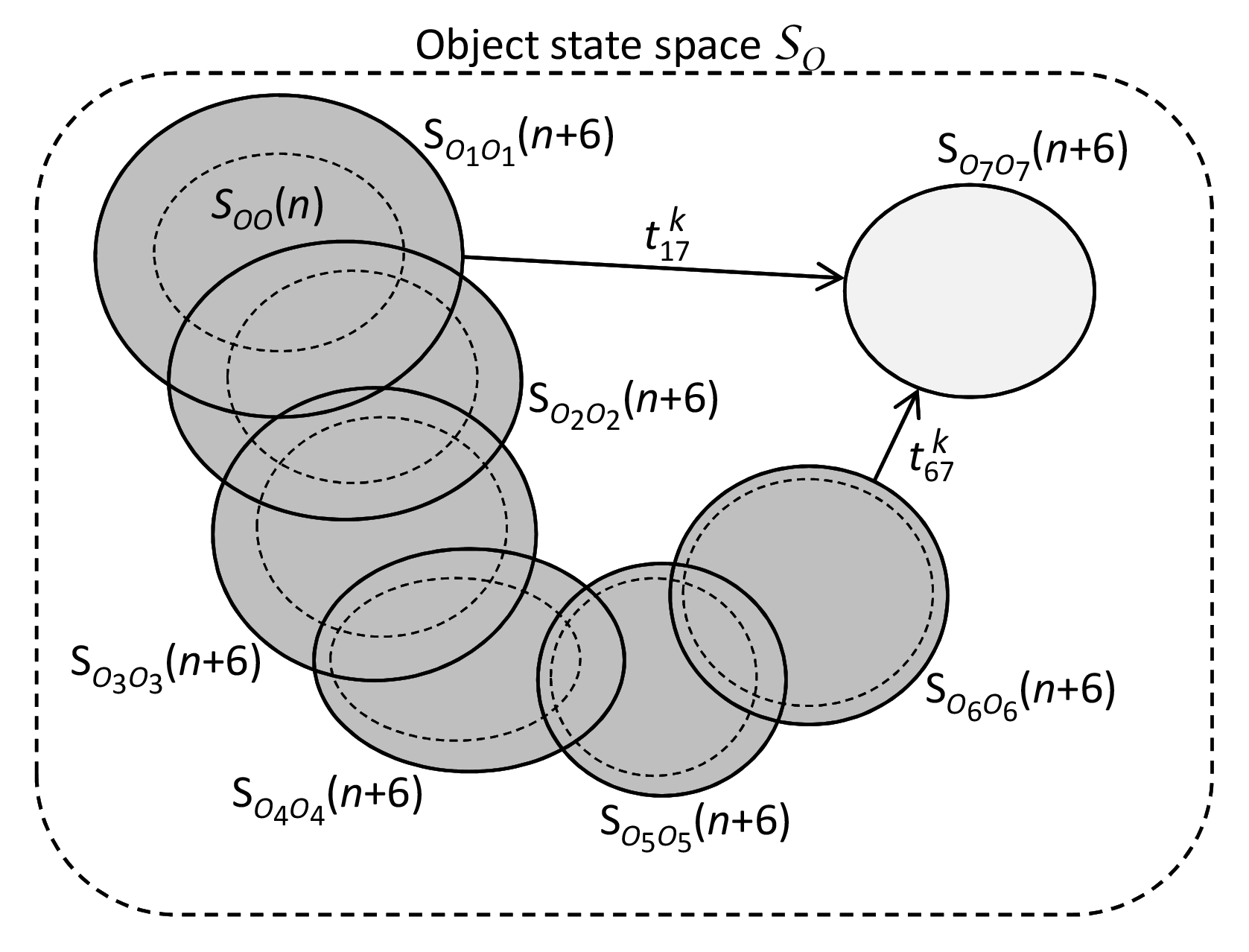}
\end{center}
\caption{The memories at time $n+6$ of the sequence of states shown in Fig. \ref {Figure1}. Each such memory is an object $O_{m}$ with state $S_{O_{m}O_{m}}(n+6)$ that refers to the same object $O$ at time $n+m-1$ with state $S_{OO}(n+m-1)$. At the given sequential time $n+6$, relational temporal distances $t_{ij}^{k}$ are defined between any two objects $O_{i}$ and $O_{j}$.}
\label{Figure8}
\end{figure}

The relation between sequential time $n$ and relational time $t$ becomes highlighted if we compare Fig. \ref{Figure8} with Fig. \ref{Figure1}. All object states shown in Fig. \ref{Figure8} are defined at the same time $n+6$ just after object $O$ in Fig. \ref{Figure1} has undergone a knowable change. This means that all other object states shown in Fig. \ref{Figure8} correspond to memories of past states of $O$. For example, we may write $S_{O_{1}O_{1}}(n+6)=M^{6}(S_{OO}(n))$ in the notation used in (\ref{perfectorder}). The fact that memories may be imperfect according to (\ref{sloppymemory}) is illustrated by the fact that remembered object states like $S_{O_{1}O_{1}}(n+6)$ (solid lines) contain the corresponding original object states like $S_{OO}(n)$ (dashed lines) as subsets. Temporal distances $t_{ij}^{k}$ are defined between pairs of the simultaneous object states shown in Fig. \ref{Figure8}, in contrast to pairs of the object states in Fig. \ref{Figure1}. All the latter states correspond to current object states ($Pr=1$) at the sequential time for which it is defined.

What value should be assigned to the temporal distance $t_{67}^{k}$ between the present object state $S_{O_{7}O_{7}}(n+6)$ in Fig. \ref{Figure8} and the memory $S_{O_{6}O_{6}}(n+6)$ of the immediately preceding state $S_{OO}(n+5)$ of the same object $O$ just before it knowably changed? The change defines a single event according to (\ref{eventdef}). Therefore there are no set of other events perceived by some subject $k$, no clock ticks, which can be placed in between these two object states, whose number would define $t_{67}^{k}$. Nevertheless, we cannot assign $t_{67}^{k}=0$, since if there were no temporal distance between a state before and after an event, there would be no temporal distances at all between any pair of object states. The only reasonable thing to do is to set $t_{67}^{k}=1$, reflecting the fact that events define the passage of time, and temporal distances $t_{ij}^{k}$ are defined by counting the number of events that subject $k$ can place between the perceptions of the object states $S_{O_{i}O_{i}}$ and $S_{O_{j}O_{j}}$.

This leads us to the question whether relational time $t$ can be treated as continuous at all. It seems that we must introduce a minimum temporal distance $t_{\min}=1$ that applies to the temporal distance between the memories of any two subsequent events in the list given in (\ref{eventorder}), defining two subsequent temporal updates $n\rightarrow n+1$ and $n+1\rightarrow n+2$. To each of these events we assign a corresponding unit increment of relational time $t$ according to the preceding paragraph, but between them nothing happens that could motivate any such increment. Actually, the words "between them" lack epistemic meaning.

\begin{figure}[tp]
\begin{center}
\includegraphics[width=80mm,clip=true]{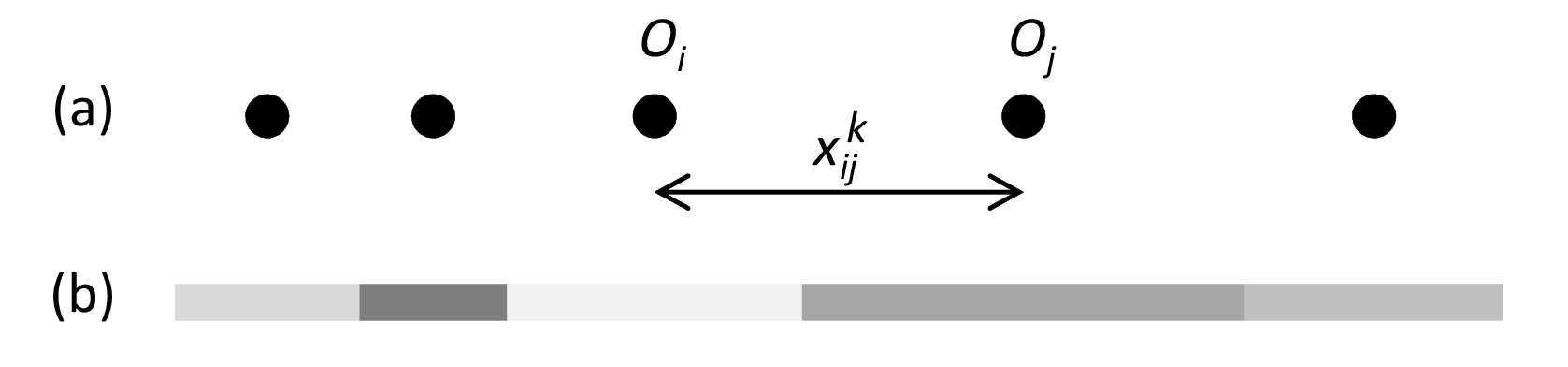}
\end{center}
\caption{The very notion that two objects $O_{i}$ and $O_{j}$ can be spatially distinguished implies that there is something in between them. This "something" corresponds to another object $O_{b}$, showing that there cannot be any closest possible pair of distinct objects, and no smallest non-zero spatial distance $x_{ij}^{k}$. This is true regardless if we (a) perceive $O_{i}$ and $O_{j}$ as spatially separated, or (b) as spatially extended.}
\label{Figure9}
\end{figure}

The situation is different when it comes to spatial distances $x_{ij}^{k}$ between two objects $O_{i}$ and $O_{j}$ with states $S_{O_{i}O_{i}}$ and $S_{O_{j}O_{j}}$, as illustrated in Fig. \ref{Figure9}. Whenever we perceive two such objects as spatially separated, so that we must assign $x_{ij}^{k}>0$, we also perceive at least one more object that is placed between them.

Let us elaborate on this point. Suppose that no object is explicitly seen between $O_{i}$ and $O_{j}$, making them a candidate of a closest possible pair. Then there are two alternatives. They may be separated by a perceived void, as shown in Fig. \ref{Figure9}(a). Then the void itself becomes the intermediate object $O_{b}$. Alternatively, $O_{i}$ and $O_{j}$ may be perceived as spatially extended and touching each other, where an internal property like brightness changes value at the interface, as shown in Fig. \ref{Figure9}(b). The necessary spatial extension of $O_{i}$ and $O_{j}$ means that $x_{ij}^{k}$ must be defined between their centers of mass, or by a similar criterion. By definition of spatial extension, we can then distinguish smaller parts $O_{b}$ between these centers of mass, between which smaller distances than $x_{ij}^{k}$ must be defined.

In either case we see that there cannot be any smallest spatial distance. Therefore the position of an object can be properly modeled as a continuous property.

\begin{figure}[tp]
\begin{center}
\includegraphics[width=80mm,clip=true]{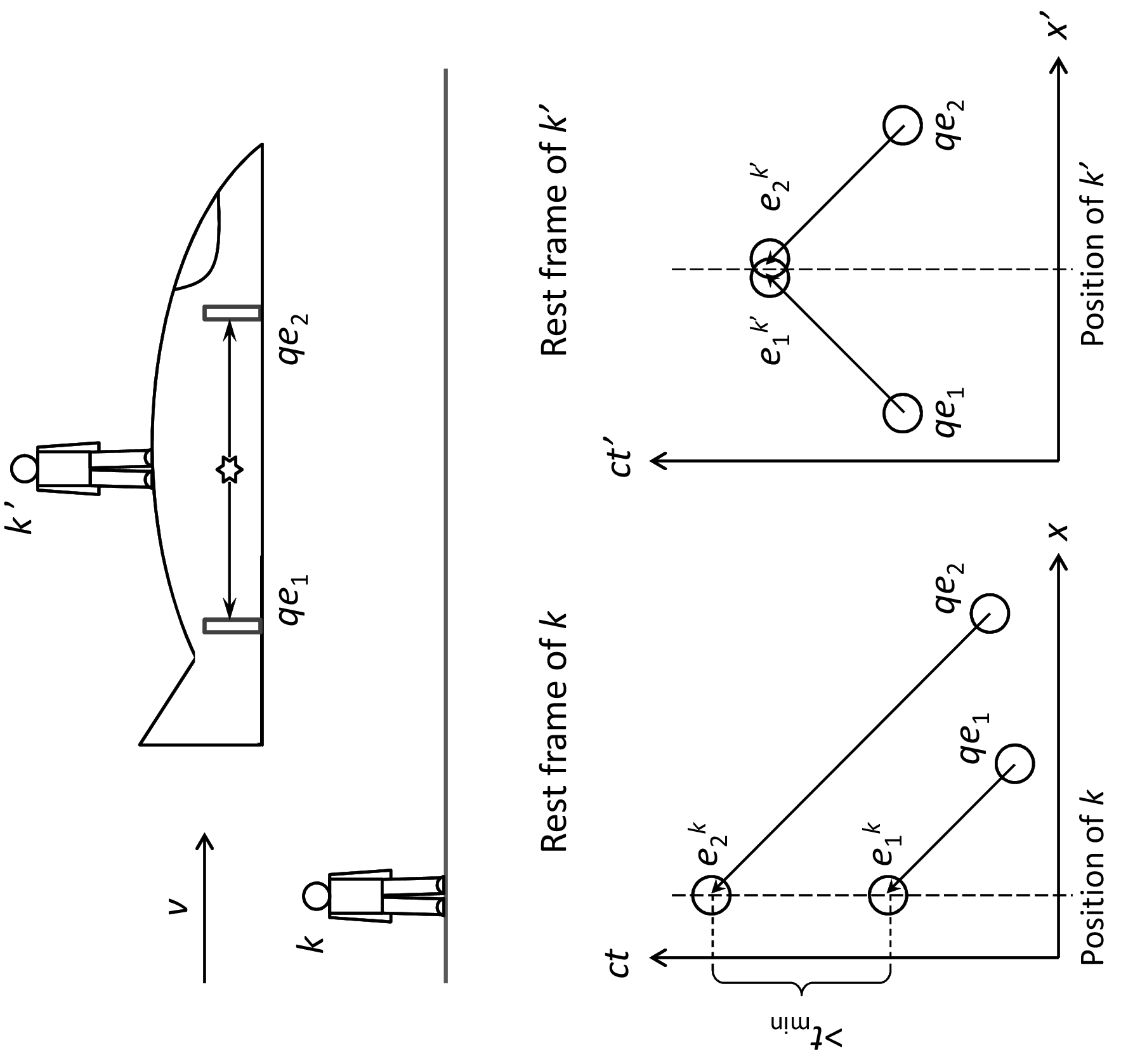}
\end{center}
\caption{The time difference between perceived events is either greater than $t_{\min}$ (like that between $e_{1}^{k}$ and $e_{2}^{k}$), or zero (like that between $e_{1}^{k'}$ and $e_{2}^{k'}$). All subjects agree on these statements. The fact that the temporal ordering of the quasievents $qe_{1}$ and $qe_{2}$ is ambiguous has no primary importance in our epistemic approach. The temporal distance between two quasievents may take any real value.} 
\label{Figure10}
\end{figure}

Figure \ref{Figure10} illustrates the fact that since space is continuous in this sense it is possible to preserve the notion of a continuous Minkowski space-time even though both sequential time $n$ and relational time $t$ apparently have to be modeled as discrete. Objects perceived by a subject $k$ are considered to be placed along the world line that is followed by the body of $k$ since that is where her perceptions arise if we want to model each subject as immersed in the same continuous space-time. The claim is then that the temporal distance $t_{12}^{k}$ measured by $k$ in her own rest frame between any two perceived objects $O_{1}$ and $O_{2}$ always fulfils $t_{12}^{k}=0$ or $t_{12}^{k}\geq t_{\min}$. (We drop the condition $t_{\min}=1$ to allow for arbitrary units.) These two objects correspond to two events $e_{1}^{k}$ and $e_{2}^{k}$ according to (\ref{eventdef}).

Subject $k$ may judge that the perceived objects correspond to two deduced quasiobjects $QO_{1}$ and $QO_{2}$ that are placed at some distance from $k$, the information of which arrives to $k$ at the speed of light. In Fig. \ref{Figure10} these two quasiobjects correspond to the events that a flash of light emitted from the middle of a space ship is reflected at a front and a rear mirror, respectively. These events are not perceived by anybody, and may thus be called two \emph{quasievents} $qe_{1}$ and $qe_{2}$. By analogy with (\ref{eventdef}) we may write

\begin{equation}
qe\leftrightarrow\{S_{QOQO}(n-1),S_{QOQO}(n),S(n-1)\}
\label{quasieventdef}
\end{equation}
with $S_{QOQO}(n-1)\cap S_{QOQO}(n)=\varnothing$.

Since spatial distances can be treated as continuous, the deduced temporal distance $t_{Q12}^{k}$ between $qe_{1}$ and $qe_{2}$ can be arbitrary small but non-zero if their separation is space-like, even though $t_{12}^{k}\geq t_{\min}$. In this example, another subject $k'$ perceives a pair of corresponding events $e_{1}^{k'}$ and $e_{2}^{k'}$ as simultaneous, so that $t_{12}^{k'}=0$. She also deduces that $t_{Q12}^{k'}=0$.

We see that from the present epistemic perspective the relativity of simultaneity applies to quasiobjects or quasievents with \emph{deduced} spatio-temporal location only, whereas the temporal ordering of \emph{perceived} events is unambiguous and universal according to (\ref{eventorder}). Also, we conclude that it is possible from the epistemic perspective to immerse all events known at a given sequential time $n$ in a Minkowski space-time, provided we distinguish between objects $O$ and quasiobjects $QO$, as well as between events $e$ and quasievents $qe$, and locate them in the space-time as indicated in Fig. \ref{Figure10}.

\section{The evolution parameter}
\label{evolp}

We experience that the world changes gradually. The shorter time that passes, the smaller change of the objects that we perceive. This is not self-evident in the present picture of time. The subjective ability to order events temporally according to (\ref{personalorder}) is simply assumed, regardless the similarity or dissimilarity of subsequent object states. However, a gradual change of the object states $S_{OO}$ is necessary in order to use (\ref{identifiable}) to define identifiable objects, and to be able to speak about the trajectory of a given object, as illustrated in Fig. \ref{Figure1}. Also, we argued that such identifiability is necessary in order to say that the world perceived at time $n+m$ is \emph{the same} as that perceived at some previous time $n$. Therefore it is essential that the evolution operators $u_{1}$ and $u_{O1}$ introduced in (\ref{sevolution}) and (\ref{objectevo}), respectively, expresses such a gradual change.

More than that, the overlapping subsequent object states shown in Fig. \ref{Figure1} make the evolution of an identifiable object $O$ seamless. It becomes impossible to tell two subsequent object states apart, but it may nevertheless be possible to perceive that the state of the object has changed after a longer period of time. For example, we see in Fig. \ref{Figure1} that the states $S_{OO}(n)$ and $S_{OO}(n+5)$ do not overlap, corresponding to the fact that they are possible to distinguish.

\begin{figure}[tp]
\begin{center}
\includegraphics[width=80mm,clip=true]{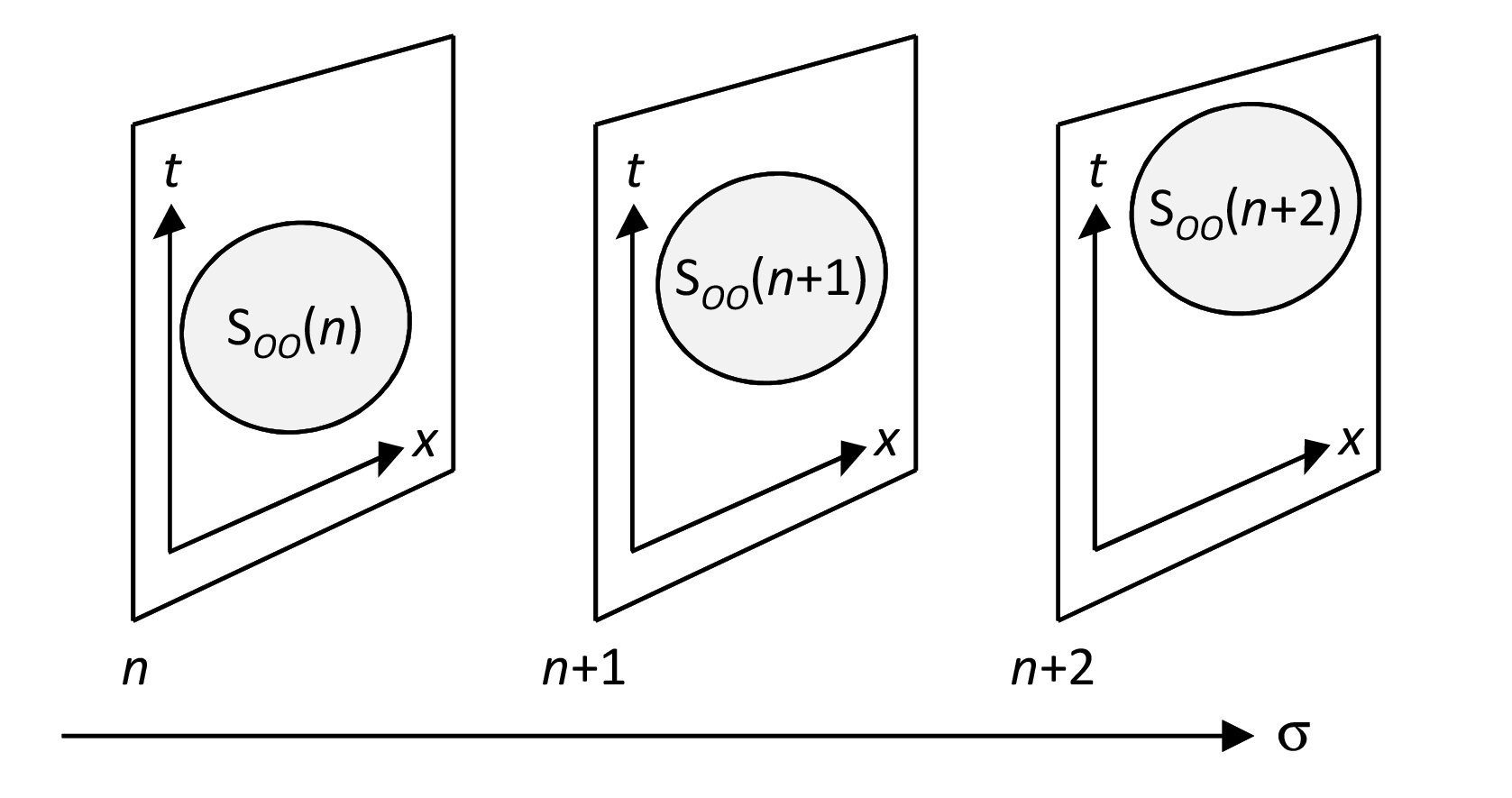}
\end{center}
\caption{The relation between the evolution parameter $\sigma$, sequential time $n$, and relational time $t$. There is an entire space-time spanned by the spatial axes $x$ and the temporal axis $t$ associated with each sequential time $n$, onto which each object state $S_{OO}(n)$ can be projected. The seamless evolution of $S_{OO}(n)$ as $n$ increases can be described using the continuous parameter $\sigma$.}
\label{Figure11}
\end{figure}

These facts make it natural to model the evolution of $O$ as if it follows a continuous trajectory. In this spirit, we may introduce a continuous operator $u_{O}(\sigma)$ which depends on an evolution parameter $\sigma\in\mathbf{\mathbb{R}}$ such that

\begin{equation}
u_{O}(\sigma_{m})=u_{Om},
\label{sigmadef}
\end{equation}
where $\sigma_{1}<\sigma_{2}<\sigma_{3}<\ldots$ and $u_{Om}$ is defined in (\ref{uomdef}). The relation between the three temporal quantities $\sigma$, $n$ and $t$ is illustrated schematically in Fig. \ref{Figure11}.

We argued in Section \ref{times} that sequential time $n$ is a transcendental quantity rather than an observable property. The same goes for $\sigma$ since it is defined via the flow of sequential time according to (\ref{sigmadef}). This means that we should not associate any Heisenberg uncertainty to its value, which is unknowable, just like the value $n$ of sequential time. Any invertible change of variables $\sigma'=f(\sigma)$ produces an equally valid evolution parameter $\sigma'$.

The evolution parameter $\sigma$ is useful to answer the question what the object $O$ will look like when we examine it the next time after an examination at time $n$, depending on how long we wait to do so. This waiting time is defined by the number $m$ of changes of \emph{other} objects $O'$ that are observed before we look at $O$ again, and $m$ may in principle be any positive integer. The adjustable $m$ and the seamless evolution of the object state make it meaningful to use $\sigma$ to express continuous evolution equations applying to any object $O$.

In contrast, it is not meaningful to express such a continuous evolution equation that applies to the entire world. All that can be said about $S(n+1)$ is a function of $S(n)$. These two states do not overlap, so that there is no sense in which the evolution of the world as a whole is seamless. Further, we cannot choose to wait an arbitrary amount $m$ of sequential time after time $n$ until we observe the world the next time, since $n+1$ corresponds by definition to the next time we look at it. The discrete mapping in (\ref{sevolution}) defined by the evolution operator $u_{1}$ is sufficient.

\section{Parameterization of physical law}
\label{parmeterlaw}

To be able to say that we understand physical law we have to be able to describe how the expected outcome of an observation depends on the variables that specify the context in which the observation takes place. In Section \ref{concepts} we introduced a general experimental context $C$. It is specified by the object state $S_{OO}(n)$ of the observed object $O$ at the time $n$ at which the context is initiated (Fig. \ref{Figure2}). This object state is specified by a set of properties $\{P_{Cj}\}$ with values $\{p_{Cj}\}$. Say that $C$ is designed to determine the value $p_{j}$ of property $P$ of the specimen $OS$ at some finite time $n+m$. Then an understanding of physical law might mean to know the function $f$ in the expression

\begin{equation}
S_{OS}(n+m-1)=f(S_{OS}(n),\{p_{Cj}\})
\label{falselaw}
\end{equation}
for each kind of specimen $OS$ and each experimental context $C$.

However, since we argue that potential knowledge is always incomplete \cite{ostborn1}, we cannot assume complete knowledge about the observed object $O$, including the specimen $OS$. This means that the knowledge about the set of property values $\{p_{Cj}\}$ is imprecise. This fuzzy knowledge cannot be coded as a set of arguments in a function. According to the principle of explicit epistemic minimalism we cannot express physical law properly if we assume a more precise knowledge about $\{p_{Cj}\}$ than we can ever get. Therefore the expression (\ref{falselaw}) is invalid. We conclude that we should not express fundamental physical law as a function of an observable property, to which we can always associate a Heisenberg uncertainty. These considerations also invalidate the expression

\begin{equation}
S_{OS}(n+m-1)=f(S_{OS}(n),S_{OA},t),
\label{falsetimelaw}
\end{equation}
where $S_{OA}$ is the state of the apparatus $OA$ at the start of the experiment (Fig. \ref{Figure2}), and $t$ is the observed time that passes between the start of the experiment and the final outcome. However, this is true as a matter of principle only. In practice we can, of course, use expressions like (\ref{falselaw}) and (\ref{falsetimelaw}), and we often do.

Regarding (\ref{falsetimelaw}) we note that if knowledge would have been complete we could have written $t=g(\{p_{Cj}\})$, since we define the context $C$ in such a way that the observer $OB$ does not interfere with the experiment after it is initiated at sequential time $n$. Figuratively speaking, she pushes a button at time $n$, and the experiment runs by itself until it reveals the outcome at some finite time $n+m$. She cannot choose during the course of the experiment when to make the observation of property $P$.

If we nevertheless want to express the outcome of an experiment as a function of some variable in a way that is correct as a matter of principle, we must use a precisely defined variable which has physical meaning, but which is not an observable property. The evolution parameter $\sigma$ fits this description. A valid expression if therefore

\begin{equation}
S_{OS}(n+m-1)=f(S_{OS}(n),S_{OA}(n),\sigma_{f}).
\label{sigmalaw}
\end{equation}

\begin{figure}[tp]
\begin{center}
\includegraphics[width=80mm,clip=true]{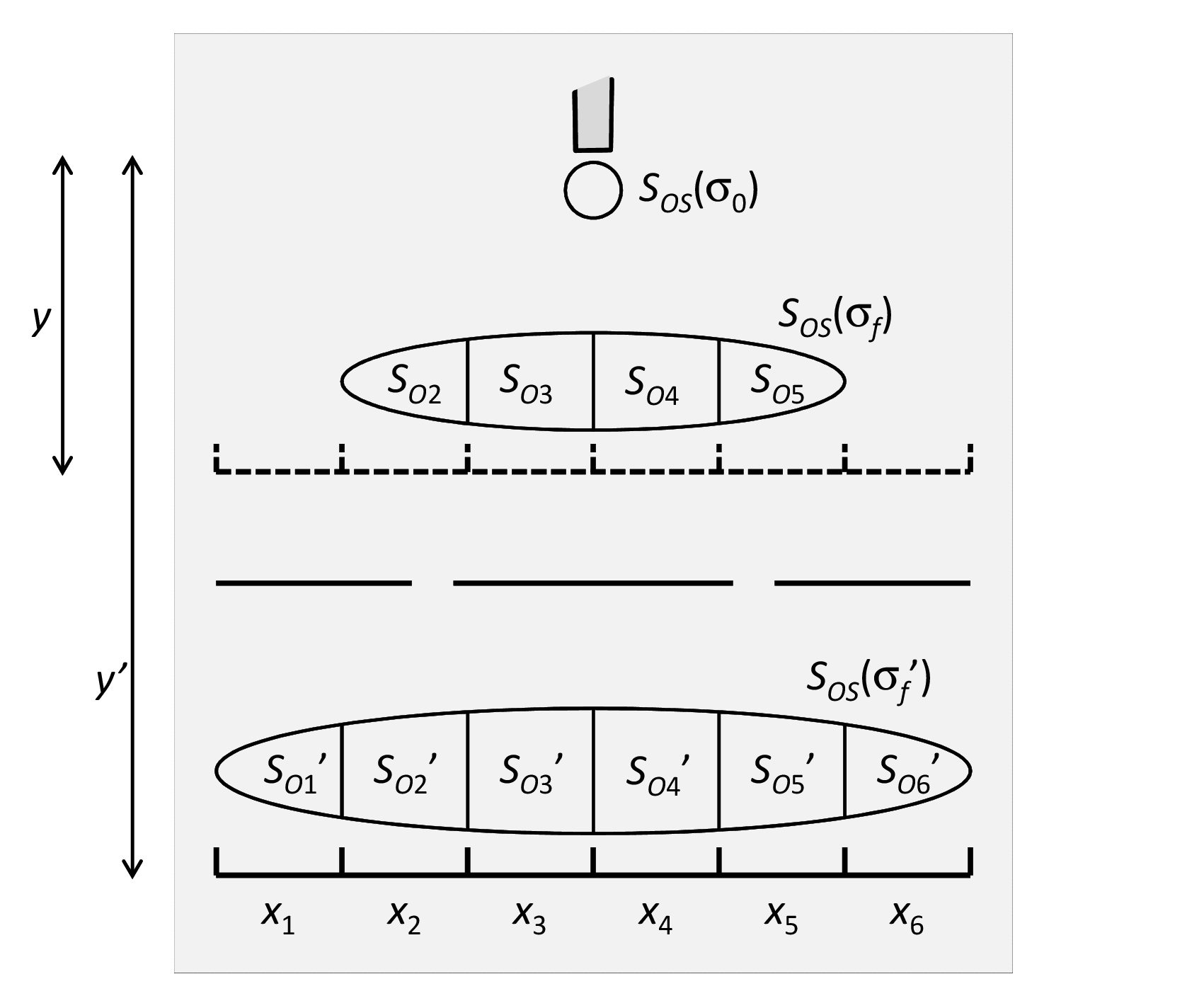}
\end{center}
\caption{Two members of a family $C(\sigma_{f})$ of observational contexts $C$, where $\sigma_{f}$ is the value of the evolution parameter $\sigma$ just before detection of the particle in a double-slit experiment at time $n+m$ after its ejection from a gun at time $n$. A change of $\sigma_{f}$ can be modeled in practice by a change of the position $y$ of the detector screen. This is possible since $d\sigma/dy>0$. The property $P$ that is observed in $C(\sigma_{f})$ is the position $x$ of the particle along the detector screen. The possible values are $p_{j}=x_{j}$. This set of values is the same in all contexts that belong to $C(\sigma_{f})$. The corresponding present alternatives $S_{Oj}(\sigma_{f})$ are defined according to Fig. \ref{Figure3}. To simplify the notation, we write $C(\sigma)$ in the main text rather than $C(\sigma_{f})$.}
\label{Figure12}
\end{figure}

By definition, $S_{OS}(n+m-1)$ is the state of the specimen just before the value $p_{j}$ of property $P$ is detected within the context $C$. Therefore the argument $\sigma_{f}$ in (\ref{sigmalaw}) is not the evolution parameter that interpolates smoothly between sequential times $n$ and $n+m$, between the initiation of the experiment and the final outcome, but always corresponds to the final value $\sigma_{f}$ of $\sigma$ at time $n+m-1$ just before detection that defines the update to time $n+m$. The situation is illustrated in Fig. \ref{Figure12}. Different values of the argument $\sigma_{f}$ in (\ref{sigmalaw}) therefore corresponds to different experimental contexts $C$. We get a family of contexts $C(\sigma_{f})$. To make the notation simpler we drop the subscript on $\sigma$ in the following, writing $C(\sigma)$. The meaning of the argument should be understood according to the preceding discussion.   

We argued in Section \ref{times} that the value of sequential time $n$ is unknowable, and that $\sigma$ is an arbitrary parametrization of the flow of sequential time that fulfils (\ref{sigmadef}). Therefore it is impossible to determine the parametrized physical law expressed in (\ref{sigmalaw}) empirically. What we can do is to mimic a change of $\sigma$ by a change of a property $P_{C}$, such as the distance $y$ between the particle gun and the detector screen in Fig. \ref{Figure12}. In so doing, we require that $d\sigma/dp_{c}>0$. We get an empirical family of contexts $C(p_{C})$ that we use to estimate physical law, as expressed in (\ref{sigmalaw}).

However, as discussed above, we cannot skip the step of introducing $\sigma$ as an argument in (\ref{sigmalaw}), sticking instead to (\ref{falselaw}). As a matter of principle, physical law transcends the present time $n$. However, to get to know it we have no choice but to use as tools the perceivable properties at time $n$, such as the observed distance $y$ between the particle gun and the detector screen in Fig. \ref{Figure12} at the start of the experiment, together with the memories or records at that time $n$ of previous choices of $y$. To put it more succint: there is a physical law that transcends the empirical evidence, but we cannot transcend the empirical evidence to learn about it.

\section{The wave function}
\label{wavef}

Assume that the set of values $\{p_{j}\}$ that are possible to observe is the same in all contexts in the family $C(\sigma)$, as illustrated in Fig. \ref{Figure12}. Then the property value states $S_{Pj}$ introduced in relation to (\ref{pvspace}) do not depend on $\sigma$. The same goes for the algebraic representations $\overline{S}_{Pj}$ of these states [Fig. \ref{Figure5})], so that the same Hilbert space $\mathcal{H}_{C}$ can be used to represent all contexts in the family $C(\sigma)$. We may therefore write

\begin{equation}
\overline{S}_{C}(\sigma)=\sum_{j}a_{j}(\sigma)\overline{S}_{Pj},
\label{sumrepf}
\end{equation}
where $\overline{S}_{C}(\sigma)\in\mathcal{H}_{C}$ is the algebraic representation of the \emph{contextual state} $S_{C}(\sigma)$. It corresponds to the potential knowledge at time $n+m-1$ (just before the observation of property $P$) about the values of the set of properties $ \{P,P',\ldots\}$ of the specimen $OS$ observed within the context $C(\sigma)$, together with the knowledge about the nature of $OS$ \cite{ostborn1}. The \emph{contextual numbers} $a_{j}$ correspond to the probability amplitudes in the conventional formulation of quantum mechanics, except for the fact that the numbers $a_{j}$ are not associated with the state of the specimen $OS$ in itself (Fig. \ref{Figure2}), but are defined within the experimental context $C$ only. This means that the probability $q_{j}$ to observe the value $p_{j}$ within the context $C(\sigma)$ fulfils $q_{j}=|a_{j}|^{2}$ according to Born's rule, where $a_{j}\in\mathbf{\mathbb{C}}$.

It should be noted that the algebraic notation in (\ref{sumrepf}) is just a convenient formal representation of the essential aspects of the potential knowledge about the experimental context $C(\sigma)$. The set of terms in the sum (\ref{sumrepf}) corresponds to the set of values $\{p_{j}\}$ of property $P$ of the specimen $OS$ that cannot be excluded at time $n+m-1$, just before $P$ is observed. The use of the symbol $+$ to separate two elements in the corresponding set $\{\overline{S}_{Pj}\}$ is motivated by the fact that it is possible to use the distributive law according to

\begin{equation}
a_{1}(a_{2}+a_{3})\overline{S}_{Pj}=(a_{1}a_{2}+a_{1}a_{3})\overline{S}_{Pj}=a_{1}a_{2}\overline{S}_{Pj}+a_{1}a_{3}\overline{S}_{Pj}
\label{distributive}
\end{equation}
for any triplet $(a_{1},a_{2},a_{3})$ of contextual numbers \cite{ostborn1}. In an analogous fashion, the fact that the property value states are mutually exclusive in the sense that $S_{OS}\subseteq S_{Pj}\Rightarrow S_{OS}\cap S_{Pj'}=\varnothing$ whenever $j\neq j'$ can be represented by the algebraic orthonormality relation

\begin{equation}
\langle\overline{S}_{Pj},\overline{S}_{Pj'}\rangle=\delta_{jj'}.
\label{orthonormal}
\end{equation}

We define the wave function $a_{P}(p,\sigma)$ such that

\begin{equation}
a_{P}(p_{j},\sigma)=a_{j}(\sigma).
\label{wfdef}
\end{equation}
The domain of $a_{P}(p,\sigma)$ is $(\{p_{1},p_{2},\ldots,p_{M}\},[0,\sigma_{\max}])$ for some arbitrary finite $\sigma_{\max}$ that depends on the parametrization, and the details of the family of contexts $C(\sigma)$. Note that we have to add the index $P$ to the wave function since it is only defined together with the property $P$ that we are about to observe. The function $a(p,\sigma)$ has no physical meaning in itself.

We may define the wave function evolution operator $u_{P}(\sigma)$ by the relation

\begin{equation}
a_{P}(p,\sigma)\equiv u_{P}(\sigma)a_{P}(p,0).
\label{evolutionequation}
\end{equation}
From a conceptual point of view, it is important to note that $u_{P}(\sigma)$ represents a map from one experimental context $C(0)$ to another context $C(\sigma)$ rather than a map from the initial state of the specimen $OS$ to a state at a later time.

\begin{figure}[tp]
\begin{center}
\includegraphics[width=80mm,clip=true]{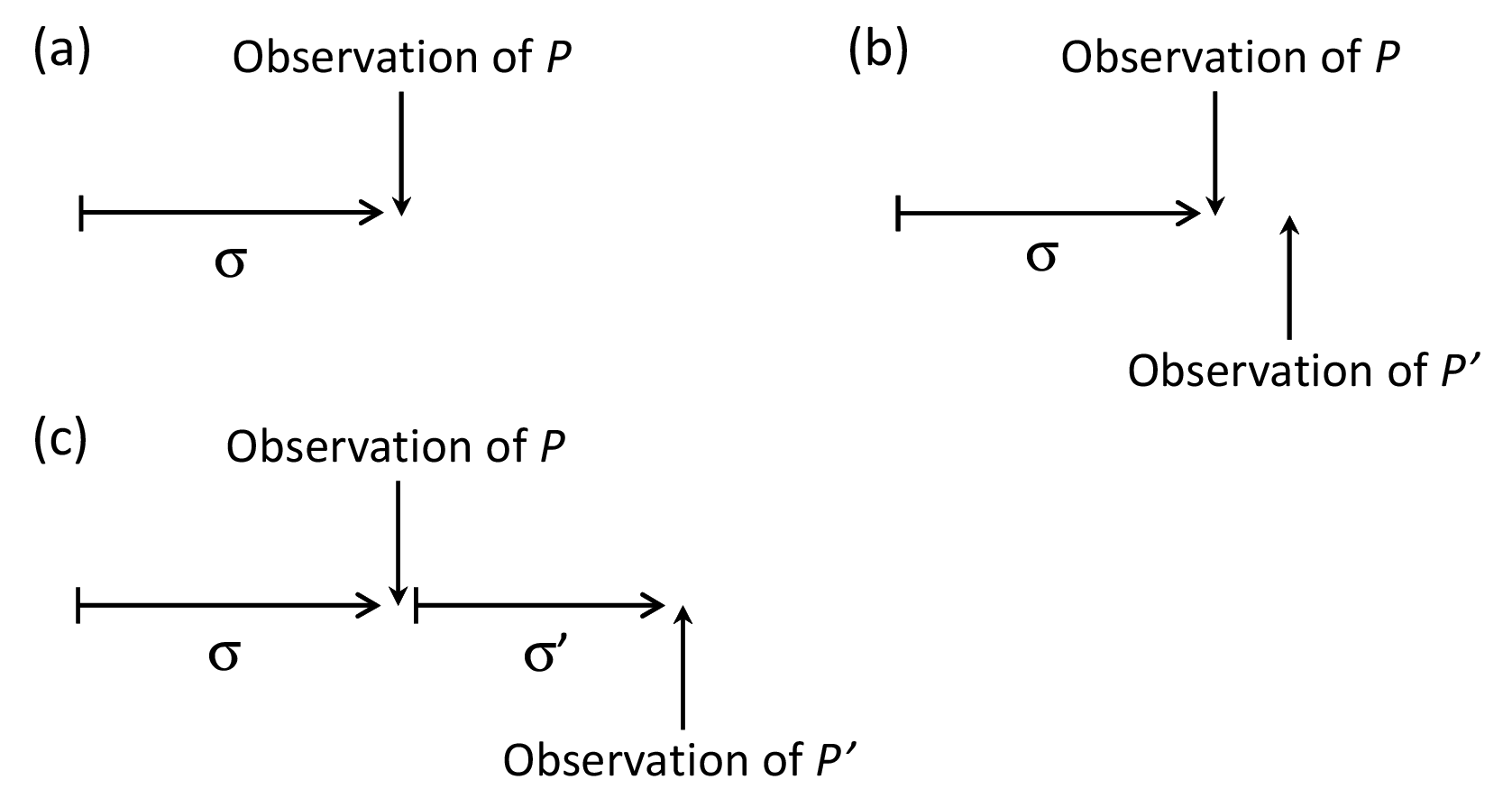}
\end{center}
\caption{(a) A family of contexts $C(\sigma)$ in which one property $P$ is observed. (b) A family of contexts $C(\sigma)$ in which two properties $P$ and $P'$ are observed in succession, and the relation between the two observations is kept fixed. (c) A family of contexts $C(\sigma,\sigma')$ in which two properties $P$ and $P'$ are observed in succession, and the relation between the two observations is varied.}
\label{Figure13}
\end{figure}

Since the wave function is defined from the set of $M$ individual amplitudes $a_{j}(\sigma)$ in the formal sum (\ref{sumrepf}) we may write

\begin{equation}
u_{P}(\sigma)a_{P}(p,0)\leftrightarrow\{u_{P}(\sigma)a_{j}(0)\}.
\label{protolinearev}
\end{equation}

For each $\sigma\in[0,\sigma_{\max}]$ the number $|a_{j}(\sigma)|^{2}$ corresponds to a probability for a predefined outcome of the experimental context $C(\sigma)$. Therefore we must have $|\sum_{j}a_{j}(\sigma)|^{2}=1$ for each such $\sigma$. This requirement corresponds to the condition that the wave function evolution operator is unitary:

\begin{equation}
|| u_{P}(\sigma)a_{P}(p,0)||=|| a_{P}(p,0) ||=1.
\label{unitaryevo}
\end{equation}

When the property $P$ is observed, the wave function $a_{P}(p,\sigma)$ is no longer defined. We may, however, consider a situation like that illustrated in Fig. \ref{Figure13}(b), where two properties $P$ and $P'$ are observed in succession, the experimental setup that determines the timing of the observation of $P$ is varied, but the part of the experimental setup that determines the relation between the observations of $P$ and $P'$ is kept fixed. Then we are considering a family of contexts $C(\sigma)$ for which we may define the wave function $a_{PP'}(p,p',\sigma)$. In that case, when $P$ is observed, the wave function collapses to another function $a_{P'}'(p')$, which may depend on the observed value $p_{j}$ of $P$. Finally, when $P'$ is observed, there is no wave function at all defined. We may also consider a situation like that in Fig. \ref{Figure13}(c), where the relation between the observations of $P$ and $P'$ is varied. Then we get a family of contexts $C(\sigma,\sigma')$ with two independent evolution parameters. We may, of course, consider even more complex families of experimental contexts, but the above discussion makes the essential points clear: a wave function is intimately tied to a specific type of experimental context, and we may introduce several evolution parameters in the same wave function.

\begin{figure}[tp]
\begin{center}
\includegraphics[width=80mm,clip=true]{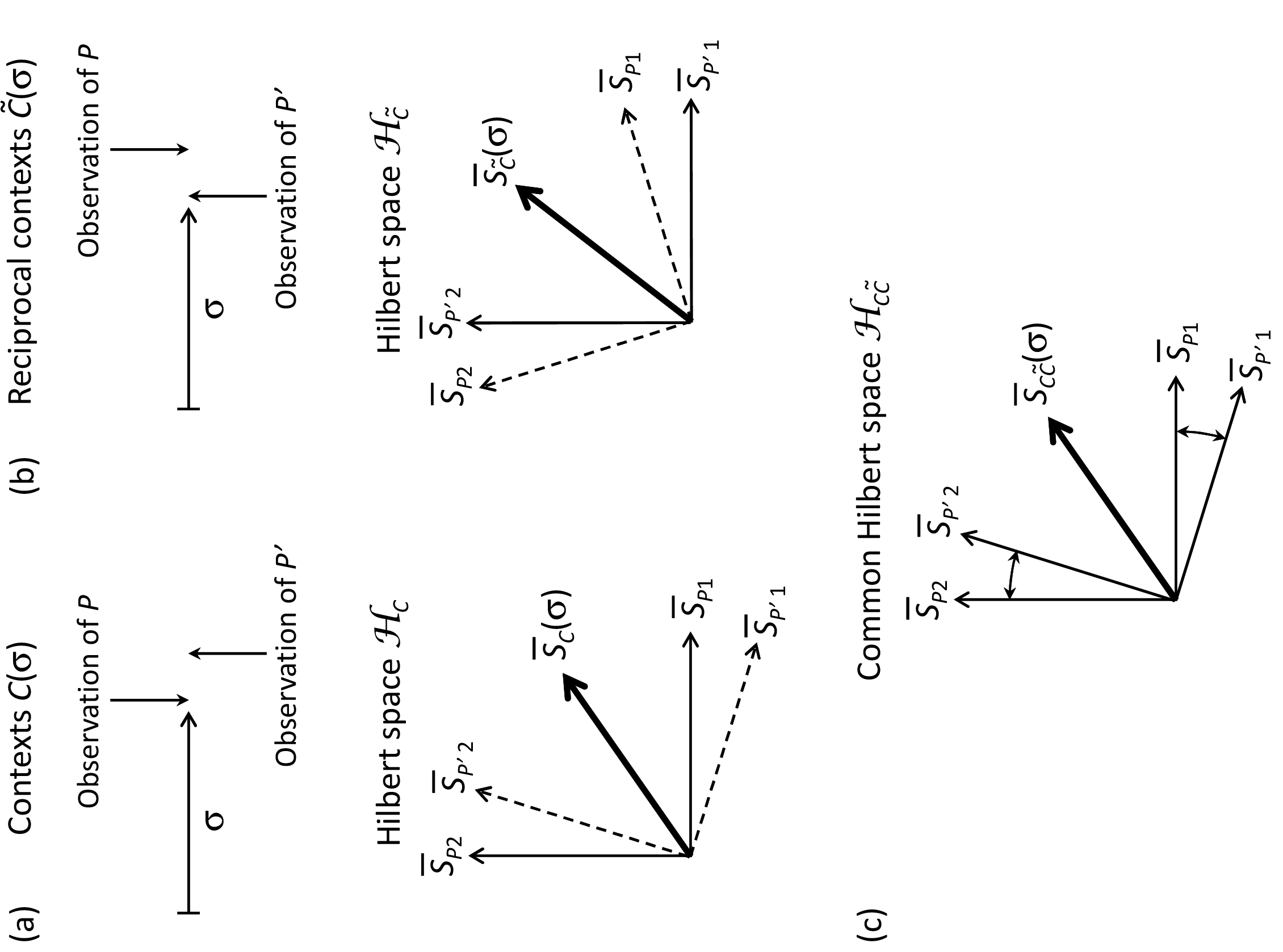}
\end{center}
\caption{(a) A family of contexts $C(\sigma)$ in which two properties $P$ and $P'$ are observed, which are not simultaneously knowable. In the corresponding Hilbert space $\mathcal{H}_{C}$ the state vector $\overline{S}_{C}(\sigma)$ represents the state just before $P$ is observed. The basis $\{\overline{S}_{Pj}\}$ is therefore primary. (b) A family of reciprocal contexts $\tilde{C}(\sigma)$ in which the order of observation is reversed. In the reciprocal Hilbert space $\mathcal{H}_{\tilde{C}}$ the state vector $\overline{S}_{\tilde{C}}(\sigma)$ represents the state just before $P'$ is observed. The basis $\{\overline{S}_{P'j'}\}$ is therefore primary. (c) We may define the joint family of contexts $C\tilde{C}(\sigma)$ and represent it in a common Hilbert space $\mathcal{H}_{C\tilde{C}}$. Here we may identify $\overline{S}_{C}(\sigma)=\overline{S}_{\tilde{C}}(\sigma)$ and the two bases $\{\overline{S}_{Pj}\}$ and $\{\overline{S}_{P'j'}\}$ get equal weight and a change of basis corresponds to a change of perspective from context $C$ to its reciprocal $\tilde{C}$ or vice versa.}
\label{Figure14}
\end{figure}

In the following discussion about evolution equations, we will make use of a particular kind of pairs of context families $C(\sigma)$ and $\tilde{C}(\sigma)$. For any $\sigma\in[0,\sigma_{\max}]$, $\tilde{C}(\sigma)$ is the \emph{reciprocal context} to $C(\sigma)$ in the sense introduced in relation to Fig. 36 in Ref. \cite{ostborn1}. To summarize the idea, we consider a context $C$ in which two properties $P$ and $P'$ that are not simultaneously knowable are observed. If the number $M$ and $M'$ of alternative observable values of $P$ and $P'$ is the same, we may represent the context $C$ in a Hilbert space $\mathcal{H}_{C}$ with dimension $D_{H}=M=M'$, in which the sets $\{\overline{S}_{Pj}\}$ and $\{\overline{S}_{P'j'}\}$ of property values state representations are two orthonormal bases (Fig. \ref{Figure14}). The reciprocal $\tilde{C}(\sigma)$ is a context where the order in which $P$ and $P'$ is observed is reversed, and the associated contextual numbers $\tilde{a}_{j}$ are such that the state vector $S_{\tilde{C}}(\sigma)$ can be identified with $S_{C}(\sigma)$ after a change of primary basis [Fig. \ref{Figure14}(b)-(c)]. We may then introduce the common state vector $\overline{S}_{C\tilde{C}}(\sigma)$ which fulfils

\begin{equation}
\overline{S}_{C\tilde{C}}(\sigma)=\sum_{j}^{M}a_{P}(p_{j},\sigma)\overline{S}_{Pj}=\sum_{j'}^{M}\tilde{a}_{P'}(p_{j'}',(\sigma)\overline{S}_{P'j'}.
\label{basischange}
\end{equation}

To any pair of properties $P$ and $P'$ observed in the context $C(\sigma)$ or its reciprocal $\tilde{C}(\sigma)$ we can associate self-adjoint operators $\overline{P}$ and $\overline{P}'$ according to (\ref{propop}). The introduction of the joint family of contexts $C\tilde{C}(\sigma)$ according to Fig. \ref{Figure14}(c) makes it possible to apply both $\overline{P}$ and $\overline{P}'$ to the same state vector $\overline{S}_{C\tilde{C}}(\sigma)$, writing

\begin{equation}\begin{array}{lllll}
\overline{P}\overline{S}_{C\tilde{C}} & = & \sum_{j}(\overline{P}_{P}a_{P})\overline{S}_{Pi} & = & \sum_{j'}(\overline{P}_{P'}\tilde{a}_{P'})\overline{S}_{P'j}\\
\overline{P}'\overline{S}_{C\tilde{C}} & = & \sum_{j}(\overline{P}_{P}'a_{P})\overline{S}_{Pi} & = & \sum_{j'}(\overline{P}_{P'}'\tilde{a}_{P'})\overline{S}_{P'j}
\label{waveops}
\end{array}\end{equation}
for a quadruplet of operators $(\overline{P}_{P},\overline{P}_{P'},\overline{P'}_{P},\overline{P}_{P'}')$, where we have suppressed the arguments for clarity. These may be called \emph{wave function property operators}. They are less general than the operators $\overline{P}$ and $\overline{P}'$ in the sense that they are defined in joint families of contexts $C\tilde{C}(\sigma)$ only. The wave function operator $\overline{P}_{P}'$ may be called the representation of the property operator $\overline{P}'$ in the basis $\{\overline{S}_{Pi}\}$ of the common Hilbert space $\mathcal{H}_{C\tilde{C}}$, and corresponding names may be given to the other three wave function operators. Since $\overline{P}$ and $\overline{P}'$ are self-adjoint operators with a complete basis in $\mathcal{H}_{C\tilde{C}}$ by definition, the same goes for the corresponding four wave function operators defined by (\ref{waveops}).

Let us turn the perspective around and consider an arbitrary self-adjoint operator $\overline{\tilde{P}}$ that acts in the Hilbert space $\mathcal{H}_{C}$ defined in a context in which property $P$ is observed, just as we did in relation to Fig. \ref{Figure5}. Then the wave function $a_{P}$ and the basis $\{\overline{S}_{Pi}\}$ are defined, and we may write

\begin{equation}
\overline{\tilde{P}}\overline{S}_{C\tilde{C}}=\sum_{j}(\overline{\tilde{P}}_{P}a_{P})\overline{S}_{Pi},
\label{oto}
\end{equation}
where $\overline{\tilde{P}}_{P}$ is the representation of $\overline{\tilde{P}}$ in this basis. We concluded in relation to Fig. \ref{Figure5} that if $\overline{\tilde{P}}$ has a complete set of $M$ eigenvectors $\{\overline{S}_{\tilde{P}j}\}$ that span $\mathcal{H}_{C}$, then it may be associated with a property $\tilde{P}$ observed within an extended context in which $\tilde{P}$ is observed after property $P$. The one-to-one correspondence between $\overline{\tilde{P}}$ and $\overline{\tilde{P}}_{P}$ expressed by (\ref{oto}) means that the same holds for $\overline{\tilde{P}}_{P}$.

In other words, to any self-adjoint operator $\overline{\tilde{P}}_{P}$ that acts on a wave function $a_{P}$ defined in a context family $C(\sigma)$ in which property $P$ is observed we can associate another property $\tilde{P}$ observed in a joint family of contexts $C\tilde{C}(\sigma)$ according to Fig. \ref{Figure14}. This is so provided that $\overline{\tilde{P}}_{P}$ has a complete set of eigenvectors $\{\overline{S}_{\tilde{P}j}\}$. This fact together with the reverse fact that we can always associate a wave function operator $\overline{\tilde{P}}_{P}$ to a property $\tilde{P}$ according to (\ref{waveops}) can be summarized in the relation

\begin{equation}
\overline{\tilde{P}}_{P}\leftrightarrow\tilde{P}.
\label{opprop}
\end{equation}
We will refer to this relation in the following when we define the properties momentum, energy and rest energy from self-adjoint operators that act on a spatio-temporal wave function.

A major goal of this paper is to carefully motivate evolution equations like the Klein-Gordon and the Dirac equation from an epistemic point of view. These differential equations specify such a spatio-temporal wave function. More precisely, they provide the probability amplitudes to observe a particle at a given position at a given time. From our perspective, this means that the observed property $P$ is the spatio-temporal position $(R,T)$, and the values $(r,t)$ of this property are clearly treated as continuous variables. In contrast, we have stressed that all properties observed within any context $C$ can take a finite set $\{p_{j}\}$ of $M$ discrete values $p_{j}$ only. We argue that it is nevertheless appropriate to consider continuous property values in evolution equations of the form (\ref{evolutionequation}) in certain cases. 

Assume that it is possible to order the set $\{p\}$ of possible values $p$ of property $P$ according to magnitude, so that we may write $\{p_{j}\}=\{p_{1},p_{2},\ldots,p_{M}\}$ with $p_{1}<p_{2}<\ldots<p_{M}$. Assume also that for each experimental context $C$, and for each pair of values $(p_{i},p_{j})$ in the corresponding set of observable values $\{p_{j}\}$, the property $P$ is such that there is always a value $p\in\{p\}$ that fulfils $p_{i}<p<p_{j}$. This is certainly true for each component of $r_{4}$ according to the discussion in relation to Fig. \ref{Figure9}. This means that for given smallest and largest observable values $p_{1}$ and $p_{M}$ there is no upper limit to $M$ in principle. Assume further that we restrict our interest to contexts for which $|a_{P}(p_{j+1},\sigma)-a_{P}(p_{j},\sigma)|/|a_{P}(p_{j},\sigma)|<<1$ for all $\sigma$ and for all $1\leq j\leq M$.

In such context families, we may then define the familiar continuous, piecewise differentiable wave function $\Psi_{P}(p,\sigma)$ according to

\begin{equation}
\Psi_{P}(p_{0j},\sigma)dp \equiv a_{P}(p_{j},\sigma),
\label{psi}
\end{equation}
for $p_{j}\leftrightarrow[p_{0j}, p_{0j}+dp)$ in the limit $M\rightarrow\infty$, where $p$ is any real number in the interval $[p_{1},p_{M}]$. This idealized wave function can be used to describe a realistic context with a finite number $M$ of observable values $p_{j}$ provided that $a_{P}(p_{j},\sigma)=\int_{p_{0j}}^{p_{0j}+\Delta p_{j}}\Psi_{P}(p,\sigma)dp$ for each $p_{j}$, where this value corresponds to an unresolved bin

\begin{equation}
p_{j}\leftrightarrow[p_{0j}, p_{0j}+\Delta p_{j})
\label{contpropv}
\end{equation}
of continuous property values. It is reasonable to assume that this condition is fulfilled when $M$ is reduced to realistic values by simply reducing the resolution power of the detector $OD$ without changing anything else in the experimental setup. However, this is not a logical necessity.

We may use the continuous wave function to formulate a continuous contextual state representation

\begin{equation}
\overline{S}_{C}(\sigma)=\int_{p\in D_{P}} \Psi_{P}(p,\sigma)dp\overline{S}_{P}(p).
\label{formalint}
\end{equation}
The meaning of the integration is just that all values of $P$ outside the support $D_{P}=[p_{1},p_{\max}]$ can be exluded as an outcome of the observation of $P$, based on the potential knowledge before the observation. It does not make sense to actually calculate the integral - there is nothing to add up. It should be seen as a purely formal representation of the contextual state just before an observation, analogous to the formal sum in (\ref{sumrepf}).

The vector $\overline{S}_{P}(p)$ in (\ref{formalint}) is defined according to the relation

\begin{equation}
\overline{S}_{Pj}\equiv\Delta p_{j}^{-1/2}\int_{p_{0j}}^{p_{0j}+\Delta p_{j}}\overline{S}_{P}(p)dp.
\label{contstate}
\end{equation}
This relation expresses the fact that given the property value state $S_{Pj}$, where the property value $p_{j}$ corresponds to the interval given in (\ref{contpropv}), we cannot exclude any of the `continuous property value states' $S_{P}(p)$.

Equations (\ref{orthonormal}) and (\ref{contstate}) imply
\begin{equation}\begin{array}{lll}
\delta_{jj'} & = & \langle\overline{S}_{Pj},\overline{S}_{Pj'}\rangle\\
& = & (\Delta p_{j}\Delta p_{j'})^{-1/2}\int_{p_{0j}}^{p_{0j}+\Delta p_{j}}\int_{p_{0j'}}^{p_{0j'}+\Delta p_{j}'}\langle\overline{S}_{P}(p),\overline{S}_{P}(p')\rangle dpdp',
\end{array}\end{equation}
which is fulfilled if and only if we identify
\begin{equation}
\langle\overline{S}_{P}(p),\overline{S}_{P}(p')\rangle=\delta(p-p').
\label{deltarel}
\end{equation}
Note that the introduction of the continuous property value state representation $\overline{S}_{P}(p)$ and the Dirac delta function is needed only in the integral representation of the contextual state $\overline{S}_{C}$, and that this integral representation is, at best, a convenient alternative to the summation representation (\ref{sumrepf}), which reflects the actual physical experimental context, with its finite number of possible outcomes.

Let us define the continuous wave function evolution operator $u_{P}^{\Psi}$ by the relation

\begin{equation}
\Psi_{P}(p,\sigma)\equiv u_{P}^{\Psi}(\sigma)\Psi_{P}(p,0).
\label{cevolutionequation}
\end{equation}
by analogy with (\ref{evolutionequation}). This evolution operator can be seen as linear in a restricted sense. Let us define the piecewise wave functions $\Psi_{Pj}$ according to

\begin{equation}
\Psi_{Pj}(p,\sigma)=\left\{
\begin{array}{ll}
\Psi_{P}(p,\sigma), & p\in[p_{0j},p_{0j}+\Delta p_{j})\\
0, & p\not\in[p_{0j},p_{0j}+\Delta p_{j})
\end{array}\right.,
\end{equation}
where $p_{0j}$ is defined in (\ref{contpropv}). We get $\Psi_{P}(p,\sigma)=\sum_{j}\Psi_{Pj}(p,\sigma)$, and

\begin{equation}
u_{P}^{\Psi}(\sigma)\sum_{j}\Psi_{Pj}(p,0)=\sum_{j}u_{P}^{\Psi}(\sigma)\Psi_{Pj}(p,0)
\label{linearev}
\end{equation}
from (\ref{protolinearev}). This linearity of $u_{P}^{\Psi}(\sigma)$ is restricted since it is only defined for the decomposition of $\Psi_{P}$ into the piecewise wave functions $\{\Psi_{Pj}\}$.

The evolution operator $u_{P}^{\Psi}$ has to be unitary for the same reason as $u_{P}$ has to be unitary, as expressed in relation to (\ref{unitaryevo}). We may therefore write

\begin{equation}
|| u_{P}^{\Psi}(\sigma)\Psi_{P}(p,0) ||=|| \Psi_{P}(p,0) ||=1,
\label{unitarycevo}
\end{equation}
where the unitarity in this case corresponds to the requirement that $\int|\Psi_{P}(p,\sigma)|^{2}dp=1$ for each $\sigma\in[0,\sigma_{\max}]$.

When we are considering the common Hilbert space $\mathcal{H}_{C\tilde{C}}$ of a context $C$ and its reciprocal $\tilde{C}$, as described in relation to Fig. \ref{Figure14}, we may write down the relations that correpond to (\ref{basischange}) and (\ref{waveops}) in the continuous case. We get

\begin{equation}
\overline{S}_{C\tilde{C}}=\int\Psi_{P}(p,\sigma)dp\overline{S}_{P}(p)=\int\tilde{\Psi}_{P'}(p',\sigma)dp'\overline{S}_{P'}(p'),
\label{contbasechange}
\end{equation}
and

\begin{equation}\begin{array}{lllll}
\overline{P}\overline{S}_{C\tilde{C}} & = & \int(\overline{P}_{P}^{\Psi}\Psi_{P})dp\overline{S}_{P}(p) & = &
\int(\overline{P}_{P'}^{\Psi}\tilde{\Psi}_{P'})dp'\overline{S}_{P'}(p')\\
\overline{P'}\overline{S}_{C\tilde{C}} & = & \int(\overline{P'}_{P}^{\Psi}\Psi_{P})dp\overline{S}_{P}(p) & = &
\int(\overline{P'}_{P'}^{\Psi}\tilde{\Psi}_{P'})dp'\overline{S}_{P'}(p')
\end{array}
\label{psiops}
\end{equation}
where we have dropped the arguments of the wave functions for notational clarity. In the following, we will see that the usual operators that correspond to momentum and energy in the Schr\"odinger picture of conventional quantum mechanics are examples of the continuous wave function operator type $\overline{P'}_{P}^{\Psi}$, as defined in (\ref{psiops}). What we will do, in effect, is to associate the properties momentum and energy to self-adjoint operators $\overline{\tilde{P}}_{R_{4}}^{\Psi}$ that appear in the derivation of evolution equations. In so doing, we make use of the continuous version

\begin{equation}
\overline{\tilde{P}}_{P}^{\Psi}\leftrightarrow\tilde{P}
\label{copprop}
\end{equation}
of the (qualified) equivalence (\ref{opprop}) between a self-adjoint operator and a property. In the continous case the crucial qualification is that the operator $\overline{\tilde{P}}_{P}^{\Psi}$ that acts on $\Psi_{P}$ must have a complete set of eigenvectors $\{\overline{S}_{\tilde{P}}(\tilde{p})\}$, as defined in (\ref{contstate}).

\section{Evolution equations}
\label{evoleq}

Consider a family of experimental contexts $C(\sigma)$ in which the property $P$ of the specimen $OS$ that is observed is the four-position $R_{4}$ with value $r_{4}=(r,ict)$, where $r=(x,y,z)$ is the spatial position. Assume that a wave function can be defined according to (\ref{wfdef}) so that we may write

\begin{equation}
a_{P}(p_{j},\sigma)=a_{R_{4}}(r_{j},t_{j},\sigma).
\end{equation}

We suppose that the context family $C(\sigma)$ is such that it can be characterized by the continuous idealization $\Psi_{R_{4}}(r,t,\sigma)$ of the above wave function, according to the discussion in relation to (\ref{psi}). We seek an evolution equation

\begin{equation}
\Psi_{R_{4}}(r,t,\sigma)\equiv u_{R_{4}}^{\Psi}(\sigma)\Psi_{R_{4}}(r,t,0)
\label{ev0}
\end{equation}
of the form (\ref{evolutionequation}). Since the evolution parameter $\sigma$ is continuous we may look for its differential counterpart

\begin{equation}
\frac{d}{d\sigma}\Psi_{R_{4}}(r,t,\sigma)=\overline{A}_{R_{4}}\Psi_{R_{4}}(r,t,\sigma).
\label{ev1}
\end{equation}

The aim of this section is to find the explicit form of the operator $\overline{A}_{R_{4}}$. It cannot depend explicitly on $\sigma$, since $\sigma$ is not a property that can be used to specify the state of an object, and the evolution depends on nothing else but the physical state according to (\ref{sevolution}) or (\ref{objectevo}). Therefore we may relate the evolution equation (\ref{ev0}) to its differential counterpart (\ref{ev1}) according to

\begin{equation}
u_{R_{4}}^{\Psi}(\sigma)=e^{\overline{A}_{R_{4}}\sigma}.
\end{equation}

The evolution operator $u_{R_{4}}^{\Psi}$ is unitary according to (\ref{unitarycevo}). Therefore we must have

\begin{equation}
\overline{A}_{R_{4}}=i\overline{B}_{R_{4}},
\label{evbop}
\end{equation}
where $\overline{B}_{R_{4}}$ has real eigenvalues. In order to fulfill (\ref{linearev}) for all conceivable decompositions of $\Psi_{R_{4}}$ into piecewise wave functions $\{\Psi_{R_{4}j}\}$ we see that $\overline{B}_{R_{4}}$ must also be a linear operator. Therefore it is self-adjoint.

We can express the expected value $r_{4}$ of $R_{4}$ just before it is observed at sequential time $n+m$ as

\begin{equation}
\langle r_{4} \rangle= \int_{\Delta}\Psi_{R_{4}}^{*}r_{4}\Psi_{R_{4}}d r_{4},
\label{mean}
\end{equation}
where $\Delta$ is the support of the wave function $\Psi_{R_{4}}$, which corresponds to the projection of the object state $S_{OS}$ onto space-time at time $n+m-1$. We have suppressed the arguments in $\Psi_{R_{4}}(r_{4},\sigma)$ for clarity. To proceed in our search for the operator $\overline{A}_{R_{4}}$ we formulate two desiderata.

\begin{enumerate}
\item The evolution equation (\ref{ev1}) should be relativistically invariant.
\item This evolution equation should allow a parametrization $\langle r_{4}\rangle(\sigma)$ so that we may write $d\langle r_{4}\rangle/d\sigma = v_{0}+\beta f(\sigma)$, where $v_{0}$ is a constant vector, and $\beta$ is a scalar such that $\beta=0$ if and only if the specimen $OS$ does not interact with any other object during the course of any of the experimental contexts in the family $C(\sigma)$.
\end{enumerate}

The second desideratum is fulfilled whenever the evolution equation is such that a free specimen follows a straight trajectory in space-time. If so, we are free to define the evolution parameter $\sigma$ so that $d\langle r_{4}\rangle/d\sigma = v_{0}$.

Equation (\ref{mean}) implies

\begin{equation}
\frac{d}{d\sigma}\langle r_{4}\rangle = i\int_{\Delta}\Psi_{R_{4}}^{*}(r_{4}\overline{B}_{R_{4}}-\overline{B}_{R_{4}}r_{4})\Psi_{R_{4}}d r_{4}
\label{meanr}
\end{equation}
where we have used the fact that $\overline{B}_{R_{4}}=-i\overline{A}_{R_{4}}$ is a self-adjoint operator $\overline{B}_{P}^{\Psi}$ according to the discussion in relation to (\ref{evbop}). 

Suppose that the wave function $\Psi_{R_{4}}(r_{4},\sigma)$ can be represented by a Fourier integral

\begin{equation}
\Psi_{R_{4}}(r_{4},\sigma)=(2\pi)^{-5/2}\int_{-\infty}^{\infty}\tilde{\Psi}_{R_{4}}(\tilde{r}_{4},\tilde{\sigma})
e^{i(\tilde{r}_{4}\cdot r_{4}+\tilde{\sigma}\sigma)}d\tilde{r}_{4}d\tilde{\sigma}.
\label{fexpansion}
\end{equation}
We may call the kernel $\tilde{\Psi}_{R_{4}}(\tilde{r}_{4},\tilde{\sigma})$ the `reciprocal wave function'. Analogously, we may call $\tilde{R}_{4}$ the `reciprocal four-position', with values

\begin{equation}
\tilde{r}_{4}\equiv(\tilde{r},\frac{i}{c}\tilde{t})
\label{reciprocalpos}
\end{equation}
where $\tilde{r}\equiv(\tilde{x},\tilde{y},\tilde{z})$, and $\tilde{\sigma}$ the `reciprocal evolution parameter'.

The qualifier 'reciprocal' is chosen since we will see in the following that the reciprocal four-position ${R}_{4}$ can be identified with a property $\tilde{P}$ observed in a reciprocal family of contexts $\tilde{C}(\sigma)$ in the sense discussed in relation to (\ref{opprop}) and (\ref{copprop}). In so doing we extend the definition of the family of contexts $C(\sigma)$ that we are considering in the present section to a family in which the four-position $P\equiv R_{4}$ is observed first, then the reciprocal four-position $\tilde{P}\equiv\tilde{R}_{4}$. Then $\tilde{C}(\sigma)$ becomes the reciprocal family of contexts in which the order of observation of these properties is reversed, as illustrated in Fig. \ref{Figure14}. In what follows, we will relate the property $\tilde{R}_{4}$ to four-momentum. 

Inserting the expansion (\ref{fexpansion}) into (\ref{meanr}), we get

\begin{equation}\begin{array}{lll}
\frac{d}{d\sigma}\langle r_{4}\rangle & = & (2\pi)^{-5}i\int^{\infty}_{-\infty}\tilde{\Psi}_{R_{4}}^{*}(\tilde{r}_{4}',\tilde{\sigma}')\tilde{\Psi}_{R_{4}}(\tilde{r}_{4},\tilde{\sigma})\times\\
&  & \left\{\int_{\Delta}
e^{-i(\tilde{r}_{4}\cdot r_{4}+\tilde{\sigma}\sigma)}(r_{4}\overline{B}_{R_{4}}-\overline{B}_{R_{4}}r_{4})
e^{i(\tilde{r_{4}}\cdot r_{4}+\tilde{\sigma}\sigma)}d r_{4}\right\}\times\\
& & d\tilde{r}_{4}d\tilde{\sigma}d\tilde{r}_{4}'d\tilde{\sigma}'
\end{array}
\label{meanr2}
\end{equation}
 
When we are considering a free specimen $OS$, the right hand side of the above equation should equal a real constant according to desideratum 2, so that the integral over $r_{4}$ within the curly brackets cannot depend on $r_{4}$ or $\sigma$. The fact that this integral cannot depend on $\sigma$ implies that the integrand cannot depend on $\sigma$, so that we must have $(r_{4}\overline{B}_{R_{4}}-\overline{B}_{R_{4}}r_{4})e^{i(\tilde{r}_{4}\cdot r_{4}+\tilde{\sigma}\sigma)}=e^{i\tilde{\sigma}\sigma}\overline{f}e^{i\tilde{r}_{4}\cdot r_{4}}$ for some operator $\overline{f}$. Since the specimen $OS$ is assumed to be free, and since the spatio-temporal position $R_{4}$ is a relational property whose value is defined in relation to an arbitrary reference frame, the constant $d\langle r_{4}\rangle/d\sigma$ must be invariant under stiff translations of the region $\Delta$ in which we know that the specimen is located. (By a `stiff' translation we mean that the shape of the region does not change.) This means that the integrand of the integral within curly brackets in (\ref{meanr2}) cannot depend on $r_{4}$ either. Therefore the operator $\overline{f}$ must be a function $f(\tilde{r}_{4},\tilde{\sigma})$ so that

\begin{equation}
(r_{4}\overline{B}_{R_{4}}-\overline{B}_{R_{4}}r_{4})e^{i(\tilde{r}_{4}\cdot r_{4}+\tilde{\sigma}\sigma)}=f(\tilde{r}_{4},\tilde{\sigma})e^{i(\tilde{r}_{4}\cdot r_{4}+\tilde{\sigma}\sigma)}.
\end{equation}
Furthermore, the imaginary unit $i$ that appears in front of the integrals at the right hand side of (\ref{meanr2}) means that $f(\tilde{r}_{4},\tilde{\sigma})$ must be imaginary. These constraints imply

\begin{equation}
\overline{B}_{R_{4}}=b_{x}\frac{\partial^{2}}{\partial x^{2}}+b_{y}\frac{\partial^{2}}{\partial y^{2}}+b_{z}\frac{\partial^{2}}{\partial z^{2}}-b_{t}\frac{1}{c^{2}}\frac{\partial^{2}}{\partial t^{2}},
\end{equation}
for some array $(b_{x},b_{y},b_{z},b_{t})$ of real, scalar constants. Since the evolution equation (\ref{ev1}) should be relativistically invariant according to desideratum 1, we must have $b=b_{x}=b_{y}=b_{z}=b_{t}$, so that

\begin{equation}
\overline{B}_{R_{4}}=-b\Box.
\label{dalembert}
\end{equation}

Equations (\ref{meanr}) and (\ref{dalembert}) imply that we may write

\begin{equation}\begin{array}{lll}
\frac{d}{d\sigma}\langle r_{4}\rangle & = & -ib\int_{-\infty}^{\infty}
\Psi_{R_{4}}^{*}\left(r_{4}\Box-\Box r_{4}\right)\Psi_{R_{4}}d r_{4}\\
& = & 2ib\int_{-\infty}^{\infty}
\Psi_{R_{4}}^{*}\left(\frac{\partial}{\partial x},\frac{\partial}{\partial y},\frac{\partial}{\partial z},\frac{i}{c}\frac{\partial}{\partial t}\right)\Psi_{R_{4}}d r_{4}.
\end{array}\end{equation}
If we insert the Fourier integral (\ref{fexpansion}) in the above expression, we get

\begin{equation}
\frac{d}{d\sigma}\langle r_{4}\rangle = -2b\langle\tilde{r}_{4}\rangle,
\label{meandr}
\end{equation}
where we have defined

\begin{equation}
\langle\tilde{r}_{4}\rangle\equiv \int_{-\infty}^{\infty}\tilde{\Psi}_{R_{4}}(\tilde{r}_{4},\tilde{\sigma})^{*}\tilde{r}_{4}\tilde{\Psi}_{R_{4}}(\tilde{r}_{4},\tilde{\sigma})d\tilde{r}_{4}d\tilde{\sigma}
\end{equation}

Since all subjects are assumed to agree on the temporal ordering of events according to (\ref{eventorder}), they also agree on the direction of time in the sense that $d\langle t\rangle/d\sigma$ has the same sign in all reference frames. It is natural to choose a parametrization so that the evolution parameter and relational time flow in the same direction:

\begin{equation}
\frac{d}{d\sigma}\langle t\rangle>0.
\label{directed}
\end{equation}
Equation (\ref{meandr}) implies that we may write

\begin{equation}
\frac{d}{d\sigma}\langle t\rangle = -\frac{2b}{c^{2}}\langle\tilde{t}\rangle,
\label{tct}
\end{equation}
according to (\ref{reciprocalpos}). This equation should hold for all allowed values of $\tilde{t}$, so that $\tilde{r}_{4}$ is always real, and

\begin{equation}
b\tilde{t}<0
\label{signrule1}
\end{equation}
according to (\ref{directed}). We have $\overline{A}_{R_{4}}=-ib\Box$ from (\ref{evbop}) and (\ref{dalembert}). If we insert this relation in (\ref{ev1}), and express $\Psi_{R_{4}}$ in terms of its Fourier integral (\ref{fexpansion}), we get

\begin{equation}
\frac{\tilde{t}^{2}}{c^{2}}=\frac{\tilde{\sigma}}{b}+\tilde{r}\cdot\tilde{r}.
\label{protoeinstein}
\end{equation}
This relation must hold for all possible values of $\tilde{r}$, in particular when $\tilde{r}=0$. Since $\tilde{t}$ is real, we must therefore have

\begin{equation}
\frac{\tilde{\sigma}}{b}> 0
\label{signrule2}
\end{equation}
for all $\tilde{\sigma}$. The sign of the parameter $b$ is arbitrary. In the following, we choose $b<0$. Then (\ref{signrule1}) and (\ref{signrule2}) lead to the convention 

\begin{equation}
\begin{array}{lll}
b & < & 0,\\
\tilde{t} & > & 0,\\
\tilde{\sigma} & < & 0.
\end{array}
\label{signconvention}
\end{equation}

The evolution operator $\overline{B}_{R_{4}}$ defined according to (\ref{ev1}) and (\ref{evbop}) acts on the wave function $\Psi_{R_{4}}$, and it is self-adjoint. The discussion in relation to (\ref{opprop}) and (\ref{copprop}) therefore suggests that $\overline{B}_{R_{4}}$ may be associated with a property $B$, and that it may therefore be identified with a wave function operator $\overline{P'}_{P}^{\Psi}$ as defined in (\ref{psiops}). Let us explore this possibility.

The set of possible values of such a property $B$ should equal the set $\{\tilde{\sigma}\}$ of all eigenvalues $\tilde{\sigma}$ to the operator $-i\frac{d}{d\sigma}$, when it acts on the set $\{\Psi_{R_{4}}\}$ of allowed wave functions that corresponds to the contextual Hilbert space $\mathcal{H}_{C}$. The eigenfunctions to $-i\frac{d}{d\sigma}$ can be written $\Psi_{R_{4}}^{(\tilde{\sigma})}=\psi(r_{4},\tilde{\sigma})e^{i\tilde{\sigma}\sigma}$, where the function $\psi(r_{4},\tilde{\sigma})$ is arbitrary. In short,

\begin{equation}\begin{array}{c}
\Psi_{R_{4}}^{(\tilde{\sigma})}=\psi_{R_{4}}(r_{4},\tilde{\sigma})e^{i\tilde{\sigma}\sigma}\\
-i\frac{d}{d\sigma}\Psi_{R_{4}}^{(\tilde{\sigma})}=\overline{B}_{R_{4}}\Psi_{R_{4}}^{(\tilde{\sigma})}=\tilde{\sigma}\Psi_{R_{4}}^{(\tilde{\sigma})}.
\end{array}
\label{eigensigma}
\end{equation}

In the same way, we might argue that there are properties $C$ and $D$ that correspond to the self-adjoint operators

\begin{equation}\begin{array}{lll}
\overline{C}_{R_{4}} & \equiv & -b\frac{1}{c^{2}}\frac{\partial^{2}}{\partial t^{2}}\\
\overline{D}_{R_{4}} & \equiv & b\left(\frac{\partial^{2}}{\partial x^{2}}+\frac{\partial^{2}}{\partial y^{2}}+\frac{\partial^{2}}{\partial z^{2}}\right).
\end{array}\end{equation}
Just like for $B$, the set of possible values of $C$ should equal the set of all possible eigenvalues $\tilde{t}^{2}$ to $\overline{C}_{R_{4}}$ as it acts in the set $\{\Psi_{R_{4}}\}$ of allowed wave functions. Likewise, the set of possible values of $D$ should be the same as the set of eigenvalues $\{\tilde{x}^{2}+\tilde{y}^{2}+\tilde{z}^{2}\}$ to $\overline{D}_{R_{4}}$. We clearly have

\begin{equation}
\overline{B}_{R_{4}}=\overline{C}_{R_{4}}+\overline{D}_{R_{4}}.
\end{equation}

This operator relation is closely related to the relativistic relation $\epsilon_{0}^{2}=\epsilon^{2}-c^{2}p^{2}$, where $\epsilon_{0}$ and $\epsilon$ are the rest energy and energy, respectively, as we will see below. However, there is a crucial difference. The property values that appear in the relativistic equation are squared, whereas there are no squares in the above operator relation that corresponds to the tentative property relation $B=C+D$. Let us explore the reason for the introduction of the squares.

We see in (\ref{eigensigma}) that for each eigenfunction $\Psi_{R_{4}}^{(\tilde{\sigma})}$ to the operator $\overline{B}_{R_{4}}$ with eigenvalue $\tilde{\sigma}$, there is another eigenfunction $\Psi_{R_{4}}^{(-\tilde{\sigma})}$ with eigenvalue $-\tilde{\sigma}$. However, (\ref{signrule2}) tells us that all property values $\tilde{\sigma}$ must have the same sign (since $b$ is a fixed constant). This means that there are eigenvalues to $\overline{B}_{R_{4}}$ that do not correspond to possible property values $\tilde{\sigma}$. Therefore $\overline{B}_{\mathbf{r}_{4}}=-id/d\sigma$ cannot be associated right away with a property $\tilde{\sigma}$, since the match between the set of eigenvalues and the set of possible property values should be perfect. However, we might accomplish such a perfect match if we restrict the set of allowed wave functions $\{\Psi_{R_{4}}\}$ on which $\overline{B}_{R_{4}}$ may act even further than those two restrictions provided by the evolution equation (\ref{ev1}), and the requirement (\ref{fexpansion}) that $\Psi_{R_{4}}$ should have a Fourier representation.

Since $\tilde{\sigma}$ cannot have any positive values, we may try to introduce another property $W$ with values $w$ such that

\begin{equation}
\tilde{\sigma}=-w^{2}.
\end{equation}
If there is such a property $W$ there should be a corresponding wave function operator $\overline{W}_{R_{4}}$ in the sense of (\ref{opprop}) and (\ref{copprop}). This operator must be self-adjoint and therefore have real eigenvalues, so that we should be able to write

\begin{equation}
\overline{B}_{R_{4}}=-\overline{W}_{R_{4}}\overline{W}_{R_{4}}.
\label{squareroot2}
\end{equation}

Using the notation $\Psi_{R_{4}}^{(\tilde{\sigma})}=\psi_{R_{4}}e^{i\tilde{\sigma}\sigma}$ in (\ref{eigensigma}) we may write

\begin{equation}
\overline{B}_{R_{4}}\psi_{R_{4}}=\tilde{\sigma}\psi_{R_{4}},
\end{equation}
since $\overline{B}_{R_{4}}=-b\Box$ does not act on $\sigma$. We get

\begin{equation}
\overline{B}_{R_{4}}\psi_{R_{4}}=-\overline{W}_{R_{4}}\overline{W}_{R_{4}}\psi_{R_{4}}=-\overline{W}_{R_{4}}w\psi_{R_{4}}(r_{4})=-w^{2}\psi_{R_{4}}(r_{4}).
\label{kleingordon1}
\end{equation}

Thus, in order to be able to interpret $\tilde{\sigma}$ as a property value and $\overline{B}_{R_{4}}$ as a wave function operator $\overline{P'}_{P}^{\Psi}$ that is associated with this property, we have to add the constraint
\begin{equation}
\overline{W}_{R_{4}}\psi_{R_{4}}(r_{4},\tilde{\sigma})=w\psi_{R_{4}}(r_{4},\tilde{\sigma})
\label{dirac1}
\end{equation}
for each function $\psi_{R_{4}}(r_{4},\tilde{\sigma})$ in the general solution

\begin{equation}
\Psi_{R_{4}}(r_{4},\sigma)=\int_{-\infty}^{\infty}\psi_{R_{4}}(r_{4},\tilde{\sigma})e^{i\tilde{\sigma}\sigma}d\tilde{\sigma}
\label{gensol}
\end{equation}
to the evolution equation
\begin{equation}
\frac{d}{d\sigma}\Psi_{R_{4}}(r_{4},\sigma)=i\overline{B}_{R_{4}}\Psi_{R_{4}}(r_{4},\sigma).
\label{ev2}
\end{equation}

What the constraint (\ref{dirac1}) does, in effect, is to restrict the set of wave functions $\{\Psi_{R_{4}}\}$ on which $\overline{B}_{R_{4}}$ may act, so that only eigenfunctions corresponding to eigenvalues with a given sign remain. We achieve the necessary perfect match between the set of eigenvalues $\tilde{\sigma}$ associated with this action and the set of possible property values.

One may ask whether there are any wave functions at all that fulfill the constraint (\ref{dirac1}). We see however, that it resembles the Dirac equation, and we will see that we can make an explicit identification when we interpret the appearing symbols properly. (Equation (\ref{kleingordon1}) can then accordingly be identified with the Klein-Gordon equation.) Thus there are indeed solutions in the form of spinors. They may expressed as elements $\Psi_{R_{4}}(r_{4},s)$ in an extended Hilbert space $\mathcal{H}_{C}$, where the extra argument $s$ denotes the discrete spinor degree of freedom with possible values $s\in\{1,2,3,4\}$. This Hilbert space is clearly spanned by the extended set of vectors $\{\overline{S}_{R_{4}}(r_{4},s)\}$, which are analogous to the vectors $\{\overline{S}_{R_{4}}(r_{4})\}$ defined according to (\ref{contstate}).

To show that $\overline{B}_{R_{4}}$ and $\overline{W}_{R_{4}}$ are wave function operators that are associated with properties $B$ and $W$ it remains to show that they possess a common complete set of eigenfunctions $\{\psi_{R_{4}}(r_{4},s,\tilde{\sigma})\}$ that corresponds to a set of vectors $\{\overline{S}_{B}(\tilde{\sigma},s)\}=\{\overline{S}_{W}(w,s)\}$ that span the extended Hilbert space. We see from (\ref{fexpansion}) that these eigenfunctions have the form $\psi_{R_{4}}(r_{4},s,\tilde{\sigma})=f(s)e^{i\tilde{r}_{4}\cdot\tilde{r}_{4}}$. These functions are also eigenfunctions to the operator

\begin{equation}
\overline{\tilde{R}}_{R_{4}}=-i\left(\frac{\partial}{\partial x},\frac{\partial}{\partial y},\frac{\partial}{\partial z},\frac{i}{c}\frac{\partial}{\partial t}\right).
\label{nabla4}
\end{equation}
which is self-adjoint with real eigenvalues $\tilde{r}_{4}$. If we manage to show that $\{\overline{S}_{\tilde{R}_{4}}(\tilde{r}_{4},s)\}=\{\overline{S}_{B}(\tilde{\sigma},s)\}=\{\overline{S}_{W}(w,s)\}$ span the extended Hilbert space, then the reciprocal four-position $\tilde{R}_{4}$ is a property together with $B$ and $W$, with associated wave function operators $\overline{\tilde{R}}_{R_{4}}$, $\overline{B}_{R_{4}}=-b \overline{\tilde{R}}_{R_{4}}\cdot\overline{\tilde{R}}_{R_{4}}$ and $\overline{W}_{R_{4}}$, respectively.

The extended, continuous Hilbert space $\mathcal{H}_{C}$ is spanned by the continuous set of vectors $\{\overline{S}_{R_{4}}(R_{4},s)\}$ defined in analogy with (\ref{contstate}). To construct a corresponding basis $\{\overline{S}_{\tilde{R}_{4}}(\tilde{R}_{4},s)\}$ that span the common continuous Hilbert space $\mathcal{H}_{C\tilde{C}}$ in the sense of Fig. \ref{Figure14}, we make the ansatz

\begin{equation}
\overline{S}_{\tilde{R}_{4}}(\tilde{r}_{4},s)=(2\pi)^{-2}\int_{\Delta}e^{i\tilde{r}_{4}\cdot r_{4}}\overline{S}_{R_{4}}(r_{4},s)d r_{4},
\label{rcontstate}
\end{equation}
where the integration is formal just like in (\ref{contstate}) and covers the region $\Delta$ in space-time in which we cannot exclude that the specimen $OS$ is located at sequential time $n+m-1$, just before the observation that determines its location more precisely.

According to this ansatz the basis vector $\overline{S}_{\tilde{R}_{4}}$ corresponding to the value $\tilde{r}_{4}$ is a superposition of all the basis vectors $\overline{S}_{R_{4}}$ corresponding to all the possible values $r_{4}$, where each of these vectors is given the same weight in the superposition. This means that no value of $R_{4}$ is preferred in relation to any given value of $\tilde{R}_{4}$. In the language used in Ref. \cite{ostborn1} in relation to Fig. 37(b), the properties $\tilde{R}_{4}$ and $R_{4}$ are treated as \emph{independent}.

According to (\ref{formalint}) we may then write

\begin{equation}\begin{array}{lll}
\overline{S}_{C\tilde{C}}(\sigma) & = & \int_{\Delta} \Psi_{R_{4}}(r_{4},s,\sigma)dr_{4}\overline{S}_{R_{4}}(r_{4},s)\\
& = & \int_{-\infty}^{\infty} \Psi_{\tilde{R}_{4}}(\tilde{r}_{4},s,\sigma)d\tilde{r}_{4}\overline{S}_{\tilde{R}_{4}}(\tilde{r}_{4},s)
\end{array}
\label{hbases}
\end{equation}
with $\Psi_{\tilde{R}_{4}}(\tilde{r}_{4},s,\sigma)=(2\pi)^{-5/2}\int_{-\infty}^{\infty}\tilde{\Psi}_{\tilde{R}_{4}}(\tilde{r}_{4},s,\tilde{\sigma})e^{i\tilde{\sigma}\sigma}d\tilde{\sigma}$, assuming in the second equality the existence of the Fourier representation (\ref{fexpansion}) of the wave function $\Psi_{R_{4}}$.

An idealized contextual state $S_{C\tilde{C}}(\sigma)$ in which the value of $R_{4}$ is known to be precisely $r_{4}'$ corresponds to the Hilbert space representation $\overline{S}_{C\tilde{C}}(\sigma)=\overline{S}_{R_{4}}(r_{4}',s)$. This equality is clearly fulfilled if and only if $\Psi_{R_{4}}(r_{4},s,\sigma)=\delta(r_{4}-r_{4}')$ according to (\ref{hbases}). In the same way we have $\overline{S}_{C\tilde{C}}(\sigma)=\overline{S}_{\tilde{R}_{4}}(\tilde{r}_{4}',s)$ if and only if $\Psi_{\tilde{R}_{4}}(\tilde{r}_{4},s,\sigma)=\delta(\tilde{r}_{4}-\tilde{r}_{4}')$. This relation is fulfilled if and only if $\Psi_{R_{4}}(r_{4},s,\sigma)\propto e^{i(\tilde{r}_{4}'\cdot r_{4}+\tilde{\sigma}'\sigma)}$ with $\tilde{\sigma}'=-b\tilde{r}_{4}'\cdot\tilde{r}_{4}'$ according to (\ref{fexpansion}) and (\ref{protoeinstein}). Therefore the statement that the set of eigenfunctions $\{e^{i(\tilde{r}_{4}\cdot r_{4}+\tilde{\sigma}\sigma)}\}$ to $\overline{\tilde{R}}_{R_{4}}$ is a complete basis in the functional space of allowed wave functions $\Psi_{R_{4}}$ is equivalent to the statement that $\{\overline{S}_{\tilde{R}_{4}}(\tilde{r}_{4},s)\}$ is a complete basis in a common Hilbert space $\mathcal{H}_{C\tilde{C}}$. 

We see in (\ref{hbases}) that $\{\overline{S}_{\tilde{R}_{4}}(\tilde{r}_{4},s)\}$ is a complete basis for $\mathcal{H}_{C\tilde{C}}$ only if the range of $\tilde{r}_{4}$ has no boundary. In order to make the basis $\{\overline{S}_{\tilde{R}_{4}}(\tilde{r}_{4},s)\}$ orthogonal we have to let the range of $r_{4}$ grow without boundary as well in the original basis $\{\overline{S}_{R_{4}}(r_{4},s)\}$. This is so since

\begin{equation}\begin{array}{c}
\langle \overline{S}_{\tilde{R}_{4}}(\tilde{r}_{4},s),\overline{S}_{\tilde{R}_{4}}(\tilde{r}_{4}',s)\rangle=\\
=(2\pi)^{-4}\int_{\Delta}e^{i(\tilde{r}_{4}'\cdot r_{4}'-\tilde{r}_{4}\cdot r_{4})}\langle\overline{S}_{R_{4}}(r_{4},s),\overline{S}_{R_{4}}(r_{4}',s)\rangle d r_{4}d r_{4}'=\\
=(2\pi)^{-4}\int_{\Delta}e^{i(\tilde{r}_{4}'\cdot r_{4}'-\tilde{r}_{4}\cdot r_{4}})\delta(r_{4}-r_{4}') d r_{4}d r_{4}'\\
\end{array}
\end{equation}
from (\ref{rcontstate}) and (\ref{deltarel}). Only in the limit that $\Delta$ grows without boundary do we get

\begin{equation}\begin{array}{lll}
\langle \overline{S}_{\tilde{r}_{4}}(\tilde{r}_{4}),\overline{S}_{\tilde{r}_{4}}(\tilde{r}_{4}')\rangle & = & (2\pi)^{-4}\int_{\Delta}e^{ir_{4}\cdot(\tilde{r}_{4}'-\tilde{r}_{4})}d r_{4}\\
& = & \delta(\tilde{r}_{4}-\tilde{r}_{4}').
\end{array}
\end{equation}
Only in this case do $B$, $W$ and $R_{4}$ become properly defined properties with associated wave function operators $\overline{\tilde{R}}_{R_{4}}$, $\overline{B}_{R_{4}}=-b \overline{\tilde{R}}_{R_{4}}\cdot\overline{\tilde{R}}_{R_{4}}$ and $\overline{W}_{R_{4}}$ as defined in (\ref{nabla4}), (\ref{dalembert}) and (\ref{squareroot2}), respectively.

In the following we therefore assume that we use such an 'unbounded' continuous representation of the experimental context. We stress, however, that any real context is always 'bounded' and discrete. Even if the experimental context is designed to observe a property $P$ whose values are 'unbounded' in principle (such as spatio-temporal position) we can always rule out some very remote property values $p$ of the given specimen beforehand. The property values $p_{j}$ that we are going to observe are always discrete in the sense that they are described by (\ref{contpropv}), with a corresponding finite set of property value states $\overline{S}_{Pj}$ given by (\ref{contstate}). A sufficient condition for the validity for the continuous idealization of a context family $C(\sigma)$ is given in relation to (\ref{psi}). We can always let the boundary of the region $\Delta$ introduced in (\ref{rcontstate}) grow without limit and still retain the required piecewise differential wave function $\Psi_{P}$. In these considerations, the property $P$ may be four-position $R_{4}$, but equally well the reciprocal four-position $\tilde{R}_{4}$ that we identified above in the idealized continuous representation, or $B$, or $W$.

The property $W$ is defined for \emph{any} specimen $OS$ that is studied in family of contexts $C(\sigma)$ such that the evolution equation (\ref{ev2}) can be defined. Its values $w$ are invariant under Lorentz transformations. It may therefore be seen as an internal property of the specimen. These qualities resemble those of the rest energy. We may therefore try to \emph{define} the rest energy $E_{0}$ so that its values $\epsilon_{0}$ fulfill

\begin{equation}
\epsilon_{0}\equiv \frac{c \hbar}{\sqrt{-b}}w=c \hbar\sqrt{\frac{\tilde{\sigma}}{b}}.
\label{restmassdef}
\end{equation}

The continuous wave function operator that corresponds to the rest energy then becomes
\begin{equation}
(\overline{E_{0}})_{R_{4}}\equiv\frac{c \hbar}{\sqrt{-b}}\overline{W}_{R_{4}},
\label{restmassop}
\end{equation}
where $\overline{W}_{R_{4}}$ is defined by the relation $\overline{W}_{R_{4}}\overline{W}_{R_{4}}=b\Box$ according to (\ref{dalembert}) and (\ref{squareroot2}).

We also try to define the four-momentum $P_{4}$ so that its values $p_{4}$ are given by

\begin{equation}
p_{4}\equiv\hbar \tilde{r}_{4}.
\label{fourmomentumdef}
\end{equation}

In that case, the continuous wave function operator that corresponds to four-momentum becomes

\begin{equation}
(\overline{P_{4}})_{R_{4}}\equiv\hbar \overline{\tilde{R}}_{R_{4}},
\label{fmodef}
\end{equation}
where $\overline{\tilde{R}}_{R_{4}}$ is given by (\ref{nabla4}).

The two definitions (\ref{restmassdef}) and (\ref{fourmomentumdef}) seem appropriate since they imply the Einstein relation $\epsilon_{0}^{2}=-c^{2}p_{4}\cdot p_{4}$ that should be fulfilled by rest energy and four-momentum. To define momentum $P$ and energy $E$ so that they conform with the conventional properties with these names, we note that we should be able to write $p_{4}=(p,i\epsilon/c)$, so that we get $\epsilon^{2}=\epsilon_{0}^{2}+c^{2}p^{2}$. Equation (\ref{reciprocalpos}) then leads us to the definitions

\begin{equation}\begin{array}{l}
p\equiv\hbar \tilde{r}\\
\epsilon\equiv\hbar\tilde{t}
\end{array}
\label{medef}
\end{equation}
with corresponding familiar continuous wave function operators

\begin{equation}\begin{array}{l}
\overline{P}_{R_{4}}\equiv-i\hbar\left(\frac{\partial}{\partial x},\frac{\partial}{\partial y},\frac{\partial}{\partial z}\right)\\
\overline{E}_{R_{4}}\equiv i\hbar\frac{\partial}{\partial t}.
\end{array}
\label{meodef}
\end{equation}
We see that momentum and energy become equivalent to the reciprocal position $\tilde{r}$ and the reciprocal time $\tilde{t}$, respectively, where Planck's constant appears as a factor of proportionality when the values of these properties are fixed within a given set of units.

With these definitions or identifications at hand, we can use (\ref{meandr}) to derive the counterpart in the present formalism to the Ehrenfest theorem:

\begin{equation}
\frac{d\langle r\rangle}{d\sigma}/\frac{d\langle t\rangle}{d\sigma}=\frac{\langle p\rangle}{\langle m\rangle},
\label{ehrenfest}
\end{equation}
where $m$ is the relativistic mass defined according to $\epsilon=mc^{2}$. The elaborate notion of time introduced in this paper makes the left hand side of this equation more involved than usual. We can recast it in more familiar form if we choose the \emph{natural parametrization}

\begin{equation}
d\langle t\rangle/d\sigma=1.
\label{naturalp}
\end{equation}
We are free to fix any such parametrization of the evolution as long as it fulfills (\ref{directed}), since $\sigma$ is not an observable property, as discussed in Sections \ref{times} and \ref{evolp}. Note, however, that (\ref{naturalp}) is not relativistically invariant, so that a given parametrization is natural in a given reference frame only. In such a frame we have

\begin{equation}
\frac{d\langle r\rangle}{d\langle t\rangle}=\frac{\langle p\rangle}{\langle m\rangle}.
\label{nehrenfest}
\end{equation}
The only difference between this relation and the ordinary Ehrenfest theorem is that the differentiation is performed with respect to the expected relational time $\langle t\rangle$ that is going to be observed, rather than to a precisely defined temporal parameter $t$.

The expected measure of time $\langle t\rangle$ until observation is always potentially known at the start of an experiment that defines a well-defined context $C$ that is described by the wave function $\Psi_{R_{4}}$. This is so since the probability of each observed time $t$ is encoded in $\Psi_{R_{4}}$ by definition. It is therefore self-consistent to treat $\langle t\rangle$ as a parameter that depends on $\sigma$. Using the definitions (\ref{restmassdef}), (\ref{fourmomentumdef}), (\ref{medef}) and (\ref{meodef}) we may therefore express the evolution equation (\ref{ev2}) as

\begin{equation}\begin{array}{lll}
i\hbar\frac{d}{d\langle t\rangle}\Psi_{R_{4}} & = & \frac{-1}{2\langle \epsilon\rangle}(\overline{E_{0}})_{R_{4}}(\overline{E_{0}})_{R_{4}}\Psi_{R_{4}}\\
& = & \frac{c^{2}}{2\langle \epsilon\rangle}\left[(\overline{P_{4}})_{R_{4}}\cdot(\overline{P_{4}})_{R_{4}}\right]\Psi_{R_{4}}\\
& = & \frac{-1}{2\langle \epsilon\rangle}\left(\overline{E}_{R_{4}}\overline{E}_{R_{4}}-c^{2}\overline{P}_{R_{4}}\cdot\overline{P}_{R_{4}}\right)\Psi_{R_{4}}.
\end{array}
\label{ev3}
\end{equation}
in the natural parametrization (\ref{naturalp}), where we have suppressed the arguments of the wave function $\Psi_{R_{4}}(r_{4},s,\langle t\rangle)$ for brevity. For a free specimen these relations correspond to

\begin{equation}
\frac{d}{d\langle t\rangle}\Psi_{R_{4}}=\frac{ic^{2}\hbar}{2\langle \epsilon\rangle}\Box\Psi_{R_{4}}.
\label{freeeveq}
\end{equation}

In the above equations we have made use of the fact that the natural parametrization (\ref{naturalp}) fixes the arbitrary parameter $b$ (that appeared first in (\ref{dalembert})) as

\begin{equation}
b=-\frac{c^{2}\hbar}{2\langle \epsilon\rangle}.
\end{equation}
This relation is easily obtained by combining (\ref{meandr}) with the definition of energy given in (\ref{medef}).

The evolution equation (\ref{freeeveq}) has the same form as Stueckelberg's evolution equation for a free particle \cite{stueckelberg2}. However, in Stueckelberg's formalism intervals $\Delta\lambda$ of the evolution parameter $\lambda$ are assigned a physical meaning; they are proportional to intervals of the proper time measured by a clock in the rest frame of the particle. In the present formalism intervals $\Delta\sigma$ of $\sigma$ have no physical meaning; they are arbitrary as long as $\sigma$ flow in the same direction as $t$, as expressed in (\ref{directed}). The evolution parameter $\sigma$ just indicates a temporal direction, whereas all physically meaningful measures of temporal intervals are intervals $\Delta t$ of relational time $t$. Stuekelberg's parameter $\lambda$ is invariant with respect to relativistic transformations of spatio-temporal coordinates, whereas the parameter $\sigma$ is immune to such transformations; they do not apply to it.   

In terms of the familiar properties defined above we may reformulate (\ref{kleingordon1}) as

\begin{equation}
(\overline{E_{0}})_{R_{4}}(\overline{E_{0}})_{R_{4}}\psi_{r_{4}}=\epsilon_{0}^{2}\psi_{R_{4}},
\label{kleingordon2}
\end{equation}
or $-c^{2}\hbar^{2}\Box\psi_{r_{4}}=\epsilon_{0}^{2}\psi_{R_{4}}$. We recognize (\ref{kleingordon2}) as the Klein-Gordon equation.

Equation (\ref{dirac1}) expresses an additional constraint on the wave function $\Psi_{R_{4}}$. In terms of the same familiar properties this equation can be expressed as

\begin{equation}
(\overline{E_{0}})_{R_{4}}\psi_{r_{4}}=\epsilon_{0}\psi_{R_{4}}
\label{dirac2}
\end{equation}
for each function $\psi_{R_{4}}(r_{4},s,\frac{e_{0}^{2}}{2\hbar\langle \epsilon\rangle})$ in the Fourier expansion

\begin{equation}
\Psi_{R_{4}}(r_{4},s,\langle t\rangle)=\int_{-\infty}^{\infty}\psi_{R_{4}}(r_{4},s,\frac{\epsilon_{0}^{2}}{2\hbar\langle \epsilon\rangle})e^{i\frac{\epsilon_{0}^{2}}{2\hbar\langle \epsilon\rangle}\langle t\rangle}d\left(\frac{\epsilon_{0}^{2}}{2\hbar\langle \epsilon\rangle}\right)
\end{equation}
of a general wave function $\Psi_{R_{4}}(r_{4},s,\langle t\rangle)$. We recognize (\ref{dirac2}) as the Dirac equation, and thanks to Dirac we know that if we write the wave function as a column vector according to $\Psi_{R_{4}}(r_{4},s,\langle t\rangle)=[\Psi_{R_{4}1}(r_{4},\langle t\rangle),\Psi_{R_{4}2}(r_{4},\langle t\rangle),\Psi_{R_{4}3}(r_{4},\langle t\rangle),\Psi_{R_{4}4}(r_{4},\langle t\rangle)]^{T}$ we can more explicitly express the rest energy operator as

\begin{equation}
(\overline{E_{0}})_{R_{4}}=c\overline{D}\cdot(\overline{P_{4}})_{R_{4}},
\end{equation}
where $\overline{D}=(\overline{D}_{1},\overline{D}_{2},\overline{D}_{3},\overline{D}_{4})$ is a vector of $4\times 4$ matrices, related to the gamma matrices according to $\overline{D}_{1}=i\gamma^{1}$, $\overline{D}_{2}=i\gamma^{2}$, $\overline{D}_{3}=i\gamma^{3}$, and $\overline{D}_{4}=\gamma^{0}$.

According to the preceding discussion the Dirac equation can be interpreted as a stationary state equation, analogous the stationary Shr\"odinger equation. The Hamiltonian $\overline{H}$ is replaced by the rest energy operator $(\overline{E_{0}})_{R_{4}}$, and the energy eigenvalue $\epsilon$ is replaced by the rest energy eigenvalue $\epsilon_{0}$. The only formal difference is that the solution $\psi$ to the Dirac equation is a function of the spatio-temporal position $r_{4}=(r,ict)$ rather than just the spatial position $r=(x,y,z)$.

Nevertheless, the interpretations of the two equations are different. Of course, the solution to the Dirac equation is not stationary in the same sense as the solution to the stationary Shr\"odinger equation. It is independent of the evolution parameter $\sigma$, but not of $t$. What (\ref{dirac2}) tells us is that the probability to observe a given position $r$ depends in a certain way on the probability to observe a given time $t$. This dependence is the same in all families $C(\sigma)$ of experimental contexts designed to observe the coordinates $r$ and $t$ of an event pertaining to a specimen $OS$ with a precisely known rest energy.

Note that this interpretation of the Dirac equation differs from the conventional one. We use to say that the solution $\Psi(r,t)$ provides the probability to observe a given position $r$ of the specimen depending on the time $t$ at which we choose to do so. Here we say that $\Psi_{R_{4}}(r,t)$ provides the probabilities to observe a given position $r$ and a given relational time $t$ at some sequential time $n+m$ that has been abstracted away from the problem.

Up until now we have focused on the case where the specimen does not interact with its environment. It is then expected to travel along a straight trajectory, so that its evolution can be parametrized as $d\langle r_{4}\rangle/d\sigma=v_{0}$ according to desideratum 2, stated at the beginning of this section. This focus is too narrow, of course. Even if we may transform a curved trajectory $d\langle r_{4}\rangle/d\sigma=v_{0}+\beta f(\sigma)$ so that it looks straight, such a coordinate transformation does not leave the form of the evolution equation invariant if it does not take the possibility of interactions into account to begin with. It has to be checked whether our main conclusions are still valid after such a generalization.

The basic form of the evolution equation (\ref{ev2}) is still valid in the general case, since we did not need to make the assumption that $\beta=0$ when we derived (\ref{ev1}) and (\ref{evbop}). The argument that lead to (\ref{dalembert}) implies that we must have $\overline{B}_{R_{4}}=-b\Box$ when $\beta=0$. Therefore we may always write

\begin{equation}
\overline{B}_{R_{4}}=-b\Box+\overline{B}_{R_{4}}^{(int)},
\label{generalb}
\end{equation}
where $\overline{B}_{R_{4}}^{(int)}$ is the interaction term. To first order in $\beta$ we may also write $\overline{B}_{R_{4}}=-b\Box+\beta\overline{B}_{R_{4}}^{(1)}$.

We are dealing with an identifiable specimen whose spatio-temporal trajectory is studied in a family of contexts $C(\sigma)$ in which $R_{4}$ is observed. The two interactions that are able to shape the observed macroscopic trajectory of such a traceable object is gravity and electromagnetism. They can be seen as gauge forces that are necessary in order to make the evolution equation invariant with respect to diffeomorphisms $r_{4}'=g(r_{4})$ and local phase changes in the wave function $\Psi_{R_{4}}(r_{4},s,\sigma)\rightarrow \Psi_{R_{4}}(r_{4},s,\sigma)e^{h(r_{4})}$. Together these forces determine the form of the interaction term $\overline{B}_{R_{4}}^{(int)}$ in the evolution operator $\overline{B}_{R_{4}}$.

It is still true in this more general setting that $\overline{B}_{R_{4}}$ cannot depend on $\sigma$, or explicitly act on it, since $\sigma$ is not a property that specifies the physical state on which the evolution operator is supposed to act, but a parameter whose numerical value is arbitrary. Therefore the general solution $\Psi_{R_{4}}(r_{4},s,\sigma)$ to (\ref{generalb}) can still be expressed as in (\ref{gensol}). Let us focus on the integrand $\psi_{R_{4}}(r_{4},s,\tilde{\sigma})e^{i\tilde{\sigma}\sigma}$. The effect of an arbitrary pair of gravitational and electromagnetic gauge transformations on this integrand is

\begin{equation}\begin{array}{lll}
\psi_{R_{4}}(r_{4},s,\tilde{\sigma})e^{i\tilde{\sigma}\sigma} & \rightarrow & \psi_{R_{4}}'(r_{4}',s,\tilde{\sigma})e^{h(r_{4}')}e^{i\tilde{\sigma}\sigma}=\\
& & \psi_{R_{4}}''(r_{4}',s,\tilde{\sigma})e^{i\tilde{\sigma}\sigma},
\end{array}
\end{equation}
so that the quantity $\tilde{\sigma}$ stays the same. This fact supports the conclusion that $\tilde{\sigma}$ can be associated with a property that is inherent to the specimen, namely its rest energy. It does not depend on the forces that act upon it, the reference frame from which we look at it, or the coordinate system that we use to describe its spatio-temporal location.

The invariance of $\tilde{\sigma}$ means that the condition $\tilde{\sigma}<0$ expressed in (\ref{signconvention}) is still meaningful and necessary. Therefore the general evolution operator should still be expressed as the square of the rest energy operator $(\overline{E_{0}})_{R_{4}}$ according to $\overline{B}_{R_{4}}=b(c \hbar)^{-2}(\overline{E_{0}})_{R_{4}}(\overline{E_{0}})_{R_{4}}$. This means that the Dirac equation (\ref{dirac2}) must still be fulfilled. We just have to generalize the definition of the rest energy operator, and make a corresponding generalization of the definition (\ref{fmodef}) of the four-momentum operator

\begin{equation}
(\overline{P_{4}})_{R_{4}}\equiv-i\hbar\left(\frac{\partial}{\partial x},\frac{\partial}{\partial y},\frac{\partial}{\partial z},\frac{i}{c}\frac{\partial}{\partial t}\right)+(\overline{P_{4}})_{R_{4}}^{(int)},
\label{generalp4}
\end{equation}
so that the operator relation $(\overline{E_{0}})_{R_{4}}(\overline{E_{0}})_{R_{4}}=c^{2}(\overline{P_{4}})_{R_{4}}\cdot (\overline{P_{4}})_{R_{4}}$ remains true, and the evolution equation can still be expressed as in (\ref{ev3}) in the natural parametrization.

The possibility to divide all property operators such as $(\overline{P_{4}})_{R_{4}}$ into a free term and an interaction term arises from the fundamental assumption that each observer can subjectively distinguish between straight and curved trajectories. This epistemic assumption may seem self-evident, but it reflects an essential part of the structure of space-time, as we see it. It is necessary in order to formulate Desideratum 2, which we used as a starting point for the discussion in this section.

\section{Some basic consequences}
\label{basiccon}

The only assumption we made about the specimen in deriving the Dirac equation is that its spatio-temporal position can be specified by a single number $r_{4}$. Thus it should hold for all non-composite minimal objects, such as all elementary fermions. This means that they must all have spin $1/2$.

What about elementary bosons? In the present approach to physics they are not seen as objects or quasiobjects, but rather as bookkeeping devices used to keep track of interactions between such objects or quasiobjects \cite{ostborn2}. Therefore elementary bosons can never be treated as a specimen in an experimental context $C$, and therefore the Dirac equation does not apply.

Whereas the stationary Schr\"odinger equation describes the wave function of a specimen with a precisely known energy, the Dirac equation describes the wave function a specimen with a precisely known rest energy. However, the precise value of this rest energy is never known. The reason is a new commutation relation involving rest energy that arises because of the two-fold description of time employed in this study. This relation has to be added to the two basic commutation relations that appear in conventional quantum mechanics.

\begin{equation}\begin{array}{rcl}
\left[x,\overline{P_{x}}\right] & = & i\hbar\\
\left[t,\overline{E}\right] & = & -i\hbar\\
\left[\sigma,(\overline{E_{0}})^{2}\right] & = & 2i\hbar\langle \epsilon\rangle.
\end{array}\label{newcom}
\end{equation}
Here the natural parametrization (\ref{naturalp}) is used in the third new relation, we define $(\overline{P_{x}})_{R_{4}}\equiv-i\hbar\partial/\partial x$, and it is understood that all quantities within brackets should be seen as continuous wave function operators that act on $\Psi_{R_{4}}$. In the same way as usual we may use these relations to derive the uncertainty relations

\begin{equation}\begin{array}{rcl}
\Delta x\Delta p_{x} & \geq & \hbar/2\\
\Delta t\Delta \epsilon & \geq & \hbar/2\\
\Delta\langle t\rangle\Delta \epsilon_{0}^{2} & \geq & \hbar\langle \epsilon\rangle,
\end{array}\label{massuncertain}
\end{equation}
where the chosen natural parametrization made it possible to replace $\Delta\sigma$ with $\Delta\langle t\rangle$ in the third relation.

In the present approach to quantum mechanics, the two first uncertainty relations are analogous and can be interpreted in the same way. Given a context $C$ designed to observe $x$ and $t$, the uncertainty of the values that we are going to see has an inverse relationship to the uncertainty of $p$ and $E$ that we would have seen if we had re-designed the context to observe these two properties before $x$ and $t$. It should be possible to identify the re-designed context with a reciprocal context, as defined in relation to Fig. \ref{Figure14}. The second relation expresses the existence of incomplete knowledge about relational time $t$, allowing for temporal interference on the same footing as spatial interference, as illustrated in Fig. \ref{Figure6}, .

The third uncertainty relation must be interpreted differently. It is analogous to the time-energy uncertainty relation in the conventional approach to quantum mechanics. The common ground is that the first quantity refers to a parameter, whereas the second refers to an observable property. The expected relational time $\langle t\rangle$ until the chosen property $P=R_{4}$ is observed at sequential time $n+m$ is known at the initiation of the experimental context $C$ at time $n$. In other words it is a function of the context. To change $\langle t\rangle$ means that we change the design of the context. We may speak of a family of contexts $C(\langle t\rangle)$ just as we speak of the family $C(\sigma)$. This means that $\Delta\langle t\rangle$ refers to the width of a distribution of different contexts, whereas the quantities $\Delta x$ and $\Delta t$ refer to the uncertainty of which property value will be observed in a given context.

\begin{figure}[tp]
\begin{center}
\includegraphics[width=80mm,clip=true]{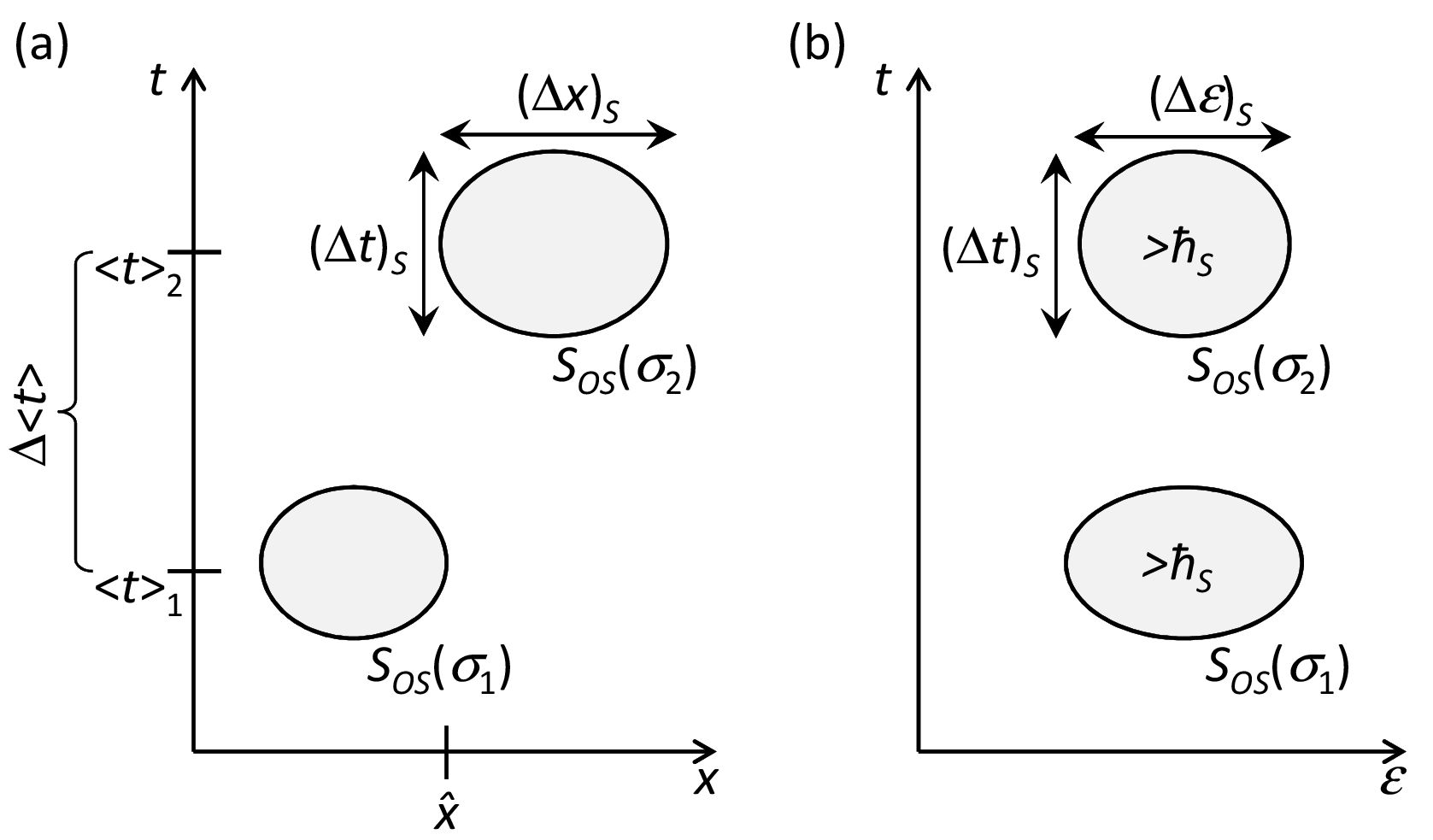}
\end{center}
\caption{(a) The state of the specimen $OS$ in two contexts $C(\sigma_{1})$ and $C(\sigma_{2})$ in the family $C(\sigma)$ just before the observation of its spatio-temporal location $R_{4}$ at time $n+m$. The knowledge about the values of $x$ and $t$ that will be observed are typically incomplete, corresponding to uncertainties $(\Delta x)_{S}>0$ and $(\Delta t)_{S}>0$. In contrast, the expected temporal location $\langle t\rangle$ of the observation is always known beforehand. If we also know that there is some property value $\hat{x}$ of $OS$ that can be observed only if $\langle t\rangle_{1}<\langle t\rangle<\langle t\rangle_{2}$, then the there is an uncertainty $\Delta\epsilon_{0}=\sqrt{\hbar\langle\epsilon\rangle/\Delta\langle t\rangle}$ of the rest energy $\epsilon_{O}$ of $OS$, where $\epsilon$ is its total energy, and $\Delta\langle t\rangle=\langle t\rangle_{2}-\langle t\rangle_{1}$. (b) The uncertainties of energy and time just before the observation of $OS$ fulfil $(\Delta t)_{S}(\Delta\epsilon)_{S}\gtrsim\hbar_{S}$, where $\hbar_{S}$ corresponds to a minimum area of the specimen state $S_{OS}$ when projected onto the energy-time plane.}
\label{Figure15}
\end{figure}

What does this mean in practice? If you know beforehand that there is a considerable chance to observe a given property value $r_{4}$ of the specimen $OS$ only in experimental contexts $C(\langle t\rangle)$ where the observation takes place within the time interval $(\langle t\rangle_{1},\langle t\rangle_{2})=(\langle t\rangle_{1},\langle t\rangle_{1}+\Delta \langle t\rangle)$, then the uncertainty of the rest energy $\Delta \epsilon_{0}$ of the specimen is always greater than $\sqrt{\hbar\langle \epsilon\rangle/\Delta\langle t\rangle}$, where $\langle \epsilon\rangle$ is its total energy [Fig. \ref{Figure15}(a)].

If we know the property value to begin with, and ask whether the specimen is still described by it at a later time, we may set $\langle t\rangle_{1}=0$ and $\Delta\langle t\rangle=T$, where $T$ is the expected life-time of the present state of the specimen. Noting that we always have $\epsilon\geq\epsilon_{0}$, we arrive at the relations

\begin{equation}
\Delta\epsilon_{0}\geq\sqrt{\hbar\epsilon/T}\geq\sqrt{\hbar\epsilon_{0}/T}.
\end{equation}

We conclude that we may have $\Delta\epsilon_{0}=0$ only if the specimen never undergoes any perceivable change at all. In particular, all moving specimens will have an uncertain rest energy. Simply put, the faster things change, the more uncertain are the rest energies of these things.

It follows that we must have $\epsilon_{0}>0$ for all specimens, since $\epsilon_{0}=0$ implies $\Delta\epsilon_{0}=0$. The reason is that the two alternatives $\epsilon_{0}=0$ and $\epsilon_{0}>0$ are incompatible for one and the same specimen. Mass-less and massive entities are qualitatively different and cannot be superposed.

The fact that the present approach to quantum mechanics implies that all specimens are massive also follows directly from the definition (\ref{restmassdef}) of rest energy, in conjunction with the inequality $\tilde{\sigma}<0$ expressed in (\ref{signconvention}). The basic reason why we cannot have $\tilde{\sigma}=0$ and thus $\epsilon_{0}=0$ is that we cannot have $\frac{d}{d\sigma}\langle t\rangle=0$ instead of (\ref{directed}). This condition corresponds to the requirement that time cannot stand still. The conclusion that all specimens therefore must have non-zero rest energy goes for all elementary fermions as well. Note again that in the present approach to physics we do not treat photons and other elementary bosons as possible specimens, which means that the argument does not apply for them.

In this approach, the state $S_{OS}$ in object state space $\mathcal{S}_{O}$ is the basic layer of physical description of a specimen $OS$ studied in an experimental context, rather than the state vector $ \overline{S}_{C}$ in contextual Hilbert space $\mathcal{H}_{C}$. Therefore we should translate the uncertainty relations (\ref{massuncertain}) to statements about $S_{OS}$.

The projection of $S_{OS}$ onto space-time typically has a boundary, as illustrated in Fig \ref{Figure12}. This simply means that we can exclude sufficiently faraway regions in space and time as locations of the specimen during the course of the experiment. Therefore the support of $\Psi_{R_{4}}$ is typically finite. The energy and momentum of the specimen is typically bounded as well, so that the projection of $S_{OS}$ onto momentum-energy also has a boundary, and a momentum-energy wave function $\Psi_{P_{4}}$ has finite support. Therefore it should be possible to define absolute measures of uncertainty $(\Delta x)_{S}$, $(\Delta t)_{S}$, $(\Delta p_{x})_{S}$, and $(\Delta \epsilon)_{S}$, which reflect the diameter of $S_{OS}$ along the corresponding axes in $\mathcal{S}_{O}$, rather than the standard deviations of the wave functions that go into (\ref{massuncertain}). Such measures are shown in Fig. \ref{Figure15}.

We may then define corresponding uncertainty relations $(\Delta x)_{S}(\Delta p_{x})_{S}\gtrsim\hbar_{S}$ and $(\Delta t)_{S}(\Delta \epsilon)_{S}\gtrsim\hbar_{S}$, where $\hbar_{S}$ is a minimum measure on the projection of $S_{OS}$ onto the plane in $\mathcal{S}_{O}$ defined by two conjugate properties. This `state space Planck's constant' $\hbar_{S}$ has to be of the same order of magnitude as $\hbar$. The fact that it is non-zero reflects our basic assumption that potential knowledge is always incomplete, so that the state of an object can never shrink to a single element $Z_{O}$ in object state space $\mathcal{S}_{O}$ \cite{ostborn1}.

The condition $\tilde{\sigma}<0$ in (\ref{signconvention}) motivates the introduction of the Dirac operator $\overline{W}_{R_{4}}$ in (\ref{squareroot2}) as the `square root' of the evolution operator $\overline{B}_{R_{4}}$, as well as the Dirac equation in (\ref{dirac1}), in order to avoid positive values of $\tilde{\sigma}$ in the Fourier expansion (\ref{fexpansion}). In (\ref{restmassop}) we identified  $\overline{W}_{R_{4}}$ as proportional to the rest energy operator. Similarly, the condition $\tilde{t}>0$ in (\ref{signconvention}) might give us the idea to introduce the `square root' of the energy operator, as well as a second Dirac-type equation, in order to ensure that energy $\epsilon=\hbar\tilde{t}$ stays positive. However, such an equation would not be Lorentz invariant, and can therefore not be part of fundamental physical law.

Instead, the problem is solved by the introduction of antimatter. The picture that antimatter corresponds to objects travelling backward in time is particularly appealing from the present perspective. The conclusion that $\epsilon>0$ can be traced back to the condition $d\langle t\rangle/d\sigma>0$. In the same way we may deduce that $\epsilon<0$ if and only if $d\langle t\rangle/d\sigma<0$. This condition provides a clear meaning to the statement that the object travels backward in time. As we subjectively feels that sequential time flows in a forward direction, corresponding to increasing values of $n$ and $\sigma$, we assign earlier relational times $t$ to the object that we track, so that it recedes further into history.

In the absence of the two aspects of time employed in this study one may ask the following question: in relation to what is the antimatter traveling backward in time? I would like to argue that whenever we try to make a picture of what is going on we \emph{implicitly} make use of those two aspects of time that we have tried to explore \emph{explicitly} in this paper. When we reverse the arrow of a world-line in a Feynman diagram we imagine that the object moves downward to smaller values of $t$ as we subjectively feel that time passes in a forward direction as usual.

\section{Discussion}
\label{discuss}

In this paper we have suggested that the `problem of time' might be dissolved if we allow more structure in the conceptualization and formalization of time. We identify two distinct but interwoven aspects of time. Sequential time $n$ is inherently directed, and corresponds to the perceived flow of time, as one event follows after another. Relational time $t$ describes the knowledge at the present point of sequential time about the temporal relations between all events perceived up until now. This means that we postulate an inherent difference between the past and the future, giving `now' a special status.

The idea behind this approach to time is the hypothesis that there is a one-to-one correspondence between the forms of perception and the form of physical law. In order to formulate physical law properly, we must therefore take all these forms of perception into account. This approach to physics seems to be different from that of Einstein. Rudolf Carnap recollected his attitude as follows \cite{carnap}.

\begin{quote}
\emph{Once Einstein said that the problem of the Now worried him seriously. He explained that the experience of the Now means something special for man, something essentially different from the past and the future, but that this important difference does not and cannot occur within physics. That this experience cannot be grasped by science seemed to him a matter of painful but inevitable resignation.}
\end{quote}

Such a resignation can be described as the acceptance that there are degrees of freedom of perception that have no counterpart in the physical world. In other words, it amounts to the belief that man is not firmly rooted in this world, but look down on it from a transcendental viewpoint that cannot be described by physics. Rather than falling prey to such a resignation, a person who believes in the power of science would try to extend the physical formalism to incorporate the experiences that worried Einstein.

Similar extensions to the physical formalism have already proven fruitful. All experiences involve both the observer and the observed, the subjective and the objective. Quantum mechanics incorporates both of these aspects of the perceived world into its postulates. These postulates dissolved several problems that seemed impossible to overcome when only the objective aspect of the world was taken into account within the classical framework, and they led to many new predictions.

With regard to time, the situation has been equally one-sided. But here we have had one established theory - quantum mechanics - which looks only at one side of the temporal coin and treats time as a parameter that formalizes the perceived flow of time and evolves the state, and another theory - relativity - which looks only at the other side of the coin and treats time as an observable that quantifies perceived relations between events. An extension of the framework is needed that take both sides into account. Just as in the case of quantum mechanics we expect that a proper such extension would dissolve old problems and lead to new predictions.

The present attempt in this direction leads to the introduction of rest energy as a quantity conjugate to the parameter $\sigma$, which is used to evolve the state in differential evolution equations. This conjugate relationship implies a new uncertainty relation, from which it follows that the rest energy of a non-stationary object cannot be exactly known. We conclude also that the rest energy of any object must be greater than zero because sequential time cannot stop flowing. The inherent directionality of this flow makes it possible to motivate the Dirac equation, and the general nature of this derivation means that it should apply to all massive pointlike objects, so that all elementary fermions must have spin $1/2$.

The line of reasoning that leads to the Dirac equation relies on the approach to quantum mechanics introduced in Ref. \cite{ostborn1}. One of the goals of that paper was to motivate a one-to-one correspondence between observable properties and certain self-adjoint operators that act in finite-dimensional Hilbert spaces that describe experimental contexts. In this paper we have tried to extend this correspondence to the case where the contexts are represented as infinite-dimensional Hilbert spaces, where properties may take a continuity of values, and the operators act on differentiable wave functions.

A claimed correspondence of this kind led to the above-mentioned identification of the rest energy in the formalism, together with its associated operator. Analogous correspondences led to the identification of energy and momentum, together with their associated operators. In this approach the association between a property and a certain self-adjoint operator does not have to be postulated, but follows from first principles. The same is true for the specific form of the familiar energy and momentum operators in the position representation.

The present approach to quantum mechanics gives a more limited role than usual to differential evolution equations such as the Dirac equation. Rather than describing the evolution of the world at large, they represent a family of experimental contexts $C(\sigma)$ in which a well-defined specimen is studied in a predefined way. The Hilbert space and the wave function becomes contextual and are defined only during the course of the experiment. Also, differential evolution equations are not seen as immediate expressions of fundamental physical law, but represent continuous idealizations of such experimental contexts, in which no more than a finite number of property values can actually be observed.

An appealing feature of the extended formalization of time introduced in this paper is that it enhances the symmetry between space-time and momentum-energy space. The liberation of relational time $t$ from the task of evolving the state means that it can be treated as an observable in quantum theory on the same footing as spatial distance $r$. The pairs of observables $(r,t)$ and $(p,\epsilon)$ become reciprocal mirror images. This symmetry will be explored in upcoming papers. One may imagine, for example, that spatio-temporal distances become discrete in bound states just as the energy spectrum.

The formalism developed in this paper has several similarities to the parametrized quantum theory of Stueckelberg, Horwitz, Piron and others \cite{stueckelberg1,stueckelberg2,fanchi1,fanchi2,horwitz1,horwitz2,land,pavsic}. In both approaches two aspects of time are distinguished, evolution equations become covariant and more symmetric, and relational (relativistic) time together with rest mass are treated as observables with associated operators. However, the evolution parameter $\sigma$ used in the present study has a different meaning than the evolution parameter $\lambda$ used by Stueckelberg his followers, as discussed in relation to (\ref{freeeveq}).

This difference can be traced back to the unconventional approach to quantum mechanics \cite{ostborn1} that is the starting point for the present treatment. The domain of validity of the wave function is limited to a given experimental context $C$, and the evolution parameter $\sigma$ parametrizes a family of such contexts $C(\sigma)$ listed in such an order that the expected time $\langle t\rangle$ passed before the finish of the experiment increases with $\sigma$. This should be contrasted to the conventional approach that is the starting point of Stueckelberg and his followers, where the wave function describes a general physical state, and the evolution parameter $\lambda$ is often related to the proper time of a particle that is part of this state. Since the proper time is a relativistic invariant, $\lambda$ is often called `the invariant evolution parameter'. In contrast, $\sigma$ cannot be called invariant, since its physical role means that relativistic transformations cannot be applied to it in the first place. 

Empirically, the present formalism is more conservative than that of Stueckelberg and his followers. The latter opens up for causal correlations between events with space-like separation, where particle trajectories may even bend backwards in time, and rest mass is not necessarily `on shell', meaning that it does not have to fulfill Einstein's relation $\epsilon_{0}^2=\epsilon^{2}-c^{2}p^{2}$ \cite{fanchi2}. In contrast, the present formalism respects the limits of the light cone, temporal ordering is unambiguous, and rest mass always fulfills $\epsilon_{0}^2=\epsilon^{2}-c^{2}p^{2}$. The latter fact follows from the way in which we identify momentum, energy and rest energy and their associated operators from abstract considerations about `properties' and `wave function property operators'. Even though the values of each of these three properties can display Heisenberg uncertainty, they are always entangled according to Einstein's relation. 

Let me conclude with a philosophical digression. Even an epistemic approach to physics such as the present one must possess an ontology that refers to a transcendental aspect of the world. Physical law itself transcends the set of perceptions that are used to deduce it. We have also argued that sequential time $n$ is transcendental in the sense that we cannot perceive all instants along the sequential time axis. We are stuck at a particular point in time $\hat{n}$ that we call `now', from which we remember past events or read records about them. From this vantage point we construct a space-time $(r,t)$ that describes the inferred structure of this network of events. Relational time $t$ becomes a property that is perceived `now'. As such it is used to give a quantitative description of a `fixed' snapshot of the world. On the other hand, the transcendental nature of sequential time $n$ and the associated evolution parameter $\sigma$ makes them suitable as tools to express transcendental evolution laws.

To accept the existence of other times $n$ requires a leap of faith. Such a leap is necessary in order to arrange these time instants sequentially, and to use this arrangement as a basis of a physical model of evolution, as we have done. A similar leap of faith is needed in order to believe in other subjects. Just as we can only access past times indirectly via perceptions at the present time, we can only access other subjects indirectly via our own perceptions. The analogy is manifest in language, via a certain congruence between the triplets of words (now, then, time) and (I, you, mind).

I would like to argue that the structure of physical law supports these leaps of faith. By taking transcendental sequential time $n$ seriously we are able to derive proper evolution equations like the Dirac equation, and give rest energy a natural place in the formalism. By assuming the need to give equal status to \emph{different} subjects observing the \emph{same} pair of events we arrive at the Lorentz invariance that seems to be fulfilled by all proper physical laws, such as the Dirac equation. It seems that physics speaks against the bleak prospects of temporal and personal solipsism.

\end{document}